\documentclass[aps,twocolumn,superscriptaddress,floatfix,nofootinbib,prx,longbibliography]{revtex4-1}

\usepackage[utf8]{inputenc}
\usepackage{lipsum}
\usepackage{graphicx}
\usepackage{amsmath,amssymb}
\usepackage{braket}
\usepackage{mathtools}
\usepackage[bookmarks=true,colorlinks=true,linkcolor=blue, urlcolor=blue, citecolor=blue]{hyperref}
\usepackage{multirow}
\usepackage{color}
\usepackage[normalem]{ulem}
\usepackage{amsfonts}
\usepackage{bbm}
\usepackage[ruled,vlined]{algorithm2e}
\usepackage{braket}

\newcommand{\bs}[1]{\boldsymbol{#1}}

\makeatletter

\definecolor{OliveGreen}{cmyk}{0.84, 0, 0.95, 0.30}

\usepackage{xcolor}
\newcommand{\andreas}[1]{\textcolor{black}{#1}}
\newcommand{\andreasnew}[1]{\textcolor{OliveGreen}{#1}}

\AtBeginDocument{\let\LS@rot\@undefined}
\makeatother

\begin{document}
\newcommand{\linkstate}[3]{(\ensuremath{#1#2;#3})}
\newcommand{\state}[1]{\ensuremath{| {#1} \rangle}}
\newcommand{\mindex}{\ensuremath{\mathbf{m}}}
\newcommand{\mindexq}{\ensuremath{{\{\vec{m_q}\}}}}
\newcommand{\flag}[1]{\textcolor{red}{#1}}
\newcommand{\avg}[1]{\ensuremath{\left[ #1 \right] }}


\title{ Statistical Mechanics Model for Clifford Random Tensor Networks and Monitored Quantum Circuits}

\author{Yaodong Li} 
\affiliation{Department of Physics, University of California, Santa Barbara, California 93106, USA}

\author{Romain Vasseur}
\affiliation{Department of Physics, University of Massachusetts, Amherst, MA 01003, USA}

\author{Matthew P. A. Fisher}
\affiliation{Department of Physics, University of California, Santa Barbara, California 93106, USA}

\author{Andreas~W.~W.~Ludwig}
\affiliation{Department of Physics, University of California, Santa Barbara, California 93106, USA}

\date{April 4, 2024}

\begin{abstract}

We study entanglement transitions in Clifford (stabilizer) random tensor networks (RTNs) and monitored quantum circuits, by introducing an exact mapping onto a (replica) statistical mechanics model. For RTNs and monitored quantum circuits with random Haar unitary gates, entanglement properties can be computed using statistical mechanics models whose fundamental degrees of freedom (`spins') are permutations, because all operators commuting with the action of the unitaries on a tensor product Hilbert space are (linear combinations of) permutations of the tensor factors (`Schur-Weyl duality'). When the unitary gates are restricted to the smaller group of Clifford unitaries, the set of all operators commuting with this action, called the {\it commutant}, will be larger, and no longer form  a group.
We use the recent full characterization of this
commutant by Gross et al.,  Comm. Math. Phys. 385, 1325 (2021)
to construct statistical mechanics  models for both Clifford RTNs and monitored quantum circuits, for  on-site Hilbert space dimensions which are powers of a prime number $p$. The elements of the commutant form the `spin' degrees of freedom of these statistical mechanics models, and we show that the Boltzmann weights are invariant under a symmetry group involving orthogonal matrices with entries
in the finite number field ${\bf F}_p$ (`Galois field') with $p$ elements. This implies that the symmetry group, and consequently
all universal properties of entanglement transitions in Clifford circuits and RTNs will, respectively, in general depend  on, and only on the prime $p$. 
We show that Clifford monitored circuits with on-site Hilbert space dimension $d=p^M$ are described by percolation in the limits $d\to \infty$ at  (a) $p=$ fixed but $M \to \infty$, and at (b) $M=1$ but $p \to \infty$. In the limit
(a) we calculate the effective central charge, and 
in the limit (b) we derive  the following {\it universal} minimal cut entanglement entropy $S_A =(\sqrt{3}/\pi)\ln p \ln L_A$ for $d=p$ large at the transition. We verify those predictions numerically, and present extensive numerical results for critical exponents at the transition in monitored Clifford circuits for prime number
on-site Hilbert space dimension $d=p$ for a variety of different values of $p$,
\andreas{finding that projective and forced measurement schemes  yield the same critical exponents,
and that they approach percolation values at large $p$.}
\andreas{We clearly establish 
multifractal scaling of the purity, reflected in a continous spectrum of critical exponents, while the typical exponent is the prefactor of the logarithm in the entanglement entropy.}
As a technical result, we generalize the notion of the Weingarten function, previously known for averages involving
the Haar measure, to averages over the Clifford group.

\end{abstract}
\maketitle

{\hypersetup{linktocpage}
\tableofcontents}

\section{Introduction}

Entanglement plays a central role in the physics of closed many-body quantum systems, in both equilibrium and non-equilibrium settings~\cite{RevModPhys.80.517,RevModPhys.81.865,RevModPhys.82.277,Calabrese_2009,LAFLORENCIE20161}. With the advent of noisy intermediate-scale quantum (NISQ) devices~\cite{Preskill2018quantumcomputingin}, the study of the dynamics of quantum information in open quantum systems has attracted a lot of attention recently. The interplay of unitary many-body quantum evolution~\cite{PhysRevX.7.031016,Nahum2018,PhysRevX.8.021013,Zhou2019,PhysRevX.8.031058,PhysRevX.8.031057} -- which generates quantum entanglement -- and the non-unitary processes generated by noisy couplings to the environments or by measurements -- which tend to reveal and destroy quantum information -- leads to a broad variety of new dynamical phases of matter and phase transitions.

A particularly interesting example that captures the competition between unitary dynamics and non-unitary processes is the so-called measurement-induced phase transition~\cite{PhysRevB.98.205136,Skinner2019}. A simple setup where such a transition occurs is provided by ``monitored'' quantum quantum circuits made up of random unitary gates, combined with local projective measurements occurring at a fixed rate. As a function of the measurement probability $p_0$, a remarkable entanglement transition occurs. At low $p_0$, the unitary dynamics can efficiently scramble quantum information into highly non-local degrees of freedom that cannot be accessed by the local measurements~\cite{Choi2020,Gullans2019,Li2020b,Fan2020}. In that regime, the system reaches a highly-entangled steady state where the entanglement entropy of a subsystem scales as its volume (volume law). In contrast, at high measurement probability $p_0$, the frequent local measurements are able to effectively collapse the wave-function of the system, reminiscent of a many-body Zeno effect, with the entanglement of subsystems scaling as their boundary (area law). This fundamentally new transition is not observable in the mixed density matrix of the system averaged over measurement outcomes, but is apparent in individual quantum trajectories of the pure state wave-function, conditional on measurement outcomes. This transition has been studied extensively in recent years,  generalized to various classes of quantum dynamics with different symmetries and dimensionality~\cite{PhysRevB.98.205136,Skinner2019,Li2019,PhysRevB.99.224307,Li2020a,10.21468/SciPostPhys.7.2.024,Gullans2019,Szyniszewski2019,Choi2020,Bao2020,JianVasseurMeasurement2019,Gullans2020,Zabalo2020,PhysRevResearch.2.023288,Ippoliti2020,Lavasani2020,Sang2020,PhysRevResearch.2.013022,PhysRevB.102.064202,Nahum2020,Turkeshi2020,Fuji2020,Lunt, Lunt2020,Fan2020,2020arXiv200503052V,Li2020b,PhysRevB.103.224210,PhysRevLett.126.060501,2021arXiv210306356L,2020arXiv201204666J,PhysRevLett.126.170503,2021arXiv210106245T,2021arXiv210209164B,2021arXiv210413372B,2021arXiv210407688B,2021arXiv210703393Z,2021arXiv210710279A,2021arXiv210804274L,PhysRevLett.126.170602,2021arXiv210609635J,2020arXiv201203857L,2021arXiv210208381B}, and realized experimentally in a trapped-ion quantum computer~\cite{2021arXiv210605881N}. 

An apparently different, but closely related entanglement transition was discovered a bit earlier in random tensor networks~\cite{VasseurPotterRTN2018,PhysRevB.102.064202,Nahum2020,2021arXiv210802225L,yang2021entanglement}. There the transition can be induced by tuning the bond dimension\footnote{Or, at a fixed sufficiently large bond dimension, by randomly diluting (=eliminating) bulk bonds with probably $p_0$, the latter serving as the parameter tuning through the transition~\cite{VasseurPotterRTN2018}}
of a state obtained at the boundary of two-dimensional random tensor networks previously formulated \andreas{deep in the volume law phase, far from the transition}
in~\cite{RTN}.
An unifying description of such entanglement transitions was proposed in Refs.~\cite{VasseurPotterRTN2018,Bao2020, JianVasseurMeasurement2019} in terms of a replica statistical mechanics model. A replica trick is needed to deal with the intrinsic non-linearities of the problem -- the entanglement transition is not visible in quantities that are linear in the density matrix of the system. The average over either Gaussian random tensors or Haar unitary gates in the replicated tensor network
can then be performed exactly, and leads to effective degrees of freedom (`spins') that are permutations of the replicas. There is a deep mathematical reason for the emergence of the permutation group here, which is known as the Schur-Weyl duality: all operators
commuting with the action of the unitary group on a tensor product Hilbert space are linear combinations of permutations of the tensor factors (replicas). This leads to an exact reformulation of the problem of computing entanglement entropies and other observables
in Haar monitored quantum circuits or random tensor networks in terms of (the replica limit of) a two-dimensional classical statistical mechanics model with permutation degrees of freedom, and ferromagnetic interactions. Even if taking the replica limit explicitly is challenging in general, this mapping ends up explaining most (if not all) of the phenomenology of entanglement transitions, which in that language become simple symmetry-breaking (ordering) transitions. In particular, it naturally explains the emergence of conformal invariance (with a dynamical exponent $z=1$) at criticality~\cite{VasseurPotterRTN2018,Li2019, JianVasseurMeasurement2019,Li2020a}. The entanglement entropy maps onto the free energy cost of `twisting' the boundary conditions in the entanglement interval, forcing the insertion of a domain wall at the boundary of the 
\andreas{entanglement interval.}
This free energy cost scales with the size of the interval in the ferromagnetic (symmetry-broken) phase of the statistical mechanics model, corresponding to the volume-law entangled phase, while it remains order one in the paramagnetic phase, corresponding to the area-law phase.  Additional simplifications occur in the limit of large on-site Hilbert space dimension, where the replica limit can be taken analytically, predicting an entanglement transition that falls into the classical percolation universality class~\cite{Bao2020,JianVasseurMeasurement2019}. 

While these analytical results were derived for Haar random unitary gates
or tensors, large scale numerical results are conveniently obtained using random Clifford unitary gates, and single-qubit measurements restricted to the Pauli group. Such Clifford circuits (and the related stabilizer tensor networks~\cite{PhysRevLett.125.241602,yang2021entanglement}) can be efficiently simulated on a classical computer in times polynomial in the system size thanks to the Gottesman–Knill theorem~\cite{PhysRevA.54.1862,1998quant.ph..7006G,PhysRevA.70.052328}, whereas the simulation times of Haar circuits usually scale exponentially with the system size.  Clifford circuits have proven useful in the study of entanglement and operator dynamics in various contexts, see {\it e.g.}~\cite{PhysRevB.92.024301}.
Numerical evidence indicates that the measurement induced transition in Clifford qubit circuits is in a different universality class from the Haar case, but a theoretical description of the Clifford transition has remained elusive. Part of the difficulty is that while the structure of the Clifford gates allows for efficient classical simulations, the very same structure makes a general theoretical formulation of averages
over Clifford gates more challenging than over featureless Haar gates. In this paper, we leverage recent results on the `commutant' of the Clifford group~\cite{GrossEtAlCMP2021}, which generalize the notion of Schur-Weyl duality to the Clifford case, to derive a replica statistical mechanics description of Clifford monitored quantum circuits and stabilizer random tensor networks for qudits with dimension $d=p^M$ where $p$ is a prime number~\cite{gottesman1999higherdim}.
The elements of the commutant replace permutations to form the `spin' degrees of freedom of the resulting 
statistical mechanics models. We show that the Boltzmann weights are invariant under a symmetry group involving orthogonal matrices with entries in the finite number field ${\bf F}_p$ (``Galois field'') with $p$ elements. This implies that the symmetry group, and all universal properties of entanglement transitions in Clifford circuits and random tensor networks will in general depend on the on-site Hilbert dimension $d=p^M$ of the circuit, but only via the prime $p$, i.e. they
are independent of the power $M$.
This also explains that Clifford circuits and random stabilizer tensor networks (i.e. Clifford random tensor networks) are in different universality classes 
than their Haar counterparts.
In particular, in the latter cases the on-site Hilbert space dimension is a short-distance
feature that does not affect their symmetry~\cite{JianVasseurMeasurement2019} and thus not the universality class (as long as $d$ is finite), in contrast to the Clifford
cases.
Our approach also allows us to derive mappings onto classical percolation in the 
limit of large on-site Hilbert space dimension $d$.
Our predictions are supported by exact numerical simulations of large monitored Clifford circuits for $d=p$ up to $p=997$, 
\andreas{and for $d = p^M$ at a few small values of $M$, which
also show that
projective and forced measurement schemes yield the same
critical exponents for Clifford circuits.}
Our approach also explains some recent numerical results on Clifford random tensor networks~\cite{yang2021entanglement}.
\andreas{Finally, we provide clear evidence that the purity of the reduced density matrix of a finite interval exhibits multifractal scaling at the transition, by numerically investigating the statistical fluctutations of the entanglement entropy about its mean. This implies different scaling exponents for  the averaged and the typical purity (the latter being equal to the prefactor of the logarithm of system size in the entanglement entropy), and is reflected in a continuous spectrum  of critical exponents associated with purity scaling.}

This paper is organized as follows. In Section~\ref{secRTN}, we focus on random tensor networks (RTNs) and extend the notion of Schur-Weyl duality to the Clifford case, relying heavily on Ref.~\cite{GrossEtAlCMP2021}. We describe the structure of the commutant and explain how the statistical mechanics model describing Haar RTNs can be extended to Clifford RTNs. In Section~\ref{SecMeasurements}, we generalize the notion of Weingarten functions to the Clifford case, and apply those results to monitored Clifford circuits. We also make concrete predictions in the limit of large on-site Hilbert space, carefully distinguishing the cases of $p$ or $M$ fixed when
$d = p^M \to \infty$. 
In Sec.~\ref{SecNumerics}, we provide numerical evidence that the universality class of the entanglement transition at qudit dimension $d=p^M$ indeed depends on $p$, but not on $M$\andreas{, report on the equality of exponents for projective and forced measurement schemes,  and establish multifractal scaling of the purity}.
Technical details and additional mathematical results are gathered in appendices.

\section{Random Tensor Networks and Schur-Weyl Duality \label{secRTN}}

\begin{figure}[b]
    \centering
    \includegraphics[width=.48\textwidth]{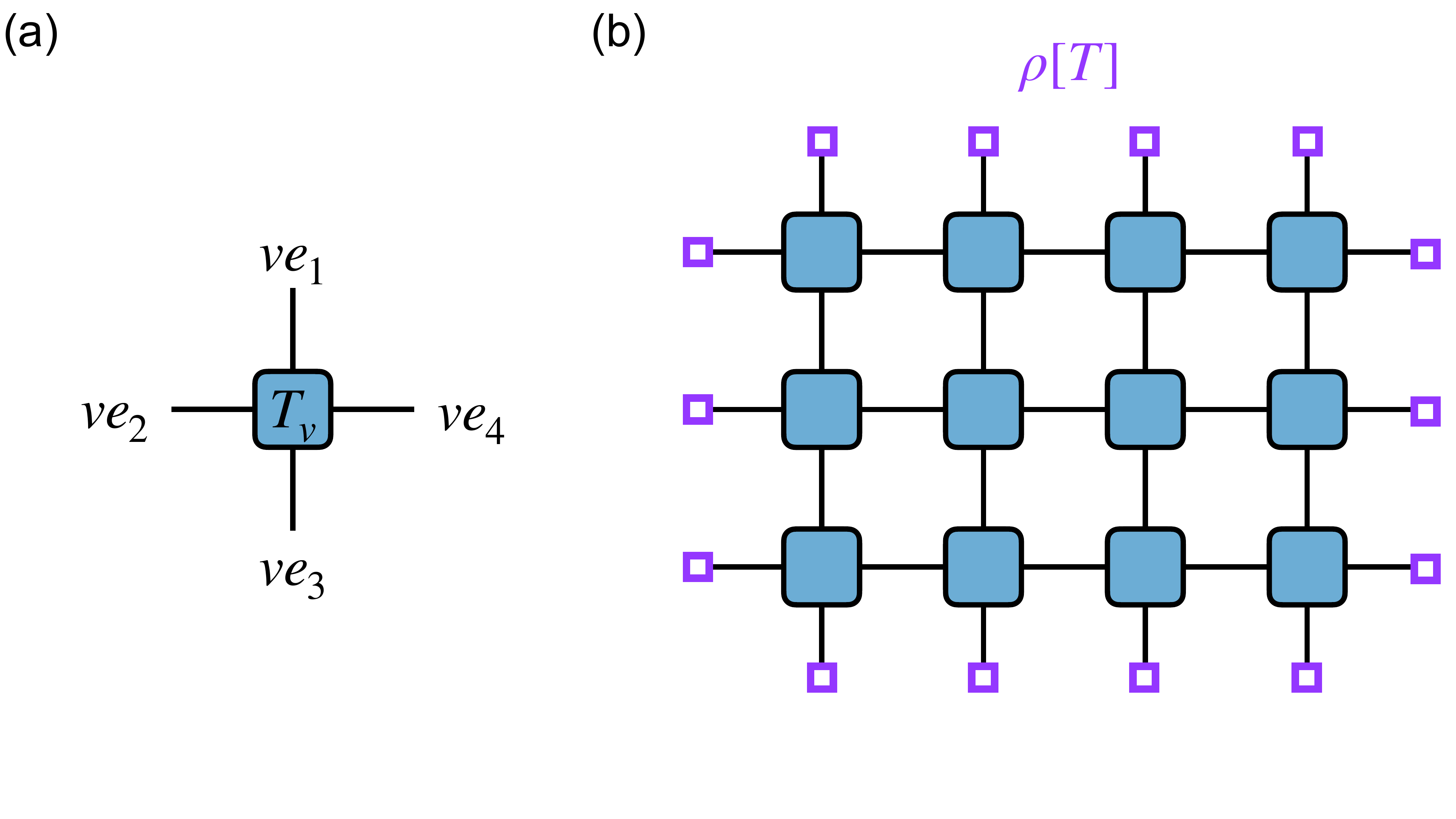}
    \caption{(a) A tensor $T_v$ at vertex $v$ with coordination number $z = 4$.
    (b) Arranging the tensors in a network (taken to be square lattice here), we obtain a ``tensor network state'' $\rho[T]$ (see  Eq.~\eqref{LabelEqRTNDensityMatrix}) by contracting indicies on all bulk edges, as implemented by projecting onto the maximally entangled state $\state{I_e}$.
    The state $\rho[T]$ is supported on the boundary qudits (highlighted by purple squares), and is in general unnormalized.
    }
    \label{fig:TN}
\end{figure}

\subsection{Review of Haar Case}
\label{LabelSubSectionReviewOfHaar}

Following~\cite{VasseurPotterRTN2018} we consider a Tensor Network with a set of vertices $v$, from each   of which 
emerge  $z$
edges $ve_1, ve_2, ..., ve_z$, where $z$ is the coordination of the vertex (assumed to be the same for all vertices), see Fig.~\ref{fig:TN}(a).
We denote the bond dimension at each edge by
$d$.
At each vertex $v$ there is   a quantum state described by a tensor
\begin{eqnarray}
\label{LabelEqHaarTensorVertex}
&& |T_v\rangle
=
\sum_{\mu_{ve_1}, \mu_{ve_2}, ..., \mu_{ve_z}=1}^d
 \ \ \  
 {\left ( T\right )}_{\mu_{ve_1}, \mu_{ve_2}, ..., \mu_{ve_z}} \times \ \  \ \
 \\ \nonumber
 && 
 \times |\mu_{ve_1}\rangle \otimes |\mu_{ve_2}\rangle \otimes ... \otimes |\mu_{ve_z}\rangle
 =
 \sum_{\mu_v=1}^D
 \ \ 
 T_{\mu_v}
 \ \ |
 \mu_v  \rangle,
\end{eqnarray}
where we introduced the shorthand 
$\mu_v=$ $(\mu_{ve_1}, \mu_{ve_2}, ..., \mu_{ve_z})$
for the  collection of  labels of the $D:=(d)^z$ basis elements at a vertex, $\mu_v =1, ...D$.
Defining a maximally entangled state on the edge $e := \langle v, v' \rangle$ connecting two adjacent vertices $v$ and $v'$,
\begin{eqnarray}
\label{LabelEqMaxEntangledEdge}
|I_e\rangle
:=
{1 \over \sqrt{d}} \sum_{\mu_{ve}, \mu_{{v'}e}=1}^d 
\delta_{\mu_{ve}, \mu_{{v'}e}}
\ \ \ 
|\mu_{ve}\rangle \otimes |\mu_{{v'}e}\rangle,
\end{eqnarray}
the (unnormalized) wave function of the Tensor Network is defined by
\begin{eqnarray}
\label{LabelEqRTNWaveFunction}
|\psi[T]\rangle
:=
\left(\otimes_e \langle I_e|
\right)
\left (\otimes_v |T_v\rangle
\right )
\end{eqnarray}
where the tensor product is over all  (bulk) edges and all vertices of the network, respectively. 
The object of interest is the (so-far still unnormalized) pure state density matrix of the network
\begin{eqnarray}
\label{LabelEqRTNDensityMatrix}
\rho[T] := 
|\psi[T]\rangle
\  \langle \psi[T]|,
\end{eqnarray}
for the remaining \andreas{(uncontracted)} degrees of freedom (`dangling legs' of the tensor network), which we will choose to be at the boundary of the  network, see Fig.~\ref{fig:TN}(b). 
Owing to the need to normalize the states of the 
wave function \andreas{at the boundary} of the network (and to describe the $n$-th R\'enyi entropies
for any integer $n$), it is necessary to work with the tensor product of $Q=nm$ copies  of  this (unnormalized) density matrix
\begin{eqnarray}
\label{LabelEqHaarReplicaPowerDensityMatrix}
\otimes^Q 
\rho[T]=
\otimes^Q
\Big (
|\psi[T]\rangle 
\  \langle \psi[T]| \Big ).
\end{eqnarray}
More precisely, our goal is to compute the $n$th entanglement Renyi entropy of a subregion $A$ of the dangling (`boundary') legs of the tensor networks
\begin{equation}
\label{LabelEqRenyiEntropy-Haar}
S_A^{(n)} = \frac{1}{1-n}  \ln \frac{{\rm tr} \rho^n_A}{({\rm tr} \rho)^n},
\end{equation}
with $\rho_A= {\rm tr}_{\overline{A}} \rho$ the reduced density matrix in the region $A$, and $\bar{A}$ the complement of $A$. The average of this quantity over the random tensors, to be defined more precisely below, can be computed from the replicated density matrix using
\begin{equation}
\label{LabelEqRenyiEntropy-Haar-Replicas}
S_A^{(n)} = \frac{1}{1-n} \underset{m \to 0}{\rm lim} \frac{1}{m} \left( ({\rm tr} \rho^n_A)^m - ({\rm tr} \rho^{\otimes Q})  \right),
\end{equation}
where $({\rm tr} \rho^n_A)^m$ and ${\rm tr} \left ( \rho^{\otimes Q}\right )$ differ only in the way the boundary legs are contracted.


The Haar Random Tensor Network is defined by letting the state
$|\mu^{(i)}_v\rangle$
 at each 
vertex $v$ in each (replica and R\'enyi) copy $i=1, ..., Q$ 
arise from a fixed basis state in the $D$-dimensional Hilbert space associated with the vertex,
 by the action of the same unitary $D\times D$ matrix\footnote{Recall that
$d$ is the bond dimension of the Hilbert space on each edge, and $D=d^z$ ($z=$coordination number)
is the dimension of the total Hilbert space associated with each vertex $v$.}
$U$  in each (replica and R\'enyi) copy, 
chosen randomly  and independently at 
each vertex  $v$ from the Haar ensemble.
Explicitly, the tensor in (\ref{LabelEqHaarTensorVertex}) will then be   of the 
form
\begin{eqnarray}
\label{LabelEq-Haar-RandomTensors}
 T_{\mu_v} = \bra{\mu_v} \hat{U} \ket{1} \coloneqq {(U)}_{\mu_v, 1}.
 \qquad \mu_v=1, ...,D.
\end{eqnarray}
The tensor factor contribution to
 (\ref{LabelEqHaarReplicaPowerDensityMatrix}) arising from the quantum states
 (\ref{LabelEqHaarTensorVertex})
 at vertex $v$ will 
 then be, using (\ref{LabelEqRTNWaveFunction}):
\begin{eqnarray}
\label{LabelEqHaarAverageOneVertex}
&&
\otimes^Q 
\Big( |T_v\rangle 
\ 
 \langle T_v|
\Big )=
\otimes^Q 
\Big( {\hat U} |1\rangle 
\ 
 \langle 1| {\hat U}^\dagger
\Big )=
\\ \nonumber
&&
=\sum_{\mu_v^{(i)},{\nu}_v^{(j)}}
{U}_{{ \mu}^{(1)}_v,1}
...
{U}_{{ \mu}^{(Q)}_v, 1}
{U^\dagger}_{1, {\nu}^{(1)}_v}
...
{U^\dagger}_{1, {\nu}^{(Q)}_v} \times
\\ \nonumber
&&
\times 
 (|\nu_v^{(1)}\rangle 
 \otimes ... \otimes |\nu_v^{(Q)}\rangle)
 \otimes
 (\langle {\mu}_v^{(1)}| 
 \otimes ... \otimes \langle {\mu}_v^{(Q)}|).
\end{eqnarray}
Specializing the familiar result\footnote{A derivation is also provided in
(\ref{LabelEqExpansionTensorProductCommutants}) of Appendix \ref{AppendixWHaar}} for the average 
$
\mathbb{E}_{U\in U(D)} \left [ ... \right ]
$
over the Haar measure $d \mu(U)$~\cite{Collins1}
\begin{eqnarray}
\nonumber
&&\int_{\rm Haar} d \mu(U) \ \ 
U_{i_1, j_1}
...
U_{i_Q, j_Q}
U^\dagger_{j'_1, i'_1}...U^\dagger_{j'_Q, i'_Q}=
\\  \nonumber
&&=\sum_{\sigma, \tau\in S_Q}
\ \ 
{\rm Wg}(\tau\sigma^{-1})
\ \ 
\delta_{i^{}_1, i'_{\sigma(1)}}
...
\delta_{i^{}_Q, i'_{\sigma(Q)}}
\delta_{j^{}_1, j'_{\tau(1)}}
...
\delta_{j^{}_Q, j'_{\tau(Q)}}
\\ 
\label{LabelEqWeingartenFunctionInComponents}
\end{eqnarray}
to the case where
$j_1=...=j_Q=1$ and $j'_1=...=j'_Q=1$, yields
%
\begin{eqnarray}
\nonumber
&&
\mathbb{E}_{U\in U(D)}
\left [U_{i_1,1}
...
U_{i_Q,1}
U^\dagger_{1, i'_1}...U^\dagger_{1,i'_Q}\right ]=
\\ \nonumber
&& ={1\over {\cal N}_H} \sum_{\tau\in S_Q}
\ \ 
\delta_{i^{}_1, i'_{\tau(1)}}
...
\delta_{i^{ }_Q, i'_{\tau(Q)}},
\end{eqnarray}
where 
${1\over {\cal N}_H} =$ 
$\sum_\sigma {\rm Wg}_D (\sigma) = \prod_{k=1}^Q (D+k-1)^{-1}$. 
This implies that the Haar average of (\ref{LabelEqHaarAverageOneVertex}) 
gives\footnote{The sum over permuations $\tau$ was replaced by the sum over $\tau^{-1}$, which makes no difference.}
a contribution coming from the vertex $v$ of the form
\begin{eqnarray} 
\label{LabelEqHaarSumRtau}
\mathbb{E}_{U\in U(D)}
\bigg [
\otimes^Q
\Big (
|T_v\rangle  \langle T_v|
\Big )
\bigg ]
=
{1\over {\cal N}_H}
\sum_{\tau \in S_Q} {\hat R}_v(\tau)
\end{eqnarray}
where
\begin{eqnarray}
\label{eqRmatrixHaar}
 {\hat R}_v(\tau)
:=
\sum_{\vec{\bs{\mu}}_v}
\ \ |{\hat \tau}({\vec{\bs\mu}_v})\rangle \ \ \langle {\vec{\bs\mu}_v}|,
\qquad \tau \in S_Q
\end{eqnarray}
and where $\vec{\bs{\mu}}_v=$
$({\mu}_v^{(1)}, {\mu}_v^{(2)}, ...,  {\mu}_v^{(Q)})$
labels an orthonormal basis of the $Q$  (replica and R\'enyi) copies of the $D$-dimensional
Hilbert space, $({{\bf C}^D})^Q$, at vertex $v$. Here
$S_Q$ denotes the permutation group of $Q$ elements, and
${\hat \tau}$ is the operator that permutes the $Q$ factors of the tensor product,
\begin{eqnarray}
\nonumber
&|\vec{\bs\mu}_v\rangle &:=
|{\mu}_v^{(1)}, {\mu}_v^{(2)}, ...,  {\mu}_v^{(Q)}\rangle,
\\  
\label{LabelEq-Vec-Boldface-Mu}
&
|{\hat \tau}(\vec{\bs\mu}_v)\rangle &:=
|{\mu}_v^{(\tau(1))}, {\mu}_v^{(\tau(2))}, ...,  {\mu}_v^{(\tau(Q))}\rangle.
\end{eqnarray}
The operator ${\hat R}_v(\tau)$ in 
(\ref{eqRmatrixHaar})
is simply the permutation operator permuting the $Q$ (replica and R\'enyi) copies
of the vector space ${\bf C}^D$ at vertex $v$.
Equation (\ref{LabelEqHaarSumRtau}) is the key result for the Haar case.
The
operator
${\hat R}_v(\tau)$ in (\ref{eqRmatrixHaar})
commutes with the action of all unitaries $U\in U(D)$ on the tensor product $({\bf C}^D)^Q$
associated with a vertex. Viewing the set of operators
${\hat R}(\tau)$
where $\tau$ runs over all permutations in the
group $S_Q$ as a basis of a (complex)  linear vector space\footnote{Which is in fact
an algebra because we can multiply the operators.}, it can be shown that this vector space of
operators forms the set of all operators commuting with the action of the unitaries
on the tensor product $({\bf C}^D)^Q$.
This set of operators
is called the {\it commutant} of action of the unitary group on the tensor product, and the stated
result is known as the famous statement of {\it Schur-Weyl
duality}.


Consider now a single edge $e$ emanating from a vertex $v$.
We label an  orthonormal basis of the $Q$  (replica and R\'enyi) copies  
of the $d$-dimensional Hilbert space of edge $e$
at vertex $v$ 
by $\vec{\mu}_{ve}=$
$(\mu_{ve}^{(1)}, \mu_{ve}^{(2)}, ..., \mu_{ve}^{(Q)})$,
corresponding to the kets
\begin{eqnarray}
|\vec{\mu}_{ve}\rangle :=
|{\mu}_{ve}^{(1)}, {\mu}_{ve}^{(2)}, ...,  {\mu}_{ve}^{(Q)}\rangle.
\end{eqnarray}
Recalling the decomposition from
(\ref{LabelEqHaarTensorVertex}),
\begin{eqnarray}
| \mu_v  \rangle=
|\mu_{ve_1}\rangle \otimes |\mu_{ve_2}\rangle \otimes ... \otimes |\mu_{ve_z}\rangle,
\end{eqnarray}
valid in each (replica and R\'enyi) copy, as well as
(\ref{LabelEq-Vec-Boldface-Mu}), we see that the operator in 
(\ref{eqRmatrixHaar}) is the tensor product of operators
\begin{eqnarray} 
\label{LabelEqHaarSumRtau-vertex-edge}
 {\hat R}_{e}(\tau),
\end{eqnarray}
defined on the edges $e$ of the vertex $v$,
where
\begin{eqnarray}
\label{eqRmatrixHaarvertex-vertex-edge}
 {\hat R}_{e}(\tau)
:=
\sum_{\vec{\mu}_{ve}}
\ \ |{\hat \tau}({\vec{\mu}_{ve}})\rangle \ \ \langle {\vec{\mu}_{ve}}|,
\qquad \tau \in S_Q
\end{eqnarray}
and ${\hat \tau}$ labels\footnote{Upon some slight abuse of notation.} 
the operator that permutes the $Q$ (replica and R\'enyi) copies  of the $d$-dimensional
Hilbert space, $({{\bf C}^d})^Q$, at an edge $e$ at vertex $v$
\begin{eqnarray}
\nonumber
&
|\vec{\mu}_{ve}\rangle &:=
|{\mu}_{ve}^{(1)}, {\mu}_{ve}^{(2)}, ...,  {\mu}_{ve}^{(Q)}\rangle
\\
\label{LabelEqDefActionOfPermutationsEdge}
&
|{\hat \tau}(\vec{\mu}_{ve})\rangle &:=
|{\mu}_{ve}^{(\tau(1))}, {\mu}_{ve}^{(\tau(2))}, ...,  {\mu}_{ve}^{(\tau(Q))}\rangle.
\end{eqnarray}
In other words, we have
\begin{eqnarray}
 {\hat R}_v(\tau)
 =
\bigotimes_{e =1}^{z}
{\hat R}_{e}(\tau)
 =
 [ {\hat R}_{e}(\tau)]^{\otimes z}.
\end{eqnarray}

Consider now two adjacent vertices $v$ and $v'$  connected by an edge $e$.
The contribution to (\ref{LabelEqHaarReplicaPowerDensityMatrix})
arising from the edge joining adjacent vertices $v$ (permutation $\tau$) and $v'$
(permutation $\sigma$)
is then from (\ref{LabelEqRTNWaveFunction}) and (\ref{LabelEqHaarReplicaPowerDensityMatrix})
\begin{eqnarray}
\label{LabelEq-Haar-edge-BoltzmanWeight}
&&
\left ( \otimes^Q \langle I_e|\right )
\ 
{\hat R}_{e}(\tau) {\hat R}_{e}(\sigma) 
\ 
\left ( \otimes^Q
|I_e\rangle
\right )=
\\ \nonumber
&&
=
{1\over d^Q}
\sum_{\vec{\mu}_{ve}, \vec{\mu}_{v'e}}
\ 
[{\hat R}_{e}(\tau)]_{\vec{\mu}_{ve}, \vec{\mu}_{v'e}}
[{\hat R}_{e}(\sigma)]_{\vec{\mu}_{v'e}, \vec{\mu}_{ve}}=
\\ \nonumber
&&
=
{1\over d^Q}
{\rm tr}
\left [
{\hat R}_{e}(\tau) {\hat R}^\dagger_{e}(\sigma) 
\right]
=
{1\over d^Q}
{\rm tr}
\left [
{\hat R}_{e}(\tau \sigma^{-1}) 
\right].
\end{eqnarray}
It is seen by inspection\cite{VasseurPotterRTN2018}
(using definition (\ref{eqRmatrixHaarvertex-vertex-edge}))
that
\begin{eqnarray}
\label{LabelEqHaarEdgeBoltzmannWeight}
{\rm tr}
\left [
{\hat R}_{e}(\tau \sigma^{-1}) 
\right]
=
e^{ \ \ln(d) \  C(\tau\sigma^{-1}) \ },
\end{eqnarray}
where $C(\tau)$ denotes the number of cycles in the permutation $\tau\in S_Q$.
Once the product over the analogous contributions from all edges and vertices 
is taken, a sum over all independent permutations $\sigma_v$  in the group
$S_Q$ at all vertices $v$  has to be performed.
This yields the (bulk) partition function of the Haar Stat Mech model~\cite{VasseurPotterRTN2018},
whose `spins' are the elements of the permutation group located at the vertices,
\begin{eqnarray}
\label{LabelEq-Haar-RTN-PartitionFunction}
&&
Z_\emptyset=\mathbb{E}_{U\in U(D)} \ {\rm tr} \rho^{\otimes Q} =
\prod_{\{\sigma_v \}} \nonumber
\ \ \exp\{ -  \sum_{\langle v, v'\rangle} \  {\cal H}(\sigma_v, \sigma_{v'}) \}, \nonumber
\\ 
&&
{\cal H}(\sigma_v, \sigma_{v'})=
\ln(d) |\sigma_v, \sigma_{v'}| :=
\ln(d) \ 
[Q-C(\sigma_v \sigma^{-1}_{v'}], \ \ \ 
\end{eqnarray}
where we can think of $|\sigma_v, \sigma_{v}|$, defined above,
as a `metric' on permutations. --
We now see  that the fact that
the degrees of freedom (the `spins') of the RTN model of random  Haar tensors are the elements of the
permutation group arises from the fact that the permutations form a basis of the commutant of the action of the
unitaries on the tensor product Hilbert space. The permutation degrees of freedom at the end of the dangling boundary legs are fixed, and are dictated by the trace under consideration. In order to compute $Z_{\emptyset}=\mathbb{E}_{U\in U(D)} \ {\rm tr} \rho^{\otimes Q}$, the boundary permutations are fixed to the identity, corresponding to contracting each replica with itself. On the other hand, one can compute $Z_{A}=\mathbb{E}_{U\in U(D)} ({\rm tr} \rho^n_A)^m $ by fixing the same identity permutation in the region ${\bar A}$, while fixing $g_{\rm SWAP} = (1 2 \dots n)^{\otimes m}$ in the region $A$ in order to compute the partial trace. 

In the sequel, we will identify, using the same logic, the degrees of freedom (i.e. the `spins') of the RTN model
for random Clifford tensors~\cite{PhysRevLett.125.241602,yang2021entanglement}. Analogously, they  will form a basis of the commutant 
of the action of the Clifford unitaries
on the $Q$-fold tensor product. Since
Clifford unitary tensors form a subgroup of all (Haar) unitary tensors, the set of operators that will
commute with their action on the tensor Hilbert space will be larger. In other words, the commutant 
will
be larger as compared to the Haar case. We will now proceed to explain the nature of this larger commutant, and thus of the spins
of the Stat. Mech. model for the RTN model of random Clifford tensors.

\subsection{Schur-Weyl Duality for the Clifford RTN model}
\label{LabelSubsectionCliffordRTN}
Throughout this subsection (and any discussion of the RTN or  monitored quantum circuits
with random Clifford unitaries in this paper) we will
take the on-site Hilbert space dimension  $d$ at each edge to be a power of a prime number $p$,
i.e. $d=p^M$ with some integer $M$.

The difference with the case of Haar unitary tensors appears in (\ref{LabelEq-Haar-RandomTensors}): In the
Clifford case the matrix $U$ is replaced by a unitary matrix, which we will denote by $V$, in the Clifford {\it subgroup} 
${\rm Cliff}(N,p)$ of all unitary $D\times D$ matrices $U(D)$, where $D = d^z = p^{z M}=p^N$, and $N=z M$. 
First, we will use this setup to describe the commutant of this action
on the $Q$-fold
tensor product space.  Second, after the description of the commutant is completed, we will consider
the generalization of the average
(\ref{LabelEqHaarSumRtau}) to 
an average over all the elements of this Clifford {\it subgroup}
${\rm Cliff}(N,p)$
acting on a fixed tensor $|T_v\rangle$ which is now chosen to be a stabilizer state.

For the first step, the key statement of Schur-Weyl duality for the Clifford group 
is that there is a specific generalization of the operators
(\ref{eqRmatrixHaar}) spanning the commutant, which are no longer characterized by permutations.

\subsubsection{Description of  properties of the Commutant}
Specifically, we will be interested in the basis $|\vec{\bs\mu}_v\rangle$ of the $Q$-fold tensor product space $({\bf C}^D)^Q$
of all $z$ edges\footnote{As in
(\ref{LabelEq-Vec-Boldface-Mu}).}
emerging from a given vertex $v$, so that
we can write
$({\bf C}^D)^Q=$
$(({\bf C}^p)^{\otimes N})^{\otimes Q}$ where $N=M \cdot z$. The $p$ basis elements of the vector space ${\bf C}^p$
can be labeled by the $p$ elements of ${\bf Z}_p :=$
${\bf Z}/(p \cdot {\bf Z})=$ $0, 1, 2, ..., (p-1)$ i.e. by the set of integers modulo $p$ (the ``computational basis''). It will
soon become important
{that, when $p$ is a prime,}
the 
\andreas{finite}
set ${\bf Z}_p$
forms a number field\footnote{Meaning that addition, multiplication and division are defined in the usual
way.} which is often  also denoted by ${\bf F}_p={\bf Z}_p$.

Let us first consider the  simplest case where $N=1$, i.e.
of the $Q$-fold tensor product $({\bf C}^p)^{\otimes Q}$, which
turns out to be the most important case, underlying all others. The same unitary $p\times p$ matrix  $V$
in the Clifford subgroup
${\rm Cliff}(1,p)$
of the unitary group $U(p)$
acts  simultaneously on each of the
$Q$ factors of the tensor product, generalizing (\ref{LabelEq-Haar-RandomTensors}) with $D$ replaced by  $p$.
Since the basis elements of ${\bf C}^p$  are labeled (as above) by elements 
$\mu\in {\bf Z}_p={\bf F}_p$, the elements of a basis
of $({\bf C}^p)^{\otimes Q}$ will be labeled by 
$Q$-dimensional (say: column-) vectors\footnote{The superscript ${}^t$ denotes the transpose.} 
with entries in ${\bf Z}_p=$ ${\bf F}_p$, written as usual
as elements of $({\bf Z}_p)^Q=$ $({\bf F}_p)^Q$, i.e. by  vectors
$\vec{\mu}=$ $(\mu^{(1)}, \mu^{(2)}, ..., \mu^{(Q)})^t$\andreas{, where
$\mu^{(1)}, ..., \mu^{(Q)} \in$ $\{0, 1, ..., (p-1)\}=$ ${\mathbf F_p}$}.
In other words, every $Q$-dimensional (column) vector with entries in ${\bf Z}_p=$ ${\bf F}_p$,
i.e. every (column) vector in  $({\bf Z}_p)^Q=$ $({\bf F}_p)^Q$, denotes a basis element of the
vector space 
$({\bf C}^p)^Q$.

The main result of \cite{GrossEtAlCMP2021} is that  in this case the generalization of
(\ref{eqRmatrixHaar}) and (\ref{eqRmatrixHaarvertex-vertex-edge}), i.e. of the operators
which in the Haar case
 formed a basis of
the commutant, i.e. of  the set (actually the vector space) of all operators commuting with the action of the unitary group on the tensor product, 
in the Clifford case 
is based on the operators
\begin{eqnarray}
\label{LabelEqClifford-r-DEF}
{\hat r}(T)
=
\sum_{(\vec{\nu}, \vec{\mu}) \in T}
\ |\vec{\nu}\rangle \ \langle \vec{\mu}|,
\end{eqnarray}
where $T$ runs over a  certain set (to be specified below) of subspaces of $({\bf Z}_p)^Q \oplus
({\bf Z}_p)^Q$. In other words, the permutation in the Haar case is now replaced by the (more general)
object $T$. The so-defined ${\hat r}(T)$ is clearly an operator acting on $({\bf C}^p)^Q$. The
set of allowed subspaces
$T$, denoted by $\Sigma_Q(p)$  and to be specified below, forms the commutant in the Clifford case.

In order to appreciate the meaning of (\ref{LabelEqClifford-r-DEF}), let us first look at a special
case: Since the set of all Clifford unitary matrices $V\in {\rm Cliff}(1,p)$ forms a subgroup of the set of all unitary matrices $U\in U(p)$, the set of all operators commuting with the action
of all {\it Clifford}
unitaries
on the
$Q$-fold tensor product of the 
Hilbert space ${\bf C}^p$  must be a larger set than the set of all those commuting with the action of {\it all}
unitaries $U$. Since we know from the previous section on the Haar  case  that the commutant
in the latter case is spanned by all permutations $\tau\in S_Q$, the operators
that appeared in (\ref{eqRmatrixHaar}) and (\ref{eqRmatrixHaarvertex-vertex-edge}) must form a subset of the set of
operators described in (\ref{LabelEqClifford-r-DEF}). Indeed, choosing the subspace of 
$({\bf Z}_p)^Q \oplus
({\bf Z}_p)^Q$ to be of the  {\it special} form
\begin{eqnarray}
\label{LabelEqSubSpace-T-for-Permutation}
T = \{ ({\hat \tau}(\vec{\mu}), \vec{\mu}) : \ \  \vec{\mu} \in({\bf Z}_p)^Q
 \}
\end{eqnarray}
reproduces precisely the previous expression from 
(\ref{eqRmatrixHaar})
and
(\ref{eqRmatrixHaarvertex-vertex-edge}),
where  in the above equation ${\hat \tau}$ acts by permuting the $Q$ rows of the column vector $\vec{\mu}$.

Before we move on to the description of the commutant
and its properties
for the Clifford case , we state an important {\it stabilization property}~\cite{GrossEtAlCMP2021}
 that says that, once the prime number $p$ is fixed,  this commutant does not depend on the dimension of the (on-site) Hilbert space on which it acts, as long as that dimension is large enough. This is a very important statement because it
says that, for a fixed prime $p$, the specific form of the commutant that we describe below is universally valid, independent of the
(sufficiently large) prime-power dimension of Hilbert space it acts on. For this purpose, we consider a general on-site
Hilbert space $({\bf C}^p)^N$ of dimension $D=p^N$ (as above), i.e. simply the $N$-fold tensor product
of the Hilbert space ${\bf C}^p$ discussed above. We now consider as before the tensor product
of $Q$ (replica, and R\'enyi) copies of this $D$-dimensional Hilbert space, i.e.
$[({\bf C}^p)^N]^{\otimes Q}=$
$({\bf C}^p)^{N\cdot Q}$,
and on each  of the $Q$ copies of the $D=p^N$-dimensional vector space $({\bf C}^p)^N$
acts the same
unitary $D\times D$ matrix $V$ in the Clifford subgroup
${\rm Cliff}(N,p)$
of 
the unitary group $U(D)$.
Now the following statement holds:

\vskip .2cm

\noindent -: {\it Stabilization property}~\cite{GrossEtAlCMP2021}

The number of linearly independent operators (which will span the commutant) 
that commute with the action of the same Clifford unitary
matrices $V\in{\rm Cliff}(N,p)$ acting \andreas{simultaneously}
on each of the $Q$ factors of the $Q$-fold tensor product,
$[({\bf C})^D]^{\otimes Q}$, where $D=p^N$ ($p=prime$), is {\it independent of $N$} as long
as $N \geq (Q-1)$.

Moreover, these linearly independent operators are simply
\begin{eqnarray}
\label{LabelEq-DEF-R-T}
{\hat R}(T) = [{\hat r}(T)]^{\otimes N}, \qquad {\rm where} \ T \in \Sigma_Q(p),
\end{eqnarray}
i.e. they are nothing but the $N$-fold tensor products of the operators introduced in
(\ref{LabelEqClifford-r-DEF}) above. Note that these operators are thus uniquely determined by the set $\Sigma_Q(p)$,
independent of the power $N$ determining the Hilbert space dimension $D=p^N$, as long as $N$ is
large enough\footnote{Note that (as briefly reviewed below) in the Stat. Mech. models for the RTN and the
quantum circuits monitored by measurements,
$Q$ is a replica (R\'enyi) index which goes to $Q\to 0$ in the RTN~\cite{VasseurPotterRTN2018}
and goes to $Q\to 1$ in the monitored quantum circuits~\cite{JianVasseurMeasurement2019}.
Thus the inequality required for stabilization will be automatically satisfied in these limits
of interest.}
(as specified above).

The key result of \cite{GrossEtAlCMP2021} is a complete characterization of the set
$\Sigma_Q(p)$
of subspaces $T$ appearing in (\ref{LabelEq-DEF-R-T}),
which describes the commutant of the Clifford action on the $Q$-fold  tensor product.
A more detailed summary  of this is given in App. \ref{LabelAppendixDescriptionAndPropertiesOfCommutant}.
Here we just list those properties of the commutant $\Sigma_Q(p)$ 
that appear directly in the formulation of the Clifford
RTN model. These are a metric on the commutant, and the invariance of this  metric
under a certain symmetry group. We now proceed to discuss these two properties.

An important property of the commutant $\Sigma_Q(p)$  is that it
contains a group as a subset (though,  $\Sigma_Q(p)$ is in general larger
than this group, and does in general not form a group). This group is the group of
all orthogonal $Q\times Q$ matrices $O$ with elements in the number field ${\bf Z}_p=$
${\bf F}_p$ which satisfy besides the ``orthogonality condition'' $O^t O = O O^t={\bs{I}}_Q$
(the $Q\times Q$ identity matrix) 
also the additional condition of ``being stochastic'', meaning that 
the column vector $1_Q$ whose entries are all the number $1 \in {\bf F}_p$, is invariant under the
action of the matrix $O$, 
as well as of its transpose,
i.e.
$O~1_Q = O^t~1_Q=1_Q$.
The set of such matrices forms a 
group
denoted by ${\cal O}_Q(p)$, the 
so-called {\it stochastic orthogonal group}.
Note that [in connection with the discussion around (\ref{LabelEqSubSpace-T-for-Permutation})]
the permutation group $S_Q$ is a subgroup of ${\cal O}_Q(p)$, namely it corresponds to
the set of those $Q\times Q$ matrices  which have only one $1 \in {\bf F}_p$ in each row and  each column;
those  matrices clearly implement  permutations.

For the formulation of the Clifford Stat Mech model we use the fact that there exists a {\it metric}
on the commutant: Namely, that for any two elements $T_a$ and $T_b$ of the commutant $\Sigma_Q(p)$
there exists a {\it metric} 
which arises from the trace
\begin{eqnarray}
\label{LabelEq-MetricRTC-Clifford}
W(T_a, T_b) := {\rm tr} \left [
{{\hat R}^\dagger}(T_a)
{\hat R}(T_b) 
\right ]= (p^{N})^{Q- |T_a, T_b|}, \ \ \ \  
\end{eqnarray}
for any $T_a, T_b \in \Sigma_Q(p)$, 
where ${\hat R}(T)$ is defined in (\ref{LabelEq-DEF-R-T}).
The metric satisfies the usual properties $|T_a, T_b| \geq 0,$
$|T_a, T_b|= |T_b, T_a|$, and $|T_a, T_b|=0$ if and only if $T_a=T_b$. {Addtional details on the definition of this metric are provided in appendix~\ref{appendixMetric}. }

The other property we need is that the metric in (\ref{LabelEq-MetricRTC-Clifford})
is invariant\footnote{See 
(\ref{LabelEqSymmetryOfMetric}) of App. \ref{LabelAppendixDescriptionAndPropertiesOfCommutant}
for details.} under the action of the direct product of two copies of the stochastic orthogonal group ${\cal O}_Q(p)\times{\cal O}_Q(p)$,
acting as
\begin{eqnarray}
\label{LabelEqInvarianceOfMetric-Clifford-RTN}
&&
|T_1, T_2| = |T'_1, T'_2|, \ \ \ {\rm where}  \ \  T_1, T_2 \in \Sigma_Q(p), 
\\ \nonumber
&&
{\rm and}
\\ \nonumber
&&
T'_1=O_L T_1 O_R^{-1}, \ T'_2=O_L T_2 O_R^{-1}, \ {\rm with} \
O_L, O_R \in {\cal O}_Q(p).
\end{eqnarray}


\subsubsection{Stat Mech model for the
Clifford~(Stabilizer)~RTN}

The construction of the Clifford RTN follows the same steps as those  reviewed in
Sect. \ref{LabelSubSectionReviewOfHaar} for Haar unitaries.
We consider the RTN where on each edge there is an
``on-site'' Hilbert space
$({\bf C}^p)^M$
of dimension $d=p^M$
of $M$ qudits, each qudit having a $p$-dimensional Hilbert space where $p$
is a prime number. (For a brief review, see e.g. App. \ref{LabelAppendixPropertiesCliffordGroupStabilizerStates}.)
In complete analogy with Sect. \ref{LabelSubSectionReviewOfHaar},
a Hilbert space $({\bf C}^p)^{M\cdot z}=$
$({\bf C}^p)^{N}$ of dimension $D=p^N$ where $N=M \cdot z$
is associated with each vertex $v$, where $z$ is the coordination
number
of a vertex.

Specifically, in the Clifford case the first line of
(\ref{LabelEqHaarAverageOneVertex}) will be replaced
by
\begin{eqnarray}
\label{LabelEqCliffordOneVertex}
\otimes^Q 
\Big( |T_v\rangle 
\ 
 \langle T_v|
\Big )=
\otimes^Q
\Bigg ( 
\left ( {\hat V} |0\rangle^{\otimes N}\right )
\ 
\left (
 \langle 0|^{\otimes N} {\hat V}^\dagger
\right )
\Bigg), \ \ \ 
\end{eqnarray}
where ${\hat V}$ is an element of the Clifford group ${\rm Cliff}(N,p)$
acting on the Hilbert space $({\bf C}^p)^N$ at vertex $v$.
This means simply that each\footnote{Recall that in the Haar case
$|1\rangle$ is just one of the $D=d^z$ states of an orthonormal basis
$|\mu_v\rangle$, $\mu_v=1, ..., D$ that can occur at a vertex $v$.}
of the states $|1\rangle$
that appear in each of the $Q$ tensor factors
in
(\ref{LabelEqHaarAverageOneVertex})
is replaced by the
state $|0\rangle^{\otimes N}\in {\bf C}^N$ which is written in the computational basis,
where $0  \in {\bf F}_p$ denotes the number zero in the finite number field ${\bf F}_p$. [See, e.g., the
paragraph preceeding that containing (\ref{LabelEqClifford-r-DEF}), or
App. \ref{LabelAppendixPropertiesCliffordGroupStabilizerStates}.]
The reason why $|0\rangle^{\otimes N}$ is chosen at each vertex is because this
state is a stabilizer state.\footnote{In fact any stabilizer state could be chosen
at a vertex $v$ - see App. \ref{LabelAppendixPropertiesCliffordGroupStabilizerStates}.}
Owing to the fact that this state is a stabilizer state, the average of 
(\ref{LabelEqCliffordOneVertex}) over the Clifford group 
at a vertex $v$ is (as shown
in \cite{GrossEtAlCMP2021})
an equal-weight superposition
over  elements of the commutant [see, e.g.,  (\ref{LabelEqSumOverClifford})
of App. \ref{LabelAppendixPropertiesCliffordGroupStabilizerStates}]:
\begin{eqnarray}
\nonumber
&&
\mathbb{E}_{{\hat V} \in {\rm Cliff}(N,p)}
\left [
\otimes^Q
\Bigg ( 
\left ( {\hat V} |0\rangle^{\otimes N}\right )
\ 
\left (
 \langle 0|^{\otimes N} {\hat V}^\dagger
\right )
\Bigg)
\right ]
=
\\  
\label{LabelEqStatMechCliffordEqualWeightSumVertex}
&&
={1\over Z_{n,p,Q} }
\sum_{T_v \in \Sigma_Q(p)} {\hat R}_v(T_v)
\end{eqnarray}
where
\begin{eqnarray}
\nonumber
{\hat R}_v(T)
=
\left [{\hat R}_e(T)\right ]^{\otimes z},
\  {\rm with} \ 
{\hat R}_e(T) = \left [{\hat r}(T) \right 
]^{\otimes M}.
\end{eqnarray}
Then, in complete analogy with (\ref{LabelEq-Haar-edge-BoltzmanWeight}), one obtains
using a maximally entangled state in an orthonormal basis on each edge, a contribution to the partition function
from that  edge of the form
\begin{eqnarray}
\nonumber
{1\over (p^M)^Q}
{\rm tr}
\left [
{\hat R}^\dagger_{e}(T_v)
{\hat R}_e(T_{v'})
\right ]
=
(p^M)^{-|T_v,T_{v'}|},
\end{eqnarray}
where $T_v$ and $T_{v'}$ denote elements of the commutant from the equal-weight sums
(\ref{LabelEqStatMechCliffordEqualWeightSumVertex})
at the two vertices $v$ and $v'$ connected by the edge $e$.
Then we obtain, 
by collecting all terms from the equal-weight sums appearing at each edge, 
in complete analogy to (\ref{LabelEq-Haar-RTN-PartitionFunction}),
the following  (bulk) partition function
of the Clifford RTN model
\begin{eqnarray}
\label{LabelEq-Clifford-RTN-PartitionFunction}
&&
Z_\emptyset=\mathbb{E}_{{\hat V} \in {\rm Cliff}(M,p)} \ {\rm tr} \rho^{\otimes Q} =
\\  \nonumber
&&
=\prod_{\{T_v\in \Sigma_Q(p) \}} 
\label{LabelEq-Clifford-RTN-Hamiltonian}
\ \ \exp\{ -  \sum_{\langle v, v'\rangle} \  {\cal H}(T_v, T_{v'}) \}, \nonumber
\\ 
&&
{\cal H}(T_v, T_{v'})=
\ln(p^M) \
|T_v, T_{v'}|.
\end{eqnarray}
Observables, including the entanglement entropy, 
are formulated in the same way as in the Haar
case. [See  (\ref{LabelEqHaarReplicaPowerDensityMatrix}), (\ref{LabelEqRenyiEntropy-Haar}),
(\ref{LabelEqRenyiEntropy-Haar-Replicas}), and the discussion below (\ref{LabelEq-Haar-RTN-PartitionFunction}).]

We recall from \cite{VasseurPotterRTN2018}\footnote{See Sect. IV F of Ref.~\cite{VasseurPotterRTN2018}.}
that the transition in the RTN model can  be driven by making 
(at each edge)
the bond dimension
random. In the Clifford case, this means that the bond dimension
is $D_{\rm bond} = p^M$ 
with $p={\rm prime}=$
fixed, where $M$ is  independently distributed at each edge according to some probability distribution. 
A simple choice for this would be random dilution, corresponding to a binary probability
distribution for $M$, taking on  two values $M=0$ and $M=M_0=$ fixed
at random. This
would drive a transition in the RTN model corresponding to the (same)  universality class
possessing  (see below) symmetry
group 
${\cal O}_Q(p)\times{\cal O}_Q(p)$ 
for any choice of a (large enough) value $M_0$.

We stress that owing to the invariance property (\ref{LabelEqInvarianceOfMetric-Clifford-RTN})
of the metric, the Clifford Stat Mech model 
is invariant under the
${\cal O}_Q(p)\times{\cal O}_Q(p)$ symmetry group {-- up to an additional $\mathbb{Z}_2$ semi-direct factor that is also present in the Haar case, and will be irrelevant to our purposes, see appendix~\ref{LabelAppendixDescriptionAndPropertiesOfCommutant}}. Note that this is a {\it different}
symmetry group\footnote{Recall that ${\cal O}_Q(p)$ denotes orthogonal matrices whose
entries are elements of the number field 
\andreasnew{${\bf F}_p$,}
which  itself has only $p$ elements.
Thus, the groups ${\cal O}_Q(p)$ are different finite groups for different primes $p$ -
they even have different order (=number of group elements).} 
for different primes $p$, but that this symmetry group is
{\it the same} for all on-site Hilbert space dimensions $d=p^M$ for a {\it fixed} prime $p$, independent
of $M$: the symmetry group depends only on the prime $p$.
This thus implies that any universal quantities occurring at continuous phase
transitions in this Stat. Mech. model will in general depend on the prime $p$, but will 
be the same for all on-site Hilbert space dimensions $d=p^M$ of the same prime
number $p$, independent of $M$. 
-- Note that this is in contrast to the Haar case, where the same universality class
occurs for the transition at any finite on-site Hilbert space dimension. In the Haar case, this must
of course be the case, because universal properties cannot depend on short-distance
(``ultraviolet'') physics - all those Stat. Mech. models have the same 
symmetry\footnote{Which we recall from \cite{VasseurPotterRTN2018} is
$S_Q\times S_Q $.}. While this feature is the same in the Clifford case
for a {\it fixed} value of the prime $p$ (i.e.,  for on-site Hilbert space dimension
$d=p^M$ and different values of $M$), RTN models possessing on-site Hilbert space
dimensions which are powers of {\it different} primes have in general
different universal critical properties at continuous phase transitions, because they
are invariant under different symmetry groups.


\section{Measurement-induced phase transition in Clifford circuits \label{SecMeasurements}}

\subsection{Setup and replica trick}

\begin{figure}[b]
    \centering
    \includegraphics[width=.49\textwidth]{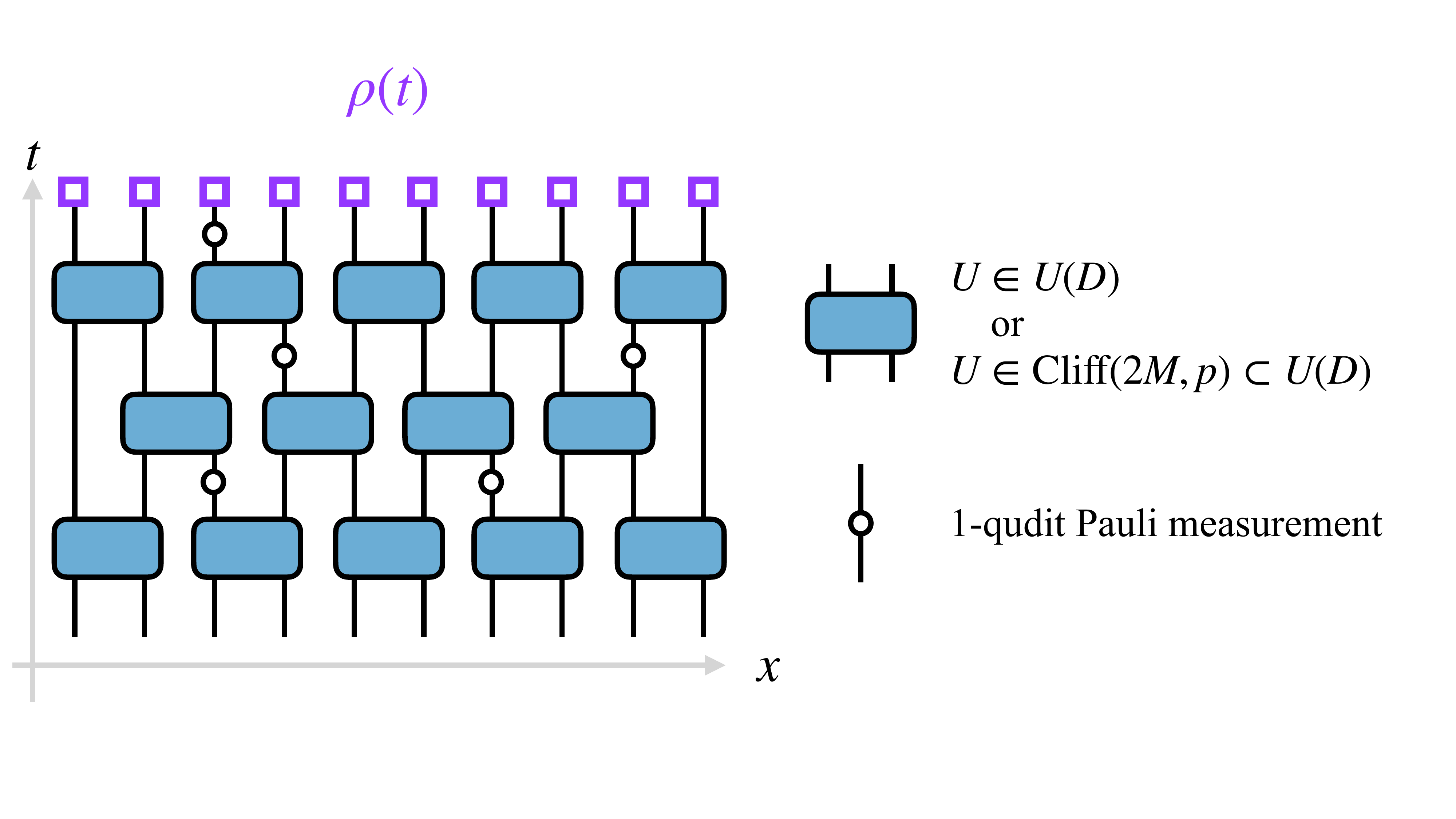}
    \caption{The architecture of the monitored quantum circuit model.
    Each blue block represents a random unitary gate, and each hollowed circle 
    represents a rank-1 projective measurement, occuring randomly with probability $p_0$.
    The dynamical state $\rho(t)$ at circuit depth $t$ is supported on qudits at the final time step (highlighted with purple squares), and is labeled by the the measurement record $\mathbf{m}$, see Eq.~\eqref{eqDefDynamicalStateCircuits}.
    We consider two cases, depending on whether the random unitaries are sampled from the Haar measure on $U(D=2d)$ (the ``random Haar circuit''), or from the uniform distribution the Clifford group $\mathrm{Cliff}(2M, p) \subset U(D)$ (the ``random Clifford circuit'').
    }
    \label{fig:circuit}
\end{figure}

We now turn to measurement-induced entanglement transitions in monitored quantum circuits. We consider a one-dimensional chain of qudits with on-site Hilbert space dimension $d$, subject to discrete-time dynamics generated by a quantum circuit with a `brick-wall' pattern, as illustrated in Fig.~\ref{fig:circuit}. Each unitary gate is acting on a pair of neighboring qudits, and will be drawn from the Clifford group (we will also briefly review the Haar case below). After each layer of the circuit (single time step), every site is either measured (projectively) with probability $p_0$, or left untouched with probability $1-p_0$. For a fixed set of unitary gates and measurement locations, the hybrid non-unitary dynamics of the system is described through the quantum channel
\begin{equation}
\label{eqDefDynamicalStateCircuits}
\rho(t) = \sum_{\mathbf{m}} K_{\mathbf{m}} \rho K^{\dagger}_{\mathbf{m}},
\end{equation}
where $\rho$ is the system's density matrix, $\mathbf{m}$ denotes measurement outcomes, and $K_{\mathbf{m}}$ are Kraus operators consisting of the random unitary gates and the projectors onto the measurement outcomes. We are interested in the entanglement properties of individual quantum trajectories $\rho_\mathbf{m} = K_{\mathbf{m}} \rho K^{\dagger}_{\mathbf{m}}$, which occur with the Born probability $p_\mathbf{m} = {\rm tr} \rho_\mathbf{m}$. Our goal is to compute the Renyi entropy of such single trajectories, averaged over measurement outcomes and random unitary gates
\begin{equation} \label{eqDefRenyiCircuits}
S_{A}^{(n)} =\mathbb{E}_{U} \sum_{\mathbf{m}} p_\mathbf{m} \frac{1}{1-n}  \ln \frac{{\rm tr} \rho^n_{A,{\mathbf{m}}}}{({\rm tr} \rho_{\mathbf{m}})^n}.
\end{equation}
Ultimately, we also want to average over measurement locations, and denote $\overline{S_{A}^{(n)}}$ the Renyi entropy averaged over measurement locations as well. As in the random tensor networks section above, we follow Refs~\cite{VasseurPotterRTN2018,Zhou2019, Bao2020,JianVasseurMeasurement2019} and use a replica trick to compute~\eqref{eqDefRenyiCircuits}:
\andreas{
\begin{equation} \label{eqDefRenyiCircuitsReplica}
\overline{S_{A}^{(n)}} =\mathbb{E}_{U} \sum_{\mathbf{m}} 
\underset{k \to 0}{\rm lim}
\frac{{\rm tr} \rho_{\mathbf{m}} }{(1-n)k}  \left( ({\rm tr} \rho^n_{A,{\mathbf{m}}})^k - ({\rm tr} \rho_{\mathbf{m}}^{\otimes k n })  \right).
\end{equation}
}
%
We will write
\andreas{$Q=nk+1$}
the number of replicas, where the additional replica compared to the random tensor networks case comes from the Born probability weighting different quantum trajectories.

\subsection{Generalized Weingarten functions}
\label{LabelSubSectionGeneralizedWeingartenFunction}

\subsubsection{Haar case}
\begin{figure}
    \centering
    \vspace{-1cm}
    \includegraphics[scale=0.38]{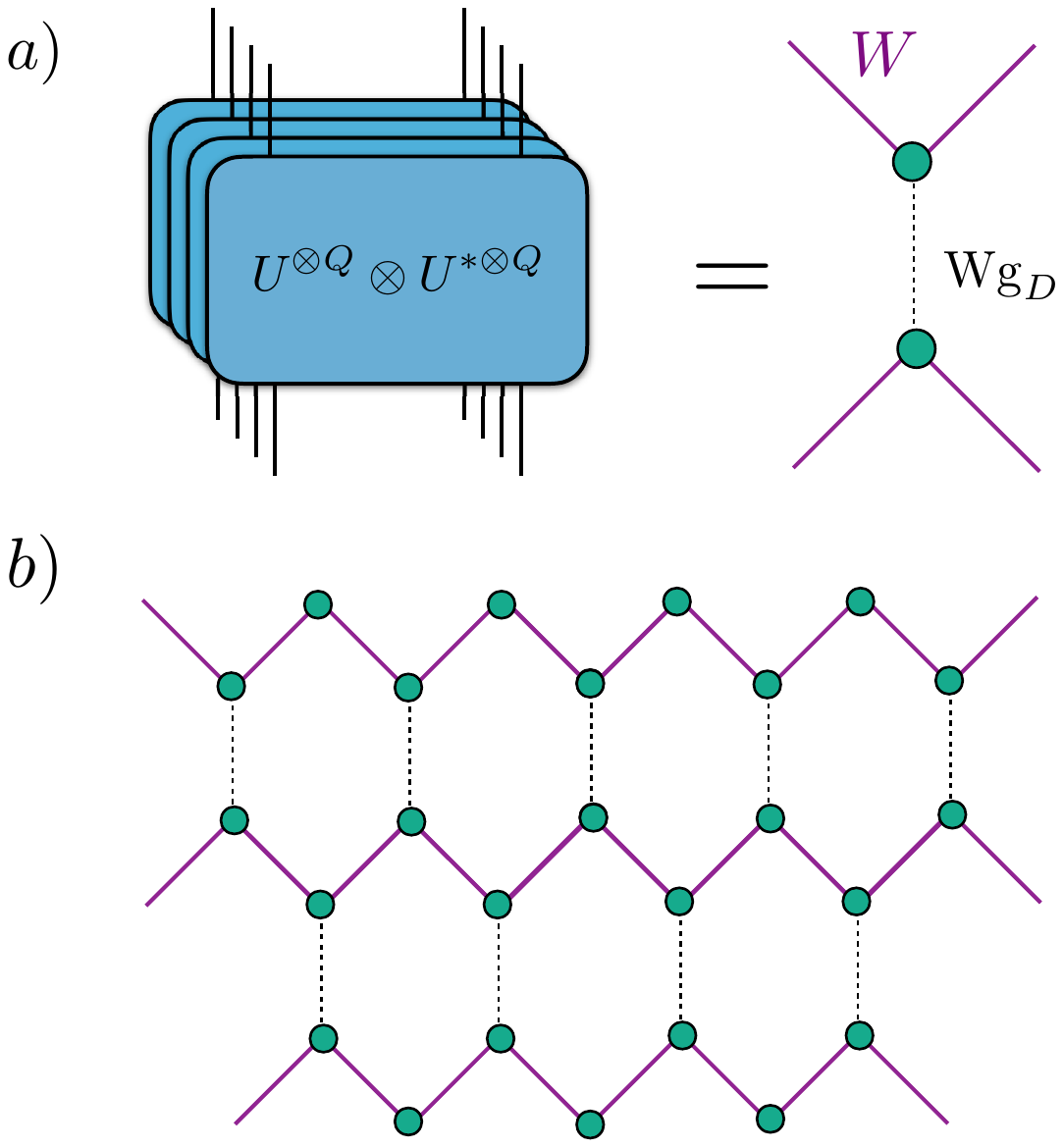}
    \vspace{-1cm}
    \caption{ Statistical mechanics model for monitored quantum circuits. (a) The average over replicated unitary gates leads to degrees of freedom belonging to the commutant, defined on vertices; Weingarten weights $\text{Wg}_{D}$, represented by the dashed line; and link weights $W$, corresponding to purple solid lines. (b) The statistical mechanics model is defined on an anisotropic honeycomb lattice.    }
    \label{fig:lattice}
\end{figure}

Let us first briefly summarize the main ingredient of the statistical mechanics model of the measurement-induced transition, with Haar random unitary gates acting on qudits. Upon replicating the model, averaging over Haar gates drawn from the unitary group $U(D=d^2)$ naturally leads to degrees of freedom living in the permutation group $S_Q$, with $Q=nk+1$ the number of replicas. This follows from the Schur-Weyl duality detailed in the previous section, which states that the permutation group $S_{Q}$ and the unitary group $U(D)$ act on $({\bf C}^D)^{\otimes Q}$ as a commuting pair. 
The Haar average of the replicated unitaries is given by (see Appendix~\ref{AppendixWHaar})
\begin{equation} \label{eqHaaraverage}
\mathbb{E}_{U\in U(D)}\left(U^{*Q}\otimes U^{Q}\right) = \sum_{\sigma,\tau \in S_{Q}} \text{Wg}_{D}(\sigma^{-1}\tau) {\hat R}_v (\sigma) \otimes {\hat R}_v (\tau),
\end{equation}
where Wg are Weingarten functions, $D=d^2$,  $ {\hat R}_v (\sigma)= {\hat R}_e (\sigma) \otimes {\hat R}_e (\sigma)$ (see eq.~\eqref{eqRmatrixHaar}) permutes the output legs of $U$ by $\sigma$, and contracts them with the corresponding legs of $U^*$ (and similarly for $\hat{R}_v(\tau)$ acting on the input legs). The tensor product in the right-hand side of eq.~\eqref{eqHaaraverage} is over output and input legs of the unitary. 
Now contrary to the previous section on random tensor networks,
the Weingarten  function will directly enter the Boltzmann weights of the statistical models. As we review in Appendix~\ref{AppendixWHaar}, those functions satisfy
\begin{equation} \label{eqWHaar}
 \sum_{\tau \in S_{Q}} \text{Wg}_{D}(\sigma^{-1}\tau) D^{C(\tau)} = \delta_{\sigma, 1},
 \end{equation}
where we recall that $C(\tau)$ denotes the number of cycles in the permutation $\tau$. 
This  implies that the Weingarten function is the inverse of the cycle counting function $ D^{C(\tau)}$.  Equation~\eqref{eqHaaraverage} naturally defines the degrees of freedom of the statistical model living on a honeycomb lattice (Fig.~\ref{fig:lattice}), where permutations live on vertices. Contracting unitary gates can be done by assigning a weight to links connecting unitaries given by 
\begin{equation} 
W(\sigma, \tau) = {\rm tr}\ \big [ {\hat R}_e^\dagger(\sigma) {\hat R}_e(\tau)\big ]   = d^{C(\sigma^{-1}\tau)}.
\end{equation}
Note the factor of $d$ here, instead of $D$, since we are focusing on a single leg of the unitary. This weight is associated to all links that were not measured. If a link was measured instead, all replicas are constrained to be in the same state, and the weight is instead $d$ after averaging over possible measurement outcomes. Those equations fully determine the weights of the statistical model in monitored Haar random circuits. Upon averaging over measurement locations and outcomes, the weight assigned to a link is therefore given by~\cite{JianVasseurMeasurement2019}
\begin{equation} 
W_p(\sigma, \tau) = (1-p_0) d^{C(\sigma^{-1}\tau)} + p_0 d.
\end{equation}
Putting these results together, we obtain an anisotropic statistical mechanics model defined on the honeycomb lattice, 
\begin{equation} \label{eqZHaar}
Z = \sum_{\lbrace g_i \in S_Q \rbrace} \prod_{ \langle ij \rangle \in G_{s}}W_p(\sigma_{i}^{-1} \tau_j) \prod_{ \langle ij \rangle \in G_{d}} \text{Wg}_{D}(\sigma_{i}^{-1} \tau_j), 
\end{equation}
where $G_s$ ($G_d$) denotes the set of solid (dashed) links on the lattice. In Fig.~\ref{fig:lattice}, the vertical (dashed) links on the honeycomb lattice represent the Weingarten functions which originated from averaging the two-site unitary gates, and the solid links keep track of the link weights originating from averaging over measurements. 

\subsubsection{Clifford case}

Let us now turn to the Clifford case, with random unitary gates drawn from the Clifford group ${\rm Cliff}(N,p) \subset U(D)$, with $D=d^2$, $N=2M$ and $d=p^M$ with $p$ prime. As in the Haar case, the average over Clifford gates is given in terms of tensor products of elements of the commutant of the Clifford group (see Appendix~\ref{AppendixWClifford})
\begin{align} \label{eqCliffordaverage}
& \mathbb{E}_{U\in {\rm Cliff}(N,p)}\left(U^{*Q}\otimes U^{Q}\right) = \notag \\
&\sum_{T_i,T_o \in \Sigma_{Q}(p)} \text{Wg}_{D}(T_i, T_0) \ {\hat R}_v(T_o) \otimes {\hat R}_v(T_i),
\end{align}
where as before, ${\hat R}_v(T_i) = {\hat R}_e(T_i) \otimes  {\hat R}_e(T_i) $ is acting on incoming legs, while $  {\hat R}_e(T_o)$ acts on outgoing legs, where the operator ${\hat R}_e =  {\hat r}^{\otimes M}$ was introduced in the previous section, see eq~\eqref{LabelEqClifford-r-DEF}.  Here the coefficients $\text{Wg}_{D}(T_i, T_0)$ are generalized Weingarten weights. The procedure to determine them follows closely the Haar case, and we show in Appendix~\ref{AppendixWClifford} that the generalization of eq.~\eqref{eqWHaar} reads
\begin{equation} 
\sum_{T_b \in \Sigma_{Q}(p)} \text{Wg}_{D}(T_a, T_b) \ 
{\rm tr}  \big [ \hat{R}_v^\dagger (T_b) \hat{R}_v (T_c)\big] = \delta_{T_a, T_c},
\end{equation}
as well as $\sum_{T_b \in \Sigma_{Q}(p)}   {\rm tr}\big [ \hat{R}_v^\dagger (T_a) \hat{R}_v (T_b)\big ] \ \text{Wg}_{D}(T_b, T_c) = \delta_{T_a, T_c}$.
In other words, the Weingarten weights (when viewed as a matrix)
are given by the inverse of the 
matrix ${\rm tr}  \big [\hat{R}_v^\dagger (T_a) \hat{R}_v (T_b)\big ]$. On the other hand, the link weights connecting two vertices are
given by
\begin{equation} \label{eqCircuitCliffordLinkWeight}
W(T_a, T_b)  = {\rm tr} \big[ {\hat R}_e^\dagger(T_a) {\hat R}_e(T_b)\big ] = d^{Q - |T_a, T_b|}, 
\end{equation}
with $d=p^M$. As in the Haar case, this weight is replaced by a factor of $d$ if a measurement occurred on that link  (upon averaging over measurement outcomes), as the measurement 
constrains all replicas to be in the same state. After averaging over measurement locations, the weight assigned to a link reads
\begin{equation} 
W_p(T_a, T_b) = (1-p_0) d^{Q - |T_a, T_b|} + p_0 d.
\end{equation}
The partition function of the statistical model in the Clifford case is then given by
\begin{equation} \label{eqZClifford}
Z = \sum_{\lbrace T_i \in \Sigma_{Q}(p) \rbrace} \prod_{ \langle ij \rangle \in G_{s}}W_p(T_{i}, T_j) \prod_{ \langle ij \rangle \in G_{d}} \text{Wg}_{D}(T_i, T_j), 
\end{equation}
with the same lattice as in the Haar case \andreas{(depicted in Fig.~\ref{fig:lattice})}.

Note that both the link and Weingarten weights are invariant under two copies of the stochastic orthogonal group ${\cal O}_Q(p)\times{\cal O}_Q(p)$, acting as $T'_a=O_L T_a O_R^{-1}, \ T'_b=O_L T_b O_R^{-1}$, with $O_L, O_R \in {\cal O}_Q(p)$. The conclusions for the Clifford RTN statistical mechanics model thus carry over to Clifford monitored quantum circuits: the Clifford Stat. Mech. model 
is invariant under a global
${\cal O}_Q(p)\times{\cal O}_Q(p)$ symmetry group. This symmetry group depends only on the prime number $p$, and not on $M$ in the on-site Hilbert space dimension $d=p^M$. This  implies that all universal quantities of the entanglement transition in Clifford monitored circuits will depend on $p$, but will be independent of $M$. This is in sharp contrast to the Haar case, where the same universality class
occurs for the transition at any finite on-site Hilbert space dimension. Although the statistical mechanics model~\eqref{eqZClifford}
describing the monitored Cliffod circuit 
has the same symmetry group as the one describing stabilizer RTNs~\eqref{LabelEq-Clifford-RTN-Hamiltonian}, both problems correspond to different replica limits: $Q \to 0$ for RTNs, and $Q \to 1$ for monitored quantum circuits.

\subsection{$M\to \infty$ limit with $p$ fixed and percolation}

While the Clifford statistical mechanics formulation is in general not analytically tractable,
the limit of large on-site Hilbert space leads to drastic simplifications. Let us first consider the link weight, given by eq.~\eqref{eqCircuitCliffordLinkWeight},
where we recall that $\left| T_a, T_b \right|$ is 
a metric satisfying $\left| T_a, T_b \right| \geq 0$, $\left| T_a, T_b \right |=\left| T_b, T_a \right |$, and $\left| T_a, T_b \right |=0$ if and only if $T_a =T_b$. This immediately implies that at large $d$, the weight for non-measured links simplifies dramatically to
\begin{equation} 
W (T_a, T_b) \sim d^Q \delta_{T_a, T_b} + \dots
\end{equation}
This corresponds to a perfectly ferromagnetic interactions that forces the elements of the commutant to be identical on unmeasured links. Since the Weingarten weights are given by the inverse of ${\rm tr}
\big[\hat{R}_v^\dagger (T_i) \hat{R}_v (T_o)\big]$, we immediately find
\begin{equation} 
\text{Wg}_{D}(T_i, T_0) \sim D^{-Q} \delta_{T_i, T_o} + \dots,
\end{equation}
with $D=d^2$. 
Note that those results hold for $d = p^M \to \infty$, independently of how the limit is taken. We now focus on the limit of $d=p^M$ with $p$ prime fixed and $M \to \infty$. (We will discuss the limit of large $p$ in the next section). Combined, those results tell us that in the $M \to \infty$ limit, the statistical mechanics weights for Clifford circuits become identical to
those of
the Haar case: 
precisely in the $M \to \infty$ limit
we obtain a Potts model, whose states are the elements 
$T_a$ of the commutant $\Sigma_{Q}(p)$. The number of states in the Potts model depends on $p$ and on the number of replicas $Q$, and is given by the dimension of the commutant~\cite{GrossEtAlCMP2021}

\begin{equation}
\label{LabelDimensionOfCommutant}
|\Sigma_{Q}(p)| =  \displaystyle\prod_{k=0}^{Q-2} \left(p^k+1 \right) = p^{\frac{(Q-1)(Q-2)}{2}} \frac{ \displaystyle\prod_{k=0}^{\infty} \left(1 + \frac{1}{p^k} \right)}{ \displaystyle\prod_{k=0}^{\infty} \left(1 + \frac{1}{p^{k+Q-1}} \right)}.
\end{equation}
(In the second equality, obtained by straightforward rewriting, the number $Q$ can be analytically
continued from integer values.)
In the replica limit $Q \to 1$, we have $|\Sigma_{Q}(p)| \to 1$, so the critical theory of the measurement induced transition is simply described by classical percolation. In particular, this predicts a diverging correlation length with a critical exponent $\nu =4/3$. 
Note that the way the replica limit is approached depends on $p$. This shows up in, ${\it e.g}$, the effective central charge $c_{\rm eff} = {\rm lim}_{Q \to 1} \frac{dc}{dQ}$ introduced in Ref.~\cite{2021arXiv210703393Z}. The central charge as a function of the number of replicas $Q$ is now given by $c(Q) = 1 - \frac{6}{x(x+1)}$ with 
$ x+1 =\pi/ \big ({ \arccos \frac{\sqrt{|\Sigma_{Q}(p)|}}{2}}\big )$.
This leads to a closed form expression for the effective central charge 
\begin{equation}
c^{M \to \infty}_{\rm eff} = \frac{5 \sqrt{3}}{8 \pi} \left( 2 \psi_{\frac{1}{p}}\left( \frac{ i \pi}{\log p^{-1}}\right) - \log \frac{p^3}{(p-1)^2} \right),
\end{equation}
where $\psi_q(z)$ is the $q$-digamma function, which is defined as the derivative of $\log \Gamma_q(z)$ with respect to $z$, where  $\Gamma_q(z)$ is the $q$-deformed Gamma function. The special case $p=2$ of this formula was reported in Ref.~\cite{2021arXiv210703393Z}.

This mapping to percolation is specific to the $M \to \infty$ limit: if $M$ is large but finite, the critical theory of the measurement-induced transition is described by the percolation CFT perturbed by a relevant perturbation (identified as the ``two-hull'' operator in Refs.~\cite{VasseurPotterRTN2018,JianVasseurMeasurement2019}), with scaling dimension $\Delta=5/4$. To see this, let us consider the Landau-Ginzburg formulation of the $|\Sigma_{Q}(p)|$-state Potts field theory in terms of the Potts order parameter field $\phi_a$ where $a = 1,...,|\Sigma_{Q}(p)|$ labels elements of the commutants, and $\sum_a \phi_a = 0$. The symmetry of the Potts theory for $M \to \infty$ is $S_{\left| \Sigma_{Q}(p) \right|}$, which is much larger than the stochastic orthogonal symmetry ${\cal O}_Q(p)\times{\cal O}_Q(p)$ of the finite $M$ case. To see this, note that ${\cal O}_Q(p) \times {\cal O}_Q(p) \subset S_{\left|{\cal O}_Q(p) \right|}$ using Cayley's theorem: the left and right actions of the group ${\cal O}_Q(p)$ on itself have a permutation representation. Note also that $S_{\left|{\cal O}_Q(p) \right|}$ is a subgroup of $S_{\left| \Sigma_{Q}(p) \right|}$ since the stochastic orthogonal group is a subset of the commutant $\Sigma_{Q}(p)$. The symmetry-breaking $S_{\left| \Sigma_{Q}(p) \right|} \to {\cal O}_Q(p)\times{\cal O}_Q(p) $ can be implemented by the perturbing the Potts model
\begin{equation} \label{eqlagrangianPotts}
{\cal L} = {\cal L}_{\rm Potts} + \sum_{g,h \in {\cal O}_Q(p)} V (g^{-1} h) \phi_{g} \phi_h,
\end{equation}
where $V$ is a class function of the group ${\cal O}_Q(p)$, ensuring a left and right stochastic orthogonal group symmetry. The perturbation has scaling dimension $\Delta=\frac{5}{4}$ at the Potts (percolation) fixed point in the replica limit, and is therefore relevant.

The fate of this theory in the IR is not known analytically, and corresponds to the generic universality class of Clifford circuits for a given $p$. For $p$ fixed and $M$ large but finite, we thus expect a crossover between percolation and the finite $d=p^M$ universality class (dependent on $p$), with the corresponding crossover length scale 
\begin{equation} \label{eq:crossover_length}
\xi(M) \sim p^{4 M/3},
\end{equation}
where we have assumed that the relevant perturbation comes with an amplitude $V \sim d^{-1}$.
Note that this crossover length scales exponentially with $M$, giving a very broad regime $L \ll \xi(M)$ described by percolation at large $M$. For length scales much smaller than $\xi(M)$, we expect to see percolation physics, and in particular, the entanglement entropy should be given by a minimal cut picture as we derive in the next section. In particular, the entanglement
entropy in that regime scales with $\ln d = M \ln p$ (see next section). However, for length scales much larger than $\xi(M)$, the physics is controlled by the infrared fixed point of the theory~\eqref{eqlagrangianPotts}, corresponding to the generic Clifford universality class for fixed prime $p$. In particular, in this regime,
the prefactor of the logarithmic scaling of the entanglement entropy at criticality will be universal, and will depend on $p$, but not on $M$. 

\subsection{Large $p$ limit and minimal cut}

Finally, we now comment on the large $p$ (prime) limit, keeping $M=1$ fixed. For each finite $p$, there is a distinct universality class, which is independent of $M$. However, the results of the previous section also indicate that the bulk theory should approach percolation as $p \to \infty$ (with $M=1$ fixed): we thus expect a series of fixed points approaching percolation as $p \to \infty$. A simple way to understand this is to consider a fixed realization of measurement locations. Each bond that is measured is effectively cut, while all other weights constrain the statistical model's ``spins'' to be the same in this limit. We thus obtain a simple percolation picture of fully-ordered (zero temperarure) ferromagnetic spin model diluted by the measurements.  As we show below, a frustrated link  costs a large energy $\sim \ln p$, leading to an effective minimal cut picture in that limit. 

Recall that computing entanglement requires computing two different partition functions $Z_A$ and $Z_{\emptyset}$, which 
differ
only by their boundary condition on the top boundary of the circuit. The boundary condition for the calculation of $Z_A$ forces a different boundary condition in region $A$, and thus introduces a domain wall (DW) near the top boundary. In the limit $d \to \infty$, the DW is forced to follow a minimal cut, defined as a path cutting a minimum number of unmeasured links (assumed to be unique for simplicity): the argument follows closely Ref.~\cite{2021arXiv210710279A} in the Haar case. Due to the uniform boundary condition in $Z_{\emptyset}$, all vertex elements in $Z_{\emptyset}$ are equal, so $Z_{\emptyset}$ is trivial and give by a single configuration of ``spins''. $Z_A$ differs from $Z_{\emptyset}$ due to the fact the DW will lead to frustrated links that contribute different weights to $Z_A$. Each frustrated unmeasured link contributes a factor ${\rm tr}\ \big[{\hat r}^\dagger(1) {\hat r}(g_{\rm SWAP})\big]/{\rm tr}\ {\hat r} (1) $ to the ratio of partition functions $Z_A/Z_{\emptyset}$, where 
\andreas{
$g_{\rm SWAP} = (1 2 \dots n)^{\otimes k}$
}
is the boundary condition enforced in the entanglement region $A$ in $Z_A$. (The permutation $g_{\rm SWAP}$ acts trivially on the last replica implementing the Born probability factor.) We find 
\andreas{
${\rm tr}\ \big[{\hat r}^\dagger(1) {\hat r}(g_{\rm SWAP})\big]/{\rm tr} \ {\hat r} (1)  = p^{k+1 -Q} = 
p^{(1-n)k} \ll 1$.
}
This corresponds \andreas{(since we are considering large $p$)} to a very large energy cost per frustrated link
\andreas{
\begin{equation} 
\Delta E = (n-1) k \ln p,
\end{equation}
}
%
so the domain wall will follow a path through the circuit minimizing the number of unmeasured links it has to cut. This leads to the expression 
\andreas{
$Z_A = p^{(1-n)k \cdot \ell_{\rm DW}}Z_{\emptyset}$,
}
with $\ell_{\rm DW}$ the number of unmeasured links that the DW crosses along the minimal cut. In the replica limit, this leads to a simple expression for the Renyi entropies 
\begin{equation} 
S^{(n)}_A =  \ell_{\rm DW}  \ln p,
\end{equation}
where this equation is valid for any given configuration of measurement locations. We will use $\overline{ \ell_{\rm DW}} $ to denote the average of $\ell_{\rm DW}$ over measurement locations, which are simply percolation configurations. This quantity has a simple scaling in percolation: it is extensive  $\overline{ \ell_{\rm DW}}   \sim L_A$ (volume law) for $p_0<p_{0,c}=1/2$, and constant $\overline{ \ell_{\rm DW}}  \sim O(1)$ (area law) for $p_0>p_{0,c}=1/2$.
At criticality, this implies a logarithmic scaling of the entanglement entropy~\cite{1986JSP....45..933C, yao1612firstpassage, Skinner2019}
\begin{equation} \label{eqMinCut}
\overline{S^{(n)}_A}  \underset{p \gg 1}{\approx} \frac{\sqrt{3}}{\pi} \ln p \ln L_A.
\end{equation}
We expect the measurement-induced transition to approach these predictions at large $p$.

\subsection{Forced measurements and RTN at large bond dimension}
\label{LabelSubsectionForcedMeasurements}

Our predictions can also be generalized to the case of {\em forced measurements}, where the given outcome for a `forced measurement' is chosen randomly in a way that is independent of the state, instead of following the Born rule. In the context of RTNs, entanglement transitions can be implemented at fixed bond dimension by randomly breaking bulk bonds~\cite{VasseurPotterRTN2018}, which can be thought of as such `forced measurements'~\cite{Nahum2020}. The corresponding statistical mechanics models have the same stochastic orthogonal group symmetry, but correspond to different replica limits $Q \to 1$ (projective measurements) {\it vs} $Q \to 0$ (forced measurements, RTNs), and are thus in general 
expected to be in different universality classes.
However, numerical results suggest that forced {\it vs.} projective measurements might be in the same universality class in the Clifford case~\cite{yang2021entanglement}, indicating that the two replica limits $Q \to 0$ and $Q \to 1$ could lead to the same critical theory.

The conclusions of the previous sections carry over to this forced measurement setup as well: in the large $D$ limit, the link and Weingarten weights simplify dramatically, the statistical mechanics model reduces to a Potts model, and the large $d=p$ limit also satisfies the minimal cut prediction~\eqref{eqMinCut}. The only difference is that the number of replicas is 
\andreas{
$Q=n k$  instead of $Q=nk+1$, 
}
reflecting the absence of Born probabilities. In particular, the replica limit corresponds to $Q \to 0$ instead of $Q\to 1$. At large $p$ and $Q$ small \andreas{(i.e.: $0<Q<1$)}, the dimension of the 
commutant reads\andreas{, using Eq.~\ref{LabelDimensionOfCommutant},}
\begin{equation}
    |\Sigma_{Q}(p)| \underset{p \to \infty}{\sim} \frac{2}{1+p^Q} p^{\frac{Q(Q+1)}{2}},
\end{equation}
which gives $ |\Sigma_{Q}(p)| \to 1$ in the replica limit $Q \to 0$. At large $p$, the replica limit is thus given by a Potts model with $|\Sigma_{Q}(p)| \to 1$ states, consistent with percolation and the minimal cut prediction.

\begin{figure}[h]
    \centering
    \includegraphics[width=.45\textwidth]{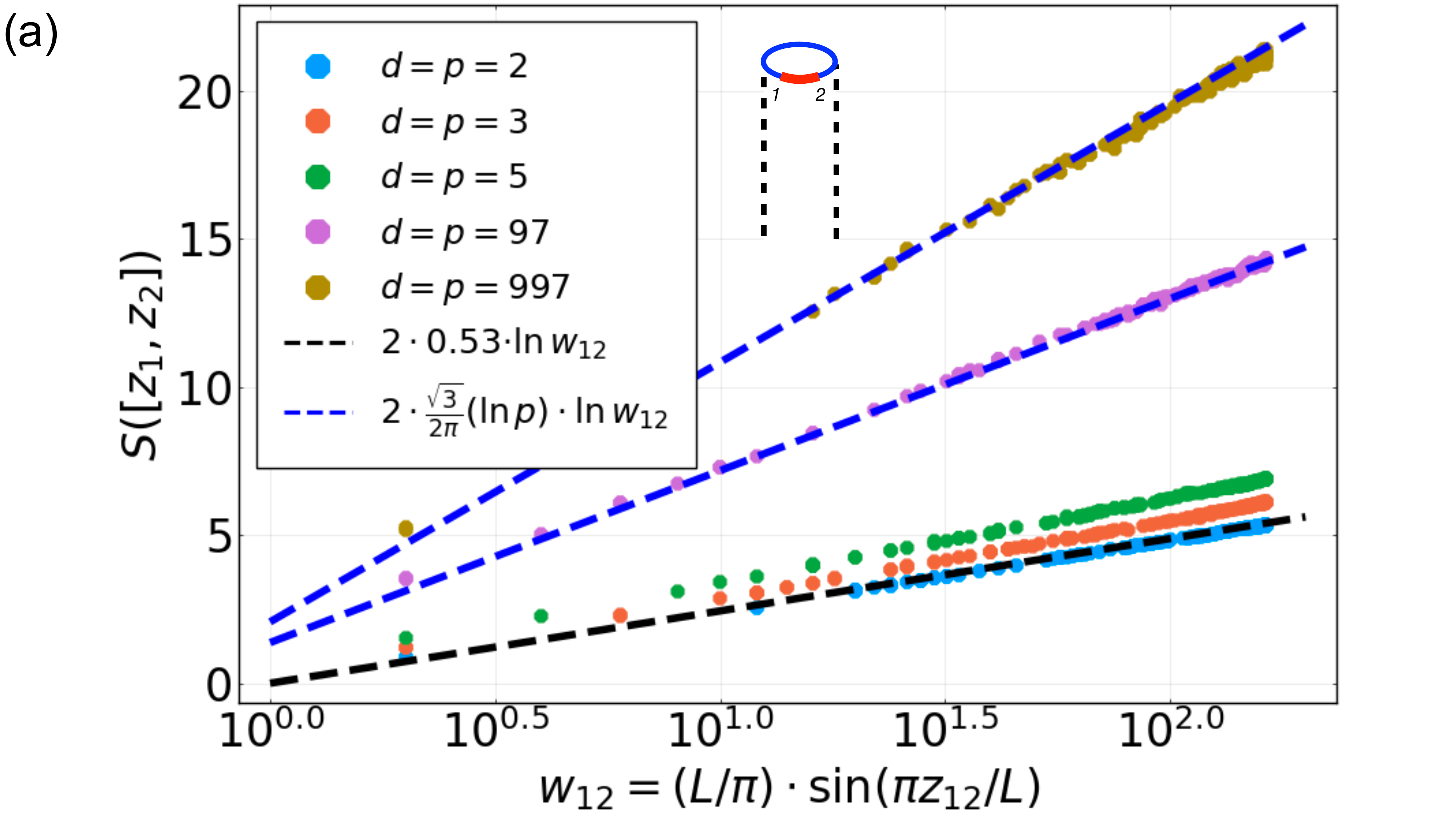}
    \includegraphics[width=.45\textwidth]{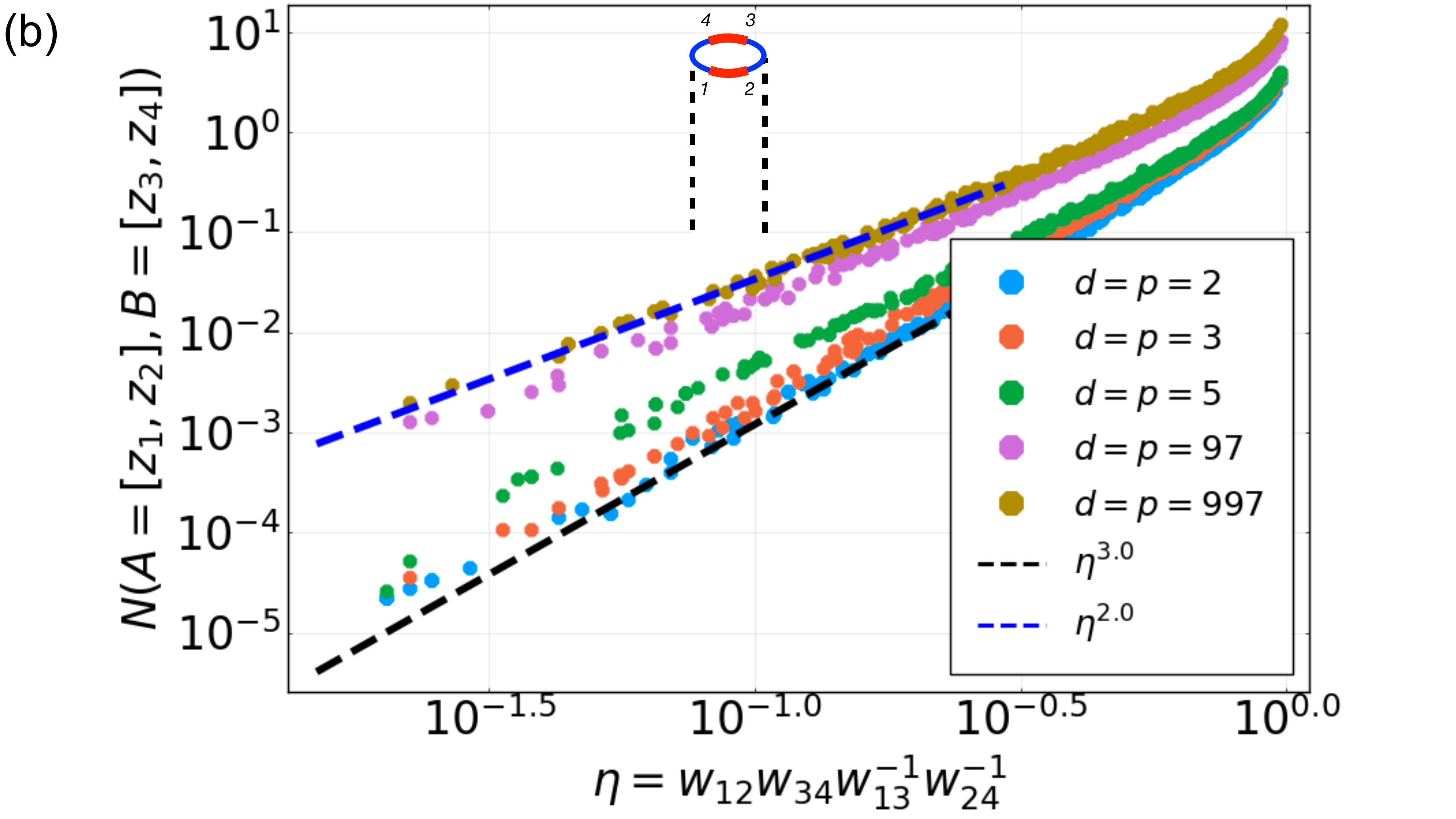}
    \includegraphics[width=.45\textwidth]{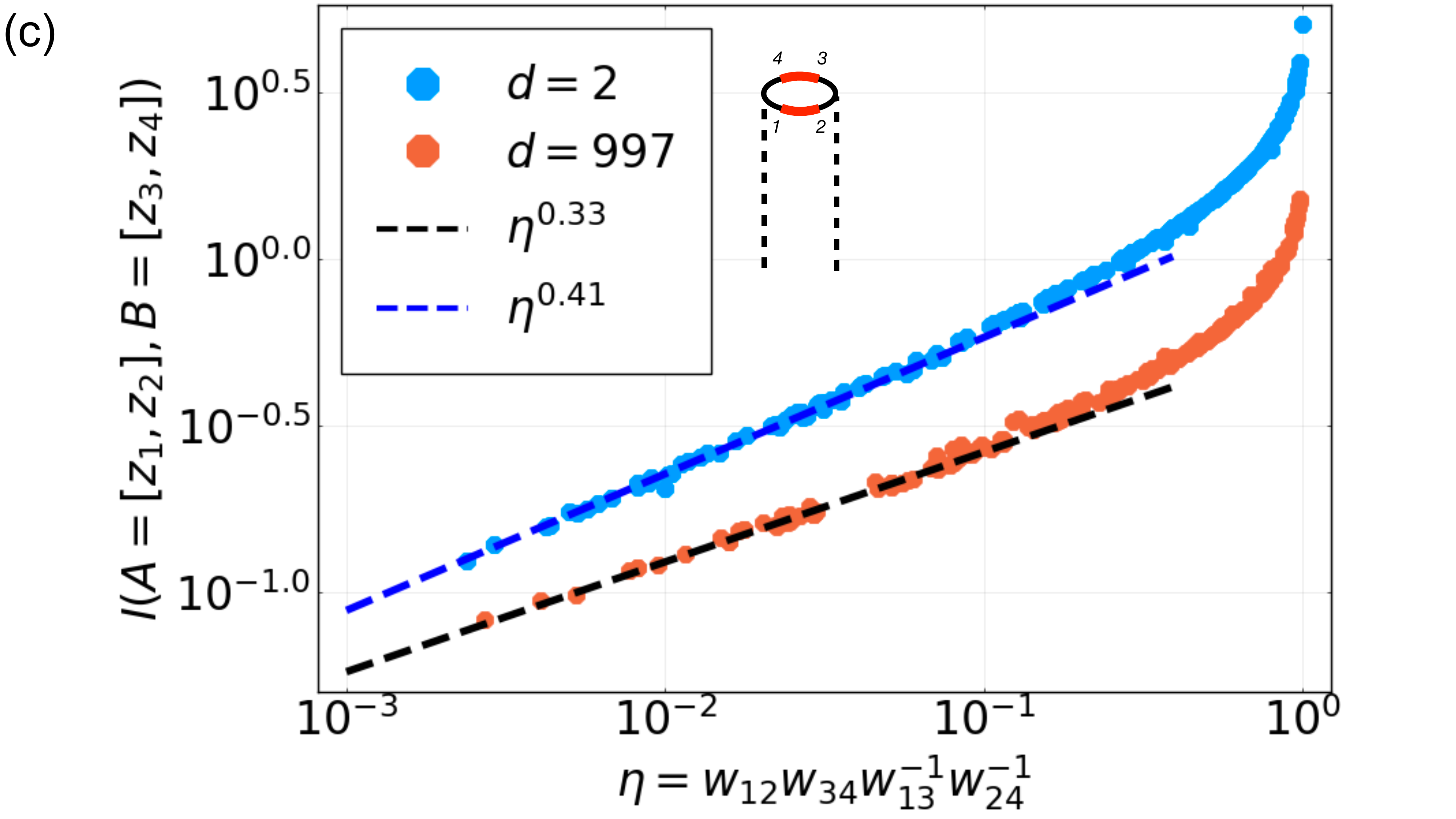}
    \caption{Numerical results for $d = p^{M=1}$.
    The fits for the critical exponents are summarized in Table~\ref{table:scaling}.
    (a) We plot $S(A = [z_1, z_2])$ against $\ln w_{12}$ (see  Eq.~\eqref{eq:SA_vs_xi}) at different values of $d=p$, where we see a clear dependence of the slope ($=2h_{a|b}$) on $p$.
    With increasing $p$, the value of $h_{a|b}$ approaches that of minimal cut percolation (see Eq.~\eqref{eqMinCut}).
    (b) The mutual negativity $N(A,B)$ between two disjoint subregions $A = [z_1, z_2]$ and $B = [z_3, z_4]$ against the cross ratio $\eta = (w_{12} w_{34}) / (w_{13} w_{24})$. 
    The powerlaw at small $\eta$ defines a critical exponent $h_{\rm MN}$, which has a clear dependence on $p$, and approaches minimal cut percolation at large $p$.
    (c) A ``localizable entanglement'' between $A$ and $B$.
    We perform a single-qudit projective measurements on each qudit outside $A \cup B$, and calculate  the mutual information $I(A,B)$ subsequently.
    The powerlaw at small $\eta$ is another critical exponent $h_{f|f}^{(1)}$, which should in general take different values at different $p$, as we illustrate.
    }
    \label{fig:d_equals_p}
\end{figure}

\section{Numerical results \label{SecNumerics}}
In this section we provide numerical evidence that supports our key conclusions from the previous sections on symmetry, namely that the universality class of the transition for qudit dimension $d = p^M$ depends on the prime number $p$, but not on the power $M$.
To this end, we consider two cases: 

(A) $d = p^M$ with fixed $M=1$ and different values of  $p$, where we expect a series of different, $p$-dependent universality classes;

(B) $d = p^M$ with fixed $p$ and a few small values of $M$, where we expect to obtain the same exponents while varying $M$.

For each value of $d$, we simulate monitored Clifford circuits of $L = 512$ qudits with periodic boundary condition, and focus on the steady state entanglement properties at their respective critical points.
The circuit thus takes the geometry of a semi-infinite cylinder, and at criticality the entanglement entropy of a region $A$ is predicted to take the following form,
\begin{align} \label{eq:SA_vs_xi}
    S(A = [z_1, z_2]) = 2 h_{a|b} \ln \left( \frac{L}{\pi} \sin \frac{\pi z_{12} }{L} \right) \coloneqq 2 h_{a|b} \ln w_{12},
\end{align}
where $h_{a|b}$ is the scaling dimension of a (primary) boundary condition changing operator in the underlying CFT.

\begin{table*}[t]
\begin{tabular}{c|c|c|c|c|c}
\hline
Data at the critical point &
$d = 2$ &
$d = 3$ &
$d = 5$ &
$d = 97$ &
$d = 997$
\\
\hline \hline
$p_{0,c}$ &  0.160    &  0.278   &   0.377   &   0.495    &  $0.500 \approx 1/2$ \\ \hline
$h_{a|b}$  &  0.53 &  0.62   &  0.66 &  $1.26 \approx \frac{\sqrt{3}}{2\pi}\cdot \ln(97)$ &  $1.90 \approx \frac{\sqrt{3}}{2\pi}\cdot \ln(997)$ \\ \hline
$h_{\rm MN}$ &  3.0  &  2.9   &   2.6   &   2.2    &  2.0 \\ \hline
$h_{f|f}^{(1)}$  &  0.41  &  -   &   -  & -  & $0.33 \approx 1/3$             \\ \hline
\end{tabular}
\caption{A comparison of the location of the critical point $p_{0,c}$ and the operator scaling dimensions $h_{a|b}, h_{\rm MN}, h_{f|f}^{(1)}$ in hybrid Clifford circuits with different prime qudit dimensions $d=p$.
}
\label{table:scaling}
\end{table*}

\begin{table*}[t]
\begin{tabular}{c|c|c|c|c|c}
\hline
Data at the critical point &
$d = 2$ &
$d = 3$ &
$d = 5$ &
$d = 23$ &
$d = 503$
\\
\hline \hline
$h_{a|b}$  &  - &  $0.53 \approx 0.48 \cdot \ln(3)$   &  $0.61 \approx 0.38 \cdot \ln(5)  $ &  $0.91 \approx 0.29\cdot \ln(23)$ &  $1.74 \approx 0.28\cdot \ln(503)$
\\ \hline
$h_{\rm MN}$ &  -  &  2.8   &   2.5  
 &   2.1    &  2.1
\\ \hline
$h_{f|f}^{(1)}$  &  -  &  0.37   &   0.36  & $0.33 \approx 1/3$ & $0.33 \approx 1/3$
\\ \hline
\end{tabular}
\caption{Exponents for Clifford RTN~\cite{yang2021entanglement} with different prime qudit dimensions $d=p$.
The exponents are all close to Clifford circuit, but different.
}
\label{table:scaling_RTN}
\end{table*}

\subsection{Different universality classes at various $d = p^{M=1}$}

Here, we take the qudit dimension to be a prime number, $d = p^{M=1}$ for $p \in \{2, 3, 5, 97, 997\}$.
In Fig.~\ref{fig:d_equals_p}(a), 
we plot the results of $S(A = [z_1, z_2])$ in the form of Eq.~\eqref{eq:SA_vs_xi}, and extract the scaling dimension $h_{a|b}$ at the respective critical points.
The locations of the critical points are determined by the best fit of $S(A = [z_1, z_2])$ to a linear function of $\ln w_{12}$, see Eq.~\eqref{eq:SA_vs_xi}.
The fits for $h_{a|b}$ and $p_{0,c}$ are collected in Table~\ref{table:scaling}.

From our fitting, the scaling dimension $h_{a|b}$ clearly depends on $p$, and increases monotonically with $p$.
It approaches the prediction from minimal cut percolation at large $p$, namely $h_{a|b} = \frac{\sqrt{3}}{2\pi} \ln p$ -- see  Eq.~\eqref{eqMinCut} and Eq.~\eqref{eq:SA_vs_xi}.
In addition, the value of $p_{0,c}$ approaches $1/2$ as $p$ is increased, again consistent with bond percolation on a square lattice.\footnote{We note in passing that the series of $p_{0,c}$ as summarized in Table~\ref{table:scaling} violates a conjectured bound~\cite{Fan2020} for all values of $d=p \ge 3$ that we accessed.}

We also include results of the ``mutual negativity'' $N(A,B)$ between two disjoint regions $A=[z_1, z_2]$ and $B=[z_3, z_4]$~\cite{Sang}.
This quantity is expected to take the following form when the cross ratio $\eta$ for the four endpoints is small,
\begin{align}
    N(A=[z_1, z_2], B=[z_3, z_4]) \propto \eta^{h_{\rm MN}}, \text{  as } \eta \to 0.
\end{align}
Here, $h_{\rm MN}$ stands for the ``mutual negativity exponent''.
In Fig.~\ref{fig:d_equals_p}(b) and Table~\ref{table:scaling}, we see that $h_{\rm MN}$ is close to $3$ for smaller values of $d=p$, and approaches $2$ as $d=p$ becomes large, as consistent with minimal cut percolation~\cite{Sang}.

To further contrast universality classes at small $d=p$ with percolation at large $d=p$, we focus on the maximum and minimum qudit dimensions in our numerics, namely $d = p \in \{2, 997\}$, and consider yet another quantity, known as a ``localizable entanglement''~\cite{Cirac2004LocalizableEntanglement}, recently introduced in the context of hybrid circuits~\cite{Li2020a}.
In this case, we again choose two disjoint regions $A=[z_1, z_2]$ and $B=[z_3, z_4]$, and perform a single-qudit projective measurement on each qudit outside $A \cup B$.
In the post-measurement state, the mutual information between $A$ and $B$ is expected to be a conformal four-point function~\cite{Li2020a}, and takes the following form when the cross ratio $\eta$ for the four endpoints is small,
\begin{align}
    I(A=[z_1, z_2], B=[z_3, z_4]) \propto \eta^{h_{f|f}^{(1)}}, \text{  as } \eta \to 0.
\end{align}
Here, $h_{f|f}^{(1)}$ is another critical exponent, which takes the value of $1/3$ in the percolation limit $d=p\to \infty$.
In our numerics (see Fig.~\ref{fig:d_equals_p}(c)), we find that $h_{f|f}^{(1)}$ fits well to $1/3 \approx 0.33$ at $d = p = 997$, but takes a \textit{distinct} value $\approx 0.41$ at $d= p = 2$~\cite{Li2020a}.

\subsection{Same universality class for $d = p$ and $d=p^2$}

\begin{figure}[t]
    \centering
    \includegraphics[width=.49\textwidth]{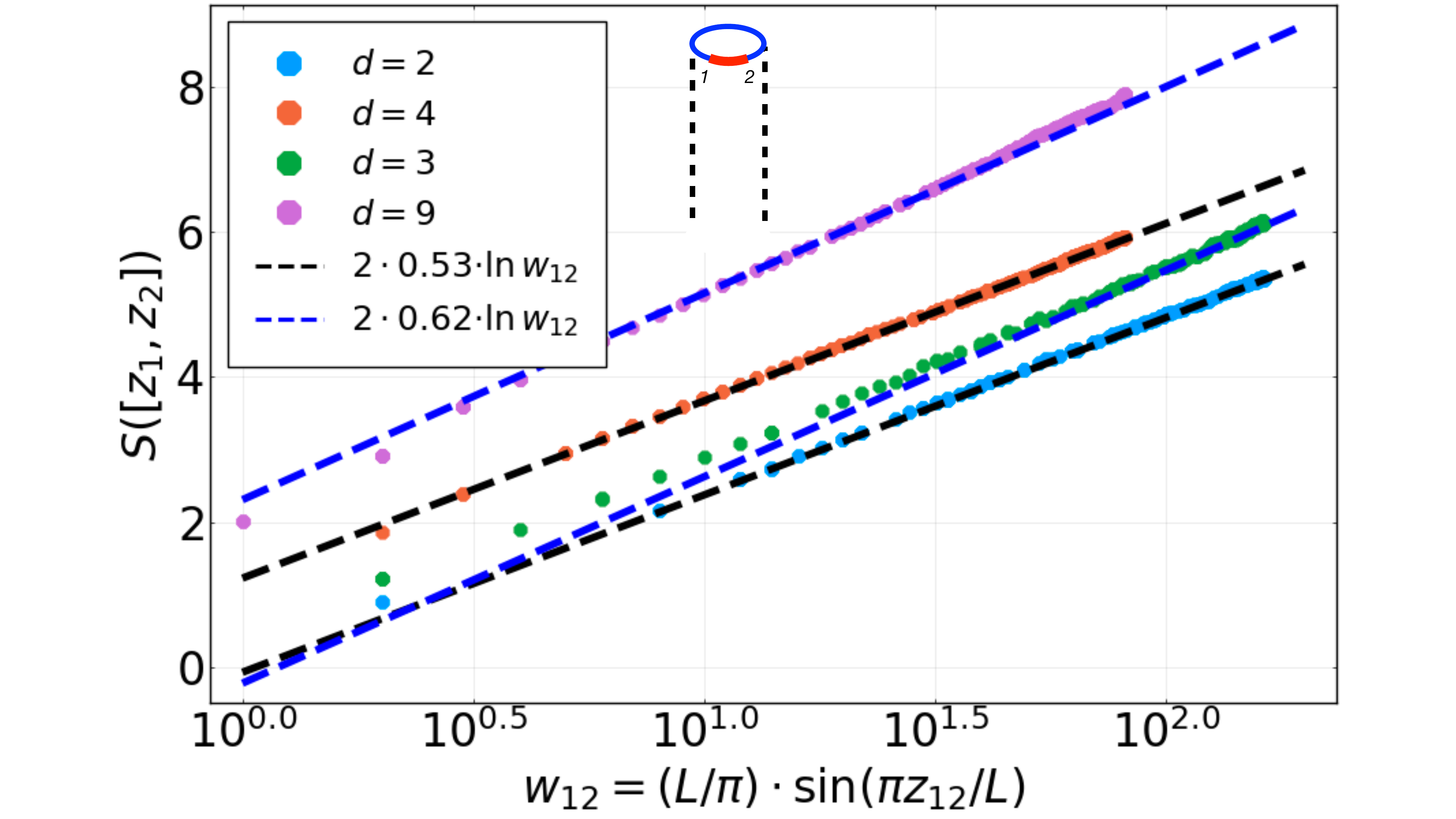}
    \caption{Numerical results for $d=p^M$, where we take $p \in \{2, 3\}$ and $M \in \{1, 2\}$.
    Comparing Fig.~\ref{fig:d_equals_p}(a) and Eq.~\eqref{eq:SA_vs_xi}, we see that $h_{a|b}$ depends on the choice of $p$, but not on $M$.
    }
    \label{fig:d_equals_p_squared}
\end{figure}

Here we compare qudit dimensions $d = p^{M}$ at a fixed $p$ but different $M$.
In order to probe the infrared fixed point, 
we choose $M = 1, 2$ so that $L \gg \xi(M)$, where $\xi(M)$ is a crossover length scale that diverges with $M$ (see Eq.~\eqref{eq:crossover_length}).
Numerically accessing the crossover from $L \gg \xi(M)$ to $L \ll \xi(M)$ is an interesting problem left for future versions of this work. 

For concreteness, we briefly describe the circuit model for $M > 1$.
At each site of the lattice we construct a  ``composite'' qudit of dimension $d=p^M$ by grouping together $M$ ``elementary'' qudits each of dimension $p$.
The circuit takes a brickwork architecture in the composite qudits, where nearest-neighbor composite qudits are coupled with random unitaries from the group $\mathrm{Cliff}(2M, p)$, and single composite qudits are subject to rank-1 projective measurements with probability $p_0$.

We compute the entanglement entropy $S(A)$ of region $A$ for $d \in \{2, 4 = 2^2\}$ and for $d \in \{3, 9 = 3^2\}$.
The results are plotted in Fig.~\ref{fig:d_equals_p_squared}, where we see that $h_{a|b}$ depends $p$ but not on the choice of $M$.
Results for $h_{a|b}$ at other values of $p$ also follow this pattern, but are not displayed.

Critical exponents for Clifford RTNs from Ref.~\cite{yang2021entanglement} are summarized in Table~\ref{table:scaling_RTN} for comparison. 
\andreas{Our}
numerical results indicate that the critical properties do not depend on the measurement scheme, {\it forced} (corresponding to the replica limit $Q \to 0$ in the statistical mechanics model) {\it vs} projective (replica limit $Q \to 1$)~\cite{yang2021entanglement}, but appear to differ (slightly) between monitored Clifford circuits and Clifford RTNs for small $p$. Since the statistical mechanics models for RTNs~\eqref{LabelEq-Clifford-RTN-PartitionFunction} and monitored circuits~\eqref{eqZClifford} have the same symmetry group, we expect the entanglement transitions in both cases to be in the same universality class for a given measurement scheme (forced or projective). The numerical discrepancy between those cases might be due to strong corrections to scaling (i.e. from irrelevant operators) in the RTN case for small $p$, since there is no transition for $p=2$ in that case.
One can, in principle, get an idea of the strength of such corrections by comparing Clifford RTNs with bond dimensions $d=p^M$ at a few small values of $M$, in a similar fashion to Fig.~\ref{fig:d_equals_p_squared}, although we leave a more thorough analysis of those corrections for future work.   

\subsection{Multifractal scaling of the purity}
\label{LabelSubSectionMultifractalScaling}



In this subsection we take the prime number to be $p=2$. Based on the discussion and the results in the rest of this  paper, 
we have no reason to expect that the general physics discussed in this subsection (multifractal scaling)  will be fundamentally different  for other prime numbers $p$,  except that numerical values for various critical exponents defined below are expected to depend  explicitly on the prime $p$ for the symmetry reasons mentioned in the other parts of this paper.

Specifically, in this subsection, we consider the purity of the normalized reduced density matrix
[compare (\ref{eqDefRenyiCircuits})]
\begin{eqnarray}
\nonumber
{\hat \rho}_{A, \mathbf{m}} =
{\rho_{A,\mathbf{m}}
\over
{\rm tr}(\rho_{A,\mathbf{m}})
}
\end{eqnarray}
of an interval $A$ at the final time-slice of the circuit.
 In every realization of circuit disorder (arising from the quantum trajectory $\mathbf{m}$, the Clifford unitaries and measurement locations, the dependence on which is completely or partially suppressed here and from now on,  for ease of notation), the spectrum of ${\hat \rho}_{A, \mathbf{m}}$
is known to be flat\footnote{all non-vanishing eigenvalues of 
${\hat \rho}_{A, \mathbf{m}}$ are equal}: Thus, in every  realization of disorder,  the $n^{\rm th}$ R\'enyi entropies $S_A^{(n)}$,
\begin{eqnarray}
\nonumber
{\rm tr} \left ( {\hat \rho}_{A, \mathbf{m}} \right )^n= 
\exp \{ - (n-1) S_A^{(n)}\},
\end{eqnarray}
 are  all equal, and we consider  without loss of generality the case $n=2$. 
[Again, the dependence
 of $S_A^{(n)}$
 on the quantum trajectory $\mathbf{m}$  and the other  types of disorder is suppressed, which is also done 
 for $G(x_1, x_2)$ defined below.] 

 Specifically, we will consider  the  disorder 
 averages\footnote{Again, the average is over quantum trajectories involving the Born rule 
 probability, over random unitaries and measurement locations.}, denoted by an overbar $\overline{ \ \left (  {. . . }^{} \right ) \ }$, 
of powers  (moments)
of the purity
 \begin{eqnarray}
 \label{LabelEqGsecondRenyiEntropy}
 G(x_1, x_2) := {\rm tr} \left ( {\hat \rho}_{A,\mathbf{m}} \right )^2
 =
 \exp \{
  - S_A^{(2)} 
\},
 \end{eqnarray}
 where $x_1$ and $x_2$ are the endpoints of the interval $A$, i.e.
 $|x_1-x_2| = |A|$: 
At the transition, these moments will all scale with certain critical exponents $X_k$
 \begin{eqnarray}
 \label{LabelEqScalingOfMoments}
\overline{
[G_(x_1, x_2) ]^k
}
\sim
{B_k\over
R_{12}^{2 X_k}},
\end{eqnarray}
where $R_{12}$ is the chord distance
\begin{eqnarray}
\nonumber
R_{12} := {L\over \pi} \sin\left (
{\pi\over L} |x_1-x_2|
\right ),
 \end{eqnarray}
$L$ is the spatial size (with  periodic boundary conditions),
and $B_k$ are non-universal amplitudes  which turn out to be of little interest to 
us\footnote{The relationship of the non-universal amplitudes $B_k$ with the universal results that we discuss below is mentioned in a footnote of Appendix \ref{LabelAppendixSectionDetailsMultifractal}.}. The moment averages in
(\ref{LabelEqScalingOfMoments}) are those of the 2-point correlation function $G(x_1, x_2)$ associated with the so-called boundary condition changing (bcc) operator `twisting' the boundary 
conditions~\cite{VasseurPotterRTN2018,JianVasseurMeasurement2019}  at the endpoints of the interval $A$.
If the dependence of the exponents $X_k$ on the index $k$ is {\it non-linear}, the 
correlation function $G(x_1, x_2)$ is said to obey {\it multifractal} 
scaling~\cite{Pixley2022Multifractality,LudwigHierarchies1990,DuplantierLudwigPRL1991}.
(See Appendix
\ref{LabelAppendixSectionDetailsMultifractal}
for a brief review.)
This means in particular that  the {\it mean}  and the {\it typical} values of $G(x_1, x_2)$ scale with {\it different} critical exponents, and  that the  entanglement transition possesses a {\it continuous spectrum of critical exponents} [corresponding to $X_k$ with continous real values of $k$, arising from non-integer moment 
powers
in (\ref{LabelEqScalingOfMoments})]. Whether multifractal scaling is present or not can be inferred from the  scaling of the statistical fluctuations of $S_A^{(2)}$ as follows.

We know from (\ref{LabelEqGsecondRenyiEntropy}) that, in every realization of disorder, the negative of the 
logarithm of $G(x_1, x_2)$ is the 2nd  R\'enyi  entropy,
\begin{eqnarray}
\nonumber
- \ln G(x_1, x_2)
=  S_A^{(2)},
\quad   \ \ \ (|x_1 - x_2| = |A|).
\end{eqnarray}
As briefly reviewed in Appendix~\ref{LabelAppendixSectionDetailsMultifractal}, a comparison of the cumulant expansion of the left hand side of Eq.~(\ref{LabelEqScalingOfMoments}) with the Taylor expansion of the exponents
$X_k$  in powers of $k$ [appearing on the right hand side of Eq.~(\ref{LabelEqScalingOfMoments})] shows that all cumulants
of the 2nd R\'enyi entropy $S_A^{(2)}$ grow linearly with the logarithm  of the chord distance, 
$\ln R_{12}$. The coefficients of proportionality are universal and equal to  $(-1)^{k-1}$ times twice the
Taylor coefficents of $X_k$ in powers of $k$,
\begin{eqnarray}
\label{LabelFirstCumulant}
\overline{
S_A^{(2)}}
\ && \ 
\sim  \ 2 \ x^{(1)} \ln R_{12}
\\ 
\label{LabelSecondCumulant}
\overline{
\left (
S_A^{(2)} - \overline{S_A^{(2)}}
\right )^2
}
\ && \ 
\sim \ -  2  \ x^{(2)} \ln R_{12}
\\ 
\label{LabelThirdCumulant}
\kappa_3[S_A^{(2)}]
\ && \ 
\sim \ 
2 \ x^{(3)} \ln R_{12}
\\ \nonumber
&& \dots
\end{eqnarray}
where $\kappa_3[S_A^{(2)}]$ denotes the 3rd cumulant of the random variable $S_A^{(2)}$, and
\begin{eqnarray}
\label{LabelTaylorExpansionXN}
X_k = k \  x^{(1)} +{k^2 \over 2!} \  x^{(2)}
+ {k^3\over 3!} \ x^{(3)} + \dots
\end{eqnarray}
Note that (\ref{LabelFirstCumulant}) is nothing but (\ref{eqDefRenyiCircuitsReplica}), expressed in terms of  $G(x_1, x_2)$ defined in (\ref{LabelEqGsecondRenyiEntropy}).

As we summarize below, we find numerically that the universal numbers $x^{(2)}$ and $x^{(3)}$, characterizing  statistical  fluctuations of the entanglement entropy about its mean, are both non-vanishing, and that at least  $x^{(2)}$ is of a magnitude which is a sizeable fraction of $x^{(1)}$, the latter characterizing the {\it averaged} entanglement entropy (\ref{eqDefRenyiCircuitsReplica}). 
It then follows from (\ref{LabelTaylorExpansionXN})
that the dependence of the exponent $X_k$ on the moment-order $k$ is non-linear. As mentioned above, this signifies multifractal scaling of the random variable
$G(x_1, x_2)$ defined in (\ref{LabelEqGsecondRenyiEntropy}), i.e. it implies (in particular) different scaling exponents for the average and typical values of this quantity: The typical exponent describes the universal prefactor factor $x^{(1)}$ multiplying the logarithm of subsystem size of the {\it averaged entanglement entropy}, as in (\ref{LabelFirstCumulant}).
The exponent $X_1$~$(\not = x^{(1)})$ describes the algebraic decay of the {\it averaged purity} [from (\ref{LabelEqGsecondRenyiEntropy}) and (\ref{LabelEqScalingOfMoments}) with $k=1$],
\begin{eqnarray}
\nonumber
\overline{ \ 
 {\rm tr} \left ( {\hat \rho}_{A,\mathbf{m}} \right )^2 \ }
 = \overline{  G(x_1, x_2)} \sim 
{B_1\over
R_{12}^{2 X_1}}.
 \end{eqnarray}

\begin{figure}[hbt!]
    \centering
    \includegraphics[width=.45\textwidth]{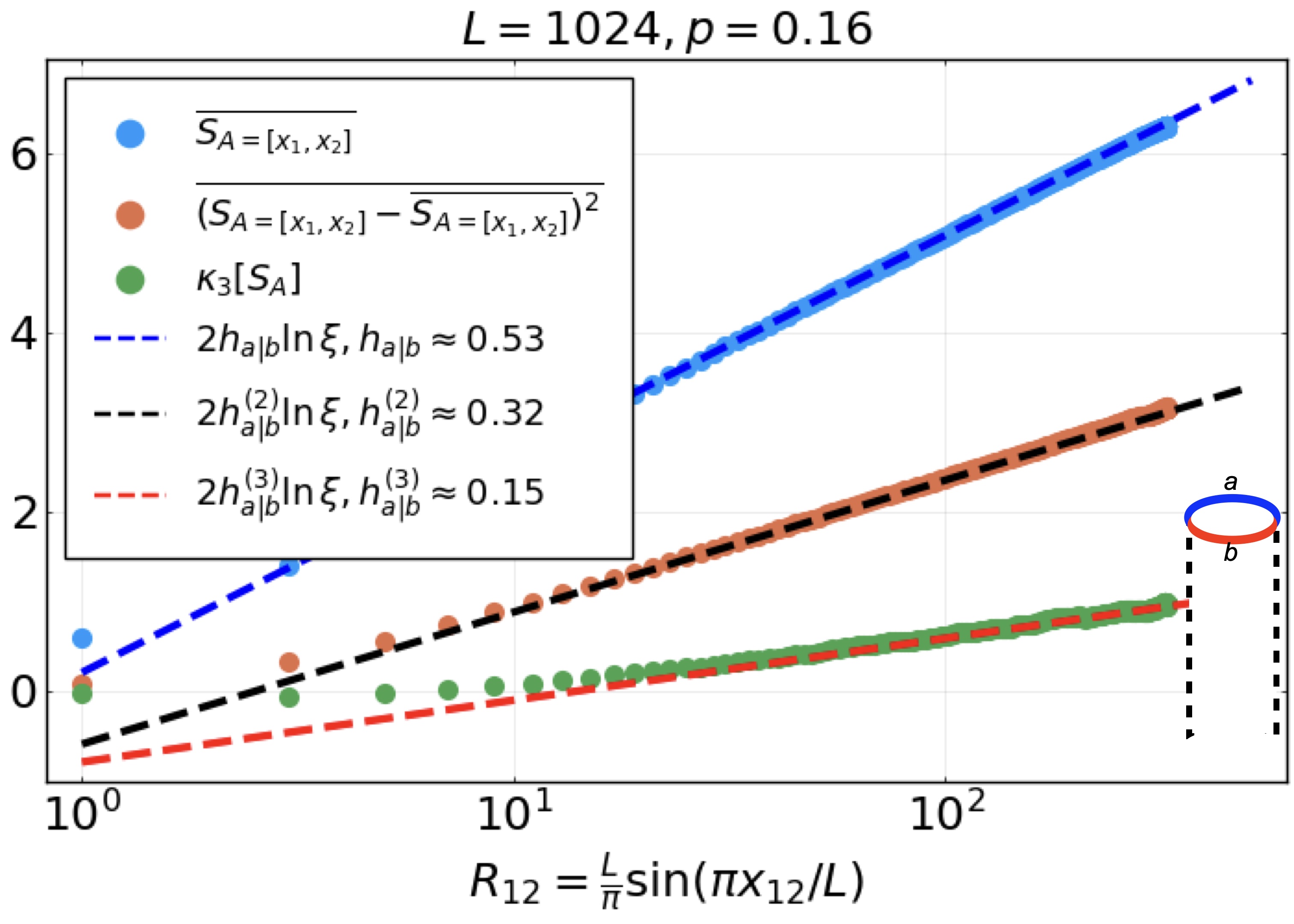}
    \caption{
    Numerical results for the first three cumulants of the second R\'enyi entanglement entropy $S_A^{(2)} =$ $S_{A=|x_1, x_2|}$: Averaged  entanglement entropy (blue) from  (\ref{LabelFirstCumulant}),
    its 2nd cumulant (red) from
    (\ref{LabelSecondCumulant}), 
    and 
    its 3rd cumulant (green) from (\ref{LabelThirdCumulant}),
    versus the chord distance, here denoted by $R_{12}$.
    The slopes of the plots yield the exponent $x^{(1)}=$ $h_{a|b} \approx$ $0.53$, 
     $-x^{(2)}=$ $h^{(2)}_{a|b} \approx$ $0.32$, and $x^{(3)}=$ $h^{(3)}_{a|b} \approx$ $0.15$ - (The notations
    $h_{a|b}$, $h^{(2)}_{a|b}$ and $h^{(3)}_{a|b}$ serve as a reminder that these are the exponents associated with the boundary condition changing (bcc) operator~\cite{VasseurPotterRTN2018,JianVasseurMeasurement2019,Li2020a}.)
    }
    \label{fig: First-and-Second-Third-Cumulant-Purity}
\end{figure}

We now proceed to discuss the numerical results. In Fig.~\ref{fig: First-and-Second-Third-Cumulant-Purity} we plot the numerical values of the first three cumulants of the entanglement entropy $S_A^{(2)}=$ $S_{A=|x_1-x_2|}$ versus the chord distance $R_{12}$ in the Figure: From the slopes of the plots  of the averaged  entanglement entropy (blue) from  (\ref{LabelFirstCumulant}), of  its 2nd cumulant (red) from (\ref{LabelSecondCumulant}), and 
 of  its 3rd cumulant (green) from (\ref{LabelThirdCumulant}),
 we read off the critical exponents $x^{(1)}=$~$h_{a|b} \approx$ $0.53$,
and $-x^{(2)}=$~$h^{(2)}_{a|b} \approx$ $0.32$, and $x^{(3)}=$~$h_{a|b}^{(3)}$~$\approx 0.15$, respectively.
(The notations
$h_{a|b}$, $h^{(2)}_{a|b}$, 
as well as
$h^{(3)}_{a|b}$, 
just serve as a reminder that these are the exponents associated with the boundary condition changing (bcc) 
operator~\cite{VasseurPotterRTN2018,JianVasseurMeasurement2019,Li2020a}, using the notation of \cite{Li2020a}.)


Given the numerical values of these exponents we 
see that the expansion term of
$X_k$ in
(\ref{LabelTaylorExpansionXN})
quadratic in $k$  is of a sizable fraction [$\approx 0.32/(2*0.53)=$~$ 0.302$] of the term linear in $k$.
Thus we conclude that  $X_k$ is clearly not a linear function of $k$ and  that the purity thus exhibits multifractal scaling. The  expansion term of $X_k$ cubic in $k$,
even though small as compared to $x^{(1)}=$~$h_{a|b}$, does not appear entirely negligible, which is presumably indicative of a $k$-dependence of $X_k$ beyond quadratic order.

In conclusion, our numerical study of the statistical fluctuations of the entanglement entropy has 
established beyond doubt multifractal scaling of the purity in our Clifford circuit for prime  $p=2$.
Moreover, as mentioned at the beginning of this subsection, we have no reason to doubt that the scaling behavior of the purity is also multifractal  for on-site  Hilbert space dimensions $d=p^M$ with other prime numbers $p$, except
that also the exponents $x^{(2)}, x^{(3)}, ...$, in addition to  $x^{(1)}=h_{a|b}$ (see TABLE \ref{table:scaling}),
 are expected to  depend explicitly on the prime  $p$, based on our results in the rest of this paper.


\section{Discussion}

In this work, we have introduced a replica statistical mechanics mapping to compute entanglement properties of stabilizer random tensor networks, and monitored Clifford quantum circuits acting on qudits with 
$d=p^M$. The degrees of freedom  (`spins') of the resulting statistical models belong to the commutant of the Clifford group, using a generalization of the concept of Schur-Weyl duality to the Clifford case. We find that the Boltzmann weights are invariant under the left and right action of the stochastic orthogonal group with entries in the finite number field~${\bf F}_p$. This implies, suprisingly, that the universal properties of the entanglement transitions in Clifford circuits and RTNs depend on the on-site Hilbert space dimension $d$ but only through the prime $p$. This is in sharp constrast with the Haar case where the symmetry group is independent of $d$, as long as it is not infinite. Our formalism also allowed us to derive exact mappings onto classical percolation and minimal cut formulas as $d=p^M$ becomes large, with different consequences for $p$ fixed and $M \to \infty$, or $M$ fixed and $p \to \infty$. 
\andreas{We confirmed our predictions using large scale stabilizer numerics, which also showed that 
projective and forced measurement schemes yield the same finite-d
critical exponents for Clifford circuits.}

Our \andreas{large-$d$ results}
here bolster a known result about stabilizer tensors.
Random stabilizer tensors with large bond dimension are perfect tensors with high probability~\cite{apel2021holographic, RTN, HaPPY_2015}.
A perfect tensor has the special property of being a unitary gate for any equal-size bipartition of its legs.
Thus, a Clifford circuit or RTN as $d \to \infty$ becomes an assembly of perfect tensors, for which the entanglement entropies are given exactly by the minimal cut through the underlying square lattice with randomly broken bonds.

\andreas{Finally, for $p=d=2$ we established that the purity of the reduced density matrix of a finite interval exhibits multifractal scaling at the transition. This implies different critical scaling exponents for the averaged and typical purity (the latter being equal to the prefactor of the logarithm of subsystem size in the entanglement entropy), and a continuous spectrum of  critical exponents associated with the scaling of (non-integer moment averages of)  the purity. The same features are expected to be present for (powers of) other primes $p$, while the corresponding exponents will vary with $p$ for the symmetry reasons discussed in this paper.}

Our approach could be generalized to other subgroups of the unitary group: formally, the main algebraic object that is needed would be the commutant of the subgroup. Once the commutant is known, most of the formulas derived in this manuscript carry over -- including the definition of the Weingarten functions {\it etc}. It would be interesting to investigate whether other universality classes of entanglement transitions can be obtained in this way. As in the Haar case, 
an analytic understanding of the replica limit of the statistical mechanics models  for finite $d$ remains a clear challenge for future work. One particularly intriguing aspect of our results is that while the statistical mechanics approach is certainly very powerful and allowed us to uncover the intricate dependence of the symmetry group on the prime number $p$, some aspects of Clifford circuits that are ``simpler'' (such as the ability to simulate them classically, or the flat entanglement spectrum of the reduced density matrix ) are not 
yet particularly transparent in the current formulation of our framework. Those special properties of the Clifford group might posssibly lead to some simplifications in the statistical mechanics formulation, that we leave for future work.

\section*{Acknowledgments}
We thank Utkarsh Agrawal, Xiao Chen, Sarang Gopalakrishnan, Michael Gullans, David Huse, Chao-Ming Jian, Vedika Khemani, Jed Pixley, Andrew Potter, Yi-Zhuang You, Justin Wilson, Zhi-Cheng Yang, and Aidan Zabalo for useful discussions and collaborations on related projects.
We acknowledge support from the Air Force Office of Scientific Research under Grant No. FA9550-21-1-0123 (R.V.), and the Alfred P. Sloan Foundation through a Sloan Research Fellowship (R.V.).
This work was supported by the Heising-Simons Foundation (Y.L. and M.P.A.F.), and by the Simons Collaboration on Ultra-Quantum Matter, which is a grant from the Simons Foundation (651440, M.P.A.F.), and was
supported in part by the National Science
Foundation under Grant No. DMR-1309667~(A.W.W.L.).
Use was made of computational facilities purchased with funds from the National Science Foundation (CNS-1725797) and administered by the Center for Scientific Computing (CSC). The CSC is supported by the California NanoSystems Institute and the Materials Research Science and Engineering Center (MRSEC; NSF DMR-1720256) at UC Santa Barbara.

\appendix

\section{Description and basic properties of the commutant $\Sigma_Q(p)$ for the Clifford case}
\label{LabelAppendixDescriptionAndPropertiesOfCommutant}

The following statements (from Ref.~\cite{GrossEtAlCMP2021}) lead to the full  characterization of the commutant $\Sigma_Q(p)$,
which is presented in statement (iv) below, together with
its properties relevant for the 
construction of the Clifford Stat. Mech. model, summarized in (v)-(vii):

\vskip .4cm

\noindent -(i): {\it (Stochastic) Orthogonal Group}

For every $O\in {\cal O}_Q(p)$ the subspace
\begin{eqnarray}
\label{LabelEq-DEF-TO}
T_O :=
\{  (O \vec{\mu}, \vec{\mu}): \ \ \vec{\mu} \in ({\bf Z}_p)^Q\}
\end{eqnarray}
is an element of $\Sigma_Q(p)$, and
any operator ${\hat r}(T)$ as introduced in (\ref{LabelEqClifford-r-DEF})
is invertible if and only if
$T$ is of the form $T=T_O$ for some $O\in {\cal O}_Q(p)$. Clearly, it follows from
(\ref{LabelEqClifford-r-DEF}) that the operator ${\hat r}(T_O)$  maps basis
element $|\vec{\mu}\rangle$ into basis element
$|O\vec{\mu}\rangle$. We will denote this operator simply by ${\hat r}(O)$,
and  we  write simply
${\hat r}(T_O)
={\hat r}(O)$.

\vskip .4cm

\noindent -(ii): {\it Left and Right multiplication by (Stochastic) Orthogonal Group}

For any subspace $T\in \Sigma_Q(p)$
as introduced in (\ref{LabelEqClifford-r-DEF}),
and  for any stochastic orthogonal matrix $ O\in {\cal O}_Q(p)$,
the objects
\begin{eqnarray}
\label{LabelEq-DEF-TO-and-OT}
OT := \{ (O\vec{\nu},\vec{\mu}):  \ (\nu, \mu)\in T\} 
\end{eqnarray}
and
\begin{eqnarray}
\nonumber
TO :=  \{ (\vec{\nu},O^t\vec{\mu}):  \ (\nu, \mu)\in T\} 
\end{eqnarray}
are themselves again elements of $\Sigma_Q(p)$, i.e.
$OT, TO \in \Sigma_Q(p)$.

\vskip .4cm

\noindent -(iii): {\it Compatibility with ${\hat r}(T)$}

The action with stochastic orthogonal matrices $O, O' \in {\cal O}_Q(p)$
on subspaces $T\in \Sigma_Q(p)$, defined in (ii) above, translates naturally into the action on the corresponding
operator ${\hat r}(T)$:
\begin{eqnarray}
\label{LabelEq-Representation-Property}
&&
{\hat r}(O) {\hat r}(T) {\hat r}(O')=
\\ \nonumber
&&
={\hat r}(OT) {\hat r}(O')=
{\hat r}(O) {\hat r}(TO')=
{\hat r}(O T O'). 
\qquad
\qquad
\end{eqnarray}

\vskip .4cm

\noindent -(iv): {\it Double Coset Decomposition of Commutant}

Statement (iii) above shows that the commutant $\Sigma_Q(p)$
is a disjoint union of {\it double cosets} with respect to the left- and right-
actions of the stochastic orthogonal group, i.e.
\begin{eqnarray}
\label{LabelEq-DoubleCosetDecomposition}
&&
\Sigma_Q(p)
=
\\ \nonumber
&&
=
{\cal O}_Q(p)\  {\bar T}_1\  {\cal O}_Q(p)
\  \cup \  ... \  \cup \ 
{\cal O}_Q(p)\  {\bar T}_k \  {\cal O}_Q(p),
\end{eqnarray}
where ${\bar T}_1, ..., {\bar T}_k$ denote representatives of different double cosets. It can be 
shown~\cite{GrossEtAlCMP2021} that $k\leq Q$. Here ${\bar T}_1=\Delta$ is the ``diagonal subspace''
\begin{eqnarray}
\label{LabelEq-DEF-Delta}
\Delta := \{ (\vec{\mu}, \vec{\mu}): \ \ \vec{\mu}\in ({\bf Z}_p)^Q \},
\end{eqnarray}
so that
${\cal O}_Q(p)\  {\bar T}_1\  {\cal O}_Q(p) ={\cal  O}_Q(p)$.

\vskip .5cm
\noindent  -(v): {\it Choice of representatives for Double Cosets}

The operators ${\hat r}(T)$ associated with elements $T$ of the commutant that are not in the first
double coset ${\cal O}_Q(p)\  {\bar T}_1 \  {\cal O}_Q(p)={\cal O}_Q(p)$ are not invertible. Representatives
${\bar T}_a$ with $a=2, ..., k$ can be chosen\footnote{upon right- and left- multiplication with suitable stochastic orthogonal matrices, and making use of Theorem 4.24 of Ref. [\cite{GrossEtAlCMP2021}]}
such that ${\hat r}({\bar T}_a)$ are 
projectors.
In particular, one can choose  double coset representatives ${\bar T}_a$, $a=2, 3, ..., k$ 
(and the notation ${\bar T}_a$ denotes from now on that such a choice has been made)
so that each is associated\footnote{The subscript ${}{CSS}$ stands for 
the so-called Calderbank-Shor-Steane codes~\cite{1996PhRvA..54.1098C, 1996RSPSA.452.2551S, PhysRevLett.77.793}
known from quantum information theory.}
with a projector $P_{CSS(N_a)}$
possessing a so-called `defect subspace' $N_a \subseteq ({\bf Z}_p)^Q$ of dimension ${\rm dim}(N_a)$, 
and
\begin{eqnarray}
\label{LabelEq-PCSS-N-Projectors}
{\hat r}({\bar T}_a)= p^{{\rm dim}(N_a)}
P_{CSS(N_a)}, \quad (a=2, 3, ..., k)
\end{eqnarray}
with trace
\begin{eqnarray}
\label{LabelEq-Trace-PCSS-N-PRojectors}
{\rm tr} \left [{\hat r}({\bar T}_a)\right]=p^{Q-{\rm dim}(N_a)}.
\end{eqnarray}
It is convenient to also introduce the notation ${\bar T}_1=\Delta$ and $N_1=0$.

\vskip .5cm

\noindent -(vi): {\it `Composition property'}

Consider two  arbitrary elements $T_1, T_2 \in \Sigma_Q(p)$ of the commutant. Each can then be reduced, upon right and left
multiplication with elements
in ${\cal O}_Q(p)$, to one of the 
elements\footnote{or to elements that are conjugate to those upon conjugation in ${\cal O}_Q(p)$, which
then have again the same right and left defect spaces, which are images under  ${\cal O}_Q(p)$ of the 
original ones.} 
${\bar T}_a$, where $a=1, 2, ..., k$, and thus can
be assigned a (possibly trivial) defect space $N_a$. Then the following composition
property holds
\begin{eqnarray}
\label{LabelEq-CompositionProperty}
{\hat r}(T_1) {\hat r}(T_2)
=p^{{\rm dim}(N_1 \cap N_2)} \ {\hat r}(T_1 T_2).
\end{eqnarray}

\vskip .5cm

\noindent -(vii): {\it Trace}

The commutant $\Sigma_Q(p)$ is equal to the disjoint union of  sets $\Sigma_Q^\ell(p)$, where $\ell=0, 1, ..., (Q-1)$,  whose elements are  those elements $T\in \Sigma_Q(p)$ with the property that ${\rm dim}(T\cap \Delta)=Q-\ell$.
Then, by the definition (\ref{LabelEqClifford-r-DEF}) of ${\hat r}(T)$ and the definition of the trace, one
has\footnote{recall (\ref{LabelEq-DEF-Delta})}
\begin{eqnarray}
\label{LabelEq-DEF-trace}
{\rm tr} [  {\hat r}(T)]=p^{Q-\ell}, 
\quad 
{\rm for \  all \  elements} \  T \in \Sigma_Q^\ell(p).
\qquad
\end{eqnarray}

\vskip .5cm

\noindent -(viii): {\it `Metric'  on an edge}

For any two elements of the commutant $\Sigma_Q(p)$, in analogy with (\ref{LabelEq-Haar-edge-BoltzmanWeight})
in the Haar case, the Boltzmann weight on an edge of the Clifford RTN is built from
\begin{eqnarray}
\label{LabelEq-Clifford-edge-Metric}
{1\over p^Q}
\ {\rm tr} \left (
\left [ {\hat r}(T_1)\right ]^\dagger  {\hat r}(T_2) 
\right )
=
\ e^{ - \ln(p) \ |T_1, T_2|}
\end{eqnarray}
where is 
$|T_1, T_2|$
a metric defined by collecting 
(\ref{LabelEq-PCSS-N-Projectors},\ref{LabelEq-Trace-PCSS-N-PRojectors},
\ref{LabelEq-CompositionProperty},\ref{LabelEq-DEF-trace}). It satisfies the usual conditions
$|T_1, T_2|=|T_2, T_1| \geq 0$, and $|T_1,T_2|=0$ if and only if $T_1=T_2$. {We refer the reader to appendix~\ref{appendixMetric} for additional details. }

As a check, let us briefly compute the metric in the case where $T_1=T_2=T$, and thereby
confirm that the above rules indeed yield the expected result $|T,T|=0$. There are two
cases to consider. In the first case, $T\in {\cal O}_Q(p)$ is  itself a group element, i.e.
an element of the first
double coset with `trivial' representative $T_1=\Delta$. In this case
the defect spaces are trivial (of dimension zero), and the trace in
(\ref{LabelEq-Clifford-edge-Metric}) is that of the identity operator
acting on the (replicated) edge Hilbert space $({\bf C}^p)^Q$, yielding $p^Q$.
This is cancelled by the prefactor in (\ref{LabelEq-Clifford-edge-Metric})
with the result that $|T,T|=0$ in this case, as expected.
In the second case, $T$ is an element of  one of the remaining (k-1) double cosets and is thus,  upon
right and left multiplication with possibly different group 
elements\footnote{These group elements disappear when forming the trace in
(\ref{LabelEq-Clifford-edge-Metric}).}
in ${\cal O}_Q(p)$, equal to
one of the `projector' representatives ${\bar T}_a$ with $a=2, ...Q$. 
In this case ${\bar T}_a=$
$P_{CSS(N_a)}$ is a 
projector satisfying $({\bar T}_a)^2
={\bar T}_a$, 
and (\ref{LabelEq-PCSS-N-Projectors},\ref{LabelEq-Trace-PCSS-N-PRojectors},
\ref{LabelEq-CompositionProperty}) 
yield for
(\ref{LabelEq-Clifford-edge-Metric})
the expression
$$
{1\over p^Q} \ p^{{\rm dim}(N_a)} \ p^{Q- {\rm dim}(N_a)}=1
$$
in agreement with the expected result $|T,T|=0$.

\vskip .4cm

\noindent -(ix): {\it Symmetry of the `Metric'}

It follows from its definition in (\ref{LabelEq-Clifford-edge-Metric}) 
and from the compatibility property (\ref{LabelEq-Representation-Property})
that the metric $|T_1, T_2|$ is invariant
under the symmetry group
${\cal O}_Q(p)\times {\cal O}_Q(p)$, 
i.e. under right and left multiplication
with  arbitrary elements of the stochastic orthogonal 
group,
\begin{eqnarray}
\label{LabelEqSymmetryOfMetric}
&&
|T_1, T_2| = |T'_1, T'_2|, \ \ \ {\rm where}  \ \  T_1, T_2 \in \Sigma_Q(p), 
\\ \nonumber
&&
{\rm and}
\\ \nonumber
&&
T'_1=O_L T_1 O_R^{-1}, \ T'_2=O_L T_2 O_R^{-1}, \ {\rm with} \
O_L, O_R \in {\cal O}_Q(p).
\end{eqnarray}

{
Just like in the Haar case, the metric is also invariant under a $\mathbb{Z}_2$ symmetry that corresponds to exchanging ket and bra. To see this, for each element of the commutant $T\in \Sigma_Q(p)$, we define $T^\dagger$ as
\begin{eqnarray}
\nonumber
{\hat r}(T)&=&{\hat r}(O_L) {\hat r}(\bar{T}){\hat r}(O_R)=
{\hat r}(O_L \bar{T}O_R),
\\  \nonumber
{\hat r}^\dagger(T)&=&{\hat r}(O^t_R) {\hat r}(\bar{T}){\hat r}(O^t_L)=
{\hat r}(O^t_R \bar{T}O^t_L) := {\hat r}(T^\dagger), 
 \\  \nonumber {\rm where} \ \ {\hat r}^\dagger(\bar{T})&=&{\hat r}(\bar{T}), \ \bar{T}^\dagger=\bar{T}.
\\ \nonumber
{\rm Thus:} \quad T&=&O_L \bar{T}O_R, \ \ 
T^\dagger:=O^t_R \bar{T}O^t_L,
\end{eqnarray}
where we have used~\eqref{LabelEq-Representation-Property},~\eqref{LabelEq-DoubleCosetDecomposition}, and~\eqref{LabelEq-PCSS-N-Projectors}. 
With this notation in hand, we establish  the following simple property of the metric,
\begin{eqnarray}
\label{LabelEqMetricTTdagger}
|T_a, T_b| = |T_a^\dagger, T_b^\dagger|,
\end{eqnarray}
which follows from eq.~\eqref{LabelEq-Clifford-edge-Metric} and
\begin{eqnarray}
\label{LabelEqElementaryDEFMetric}
{\rm tr} \Bigl ( [{\hat r}(T_a)]^\dagger \ {\hat r}(T_b) \Bigr )
&=& \sum_{({\vec \nu}_a, {\vec \mu}_a) \in T_a} \sum_{({\vec \nu}_b, {\vec \mu}_b) \in T_b}  \delta_{{\vec \nu}_a,{\vec \nu}_b}
\delta_{{\vec \mu}_a,{\vec \mu}_b} \notag \\
&=& 
{\rm tr} \left ( {\hat r}(T_a) \ \left [{\hat r}(T_b)\right ]^\dagger \right ),
\end{eqnarray}
using eq.~\eqref{LabelEqClifford-r-DEF}.
We note that 
eq.~(\ref{LabelEqElementaryDEFMetric}), {in conjuction with
(\ref{LabelEq-Clifford-edge-Metric})},
also provides an 
{explicit} elementary definition of the metric.} 

{It is straightforward to show that the Weingarten functions defined in Appendix~\ref{AppendixWClifford} below also 
{satisfy\footnote{{which follows from (\ref{LabelEqCliffWeinLeftInverse}) and  the paragraph below this equation, upon making use of (\ref{LabelEqMetricTTdagger})}}}
$Wg_D(T^\dagger_a, T^\dagger_b) = Wg_D(T_a, T_b)$, which also implies that they are 
real\footnote{{which follows  from (\ref{LabelEqExpansionTensorProductCommutantsClifford}) due to the manifest Hermiticity of ${\hat Y}$ defined in (\ref{LabelEqDEF-Y-hat}) [analogous to the Haar case as discussed in App (\ref{AppendixWHaar})].}}. Thus both principal ingredients of the Clifford statistical mechanics model for 
{monitored circuits and RTNs\footnote{{RTNs in the ``forced measurement'' formulation discussed in Sect.~\ref{LabelSubsectionForcedMeasurements} also require the use  the Weingarten function}}},
namely the Weingarten function $Wg_D(T_i, T_j)$ as well as the metric $|T_i, T_j|$, are invariant under the $\mathbb{Z}_2$ symmetry which sends  $T_i \to T_i^\dagger$ at all sites. }

\section{Basic properties of the Clifford Group, Stabilizer States, etc.}
\label{LabelAppendixPropertiesCliffordGroupStabilizerStates}

\noindent - (1)
Single qudit ($d=p$=prime), Hilbert space ${\cal H}_1 = {\bf C}^p$, computational basis
$|q\rangle$ with $q\in \{ 0, ..., (p-1)\}={\bf Z}_p$. {\it Pauli operators}
\begin{eqnarray}
\nonumber
&&
{\hat X} |q\rangle = |q+1\rangle, \quad {\hat Z} |q \rangle = \omega^q |q \rangle,
\quad (\omega=e^{2 \pi i q/d})
\\ \nonumber
&&
{\hat Y} := \tau {\hat X}^\dagger {\hat Z}^\dagger, \quad ({\rm where} \ \tau = {\rm some \ suitable \ phase});
\\ 
\label{LabelEqGeneratorsOfPauliGroup}
&&
{\rm Then:} \ \ {\hat X} {\hat Y} {\hat Z} = \tau {\bf I}
\end{eqnarray}

\vskip .2cm

\noindent - (2) {\it Hilbert space} for $n$ qudits (d=p=prime) ${\cal H}_n=({\bf C}^p)^n$,
computational basis $|{\bs q}\rangle=$ $(q_1, ..., q_n)\in ({\bf Z}_p)^n$.

\vskip .2cm

\noindent - (3)  {\it Pauli Group} ${\cal P}_n$ is the group generated by the tensor product of all Pauli operators
on the $n$ qudits, or, equivalently, owing to (\ref{LabelEqGeneratorsOfPauliGroup}), by
the group generated by the
tensor product of the operators $\tau {\hat {\bf I}},{\hat X}, {\hat Z}$.

\vskip .2cm

\noindent - (4) {\it Clifford group} ${\rm Cliff}(n, p)$ of the Hilbert space ${\cal H}_n=({\bf C}^p)^n$
of $n$ qudits is the normalizer of the Pauli group in the unitary group acting on
${\cal H}_n=({\bf C}^p)^n$, i.e. the group of all unitary operators ${\hat V}$ acting on ${\cal H}_n=({\bf C}^p)^n$
satisfying ${\hat V} {\cal P}_n {\hat V}^\dagger \subset {\cal P}_n$, up to phases.

\vskip .2cm

\noindent - (5) For a subgroup of the Pauli group, ${\cal S} \subset {\cal P}_n$ [one
that does not contain any (non-trivial) multiple of the identity operator] the operator
\begin{eqnarray}
\nonumber
{\hat {\bs P}}_{\cal S} := {1\over |{\cal S}|} \sum_{{\hat P} \in {\cal S}}
{\hat P},
\end{eqnarray}
is an orthogonal projector onto a subspace\footnote{$V_{\cal S}$ is called a ``stabilizer code''.} $V_{\cal S} \subset {\cal H}_n$ of dimension $p^n/|{\cal S}|$. When $|{\cal S}|=p^n$, the projection is onto a single state
$|{\cal S}\rangle$ called a  {\it stabilizer state} (i.e.
$V_{\cal S}={\bf C} |{\cal S}\rangle$), which is thus the unique (up to scalars) eigenvector of all Pauli operators
${\hat P}$ in ${\cal S}$,
\begin{eqnarray}
\nonumber
{\hat P} |{\cal S}\rangle = |{\cal S}\rangle, \quad ({\rm for \ all}\ {\hat P}\in {\cal S})
\end{eqnarray}
The (finite) {\it set of all (pure) stabilizer states} in ${\cal H}_n=({\bf C}^p)^n$ is denoted by ${\rm Stab}(n, p)$.
The number of stabilizer states~\cite{Gro06}
in ${\rm Stab}(n, p)$ is
$=p^n \prod_{i=1}^n (p^i +1)$.

\vskip .2cm

\noindent -  (6) For every stabilizer state $|{\cal S}\rangle$ in ${\cal H}_n=({\bf C}^p)^n$
there exists
some Clifford unitary ${\hat V}\in {\rm Cliff}(n,p)$ such that
\begin{eqnarray}
\nonumber
|{\cal S}\rangle = {\hat V} \left ( |0\rangle^{\otimes n}\right ),
\ \  [{\rm where} \ |0\rangle = |q =0\rangle, \ {\rm with} \  q=0 \in {\bf F}_p].
\end{eqnarray}
It follows from this statement that the set of stabilizer states 
${\rm Stab}(n, p)$ of the qudit Hilbert space 
${\cal H}_n=({\bf C}^p)^n$ is a {\it single orbit} of the action under 
the Clifford group ${\rm Cliff}(n, p)$.

\vskip .2cm

\noindent - (7) There is a simple way to see that the state
$|0\rangle^{\otimes n}
\in {\cal H}_n=({\bf C}^p)^n$, is a stabilizer state, i.e. an element of 
the set ${\rm Stab}(n, p)$ of stabilizer states.

The argument uses the notation of {\it Weyl operators}
\begin{eqnarray}
\nonumber
{\hat W}_{\bf x}= {\hat W}_{\bf p, q}=
\tau^{-{\bf p}\cdot{\bf q}}
\left ({\hat Z}^{p_1} {\hat X}^{q_1}\right) 
\otimes ... \otimes
\left ({\hat Z}^{p_n} {\hat X}^{q_n}\right),
\end{eqnarray}
where ${\bf x}=({\bf p, q}) \in {\cal V}_n := ({\bf Z}_p)^{2n}$ (``phase space'').
Each Weyl operator is
an element of the Paul group
${\cal P}_n$; moreover, each element of the Pauli group is, up to a  phase, equal to a Weyl operator.
It is obvious
[using
(\ref{LabelEqGeneratorsOfPauliGroup})]
that the state
$|0\rangle^{\otimes n}\in {\cal H}_n$ is the simultaneous eigenvector of the set of all Weyl operators
${\hat W}_{\bf x}$
with
${\bf x}=({\bf p, q})$ where ${\bf q=0}$, i.e.
\begin{eqnarray}
\label{LabelEqWp0StabilizerGroup}
{\hat W}_{\bf p, 0}|0\rangle^{\otimes n}
=|0\rangle^{\otimes n}.
\end{eqnarray}
This particular set of these Weyl operators thus represents the subgroup ${\cal S}\subset {\cal P}_n$
of the Pauli group, which ``stablizes'' the state $|0\rangle^{\otimes n}$. Note that the
order of this subgroup of the Pauli group is 
$|{\cal S}|=p^n$, as required.

\vskip .2cm

\noindent - (8) Futhermore, it is now possible to show that ${\hat V}|0\rangle^{\otimes n}$,
where ${\hat V} \in {\rm Cliff}(n,p)$ is an arbitrary Clifford unitary operator in ${\cal H}_n$,
is also a ``stablilizer state'', i.e. an element of 
the set ${\rm Stab}(n, p)$ of stabilizer states. To see this, rewrite
(\ref{LabelEqWp0StabilizerGroup}) as
\begin{eqnarray}
\nonumber
\left ( {\hat V} {\hat W}_{\bf p, 0} {\hat V}^\dagger \right ) \ \left ( {\hat V}|0\rangle^{\otimes n}\right )
= \left ( {\hat V} |0\rangle^{\otimes n} \right ),
\end{eqnarray}
which says that the group of $p^n$ operators
$\left ( {\hat V} {\hat W}_{\bf p, 0} {\hat V}^\dagger \right )$
``stablizes'' the state $\left ( {\hat V}|0\rangle^{\otimes n}\right )$.
Owing to Eq. (2.8) of \cite{GrossEtAlCMP2021} we know that
\begin{eqnarray}
\nonumber
{\hat V} {\hat W}_{\bf p, 0} {\hat V}^\dagger 
=
\omega^{
f_{\hat V}({\bf p, 0})}
\ {\hat W}_{\Gamma_{\hat V}({\bf p,0})},
\end{eqnarray}
where $\Gamma_{\hat V}({\bf p,0}) \in Sp(2n, p)$ and $f_{\hat V}({\bf x})$
is a suitable function on phase space ${\cal V}_n$ [defined under item (7) above -
$Sp(2n, p)$ 
is the symplectic group with entries in the field ${\bf F}_p$.]
Then since,
as also mentioned under item (7) above, every Weyl operator
 is an element of the Pauli group, the group $\{
 {\hat V} {\hat W}_{\bf p, 0} {\hat V}^\dagger:
 {\bf p}\in ({\bf Z}_p)^n
 \}$
forms a subgroup of the Pauli group ${\cal P}_n$
and can be identified with the ``stabilizer'' subgroup of the
stablizer state ${\hat V}|0\rangle^{\otimes n}$.

\vskip .2cm

\noindent - (9) {\it Projective Clifford Group, Pauli Group, and orders of groups}

Since we are interested in acting with elements of the Clifford group 
${\hat V}\in{\rm Cliff}(n,p)$ on a density matrix of a pure state $|\psi\rangle$,
i.e. ${\hat V} |\psi\rangle \ \langle \psi| {\hat V}^\dagger$, we can work with elements of the Clifford group 
modulo phases. We define the projective Clifford group as the Clifford group modulo phases, i.e.
${\overline{\rm Cliff}}(n,p):={\rm Cliff}(n,p)/U(1)$. Similarly, we call ${\overline{\cal P}}_n$
the Pauli group modulo phases. It can then be shown~\cite{GrossEtAlCMP2021,Gro06} 
that
\begin{eqnarray}
\nonumber
{\overline{\rm Cliff}}(n,p)/{\overline{\cal P}}_n={\rm Sp}_{2n}(p),
\end{eqnarray}
where, again, the latter is the symplectic group with entries in the field ${\bf F}_p$. Since the dimension of this
symplectic group is 
known~\cite{Gro06}
to be
$|{\rm Sp}_{2n}(p)|=$
$p^{n^2} \prod_{i=1}^n (p^{2i}-1)$, we know that the order of the projective Clifford group is
\begin{eqnarray}
\label{LabelEqOrderProjectiveCliffordGroup}
&& |{\overline{\rm Cliff}}(n,p)|
=p^{2n} p^{n^2} \prod_{i=1}^n (p^{2i} -1),
\\ \nonumber
&&
\end{eqnarray}
where we used the fact (see item (7) above) that the projective Pauli Group is simply the set of all Weyl operators (of which there
are as many as elements in ``phase space'' ${\cal V}_n$, defined above).

\vskip .2cm

\noindent -(10) {\it Average over the Clifford Group}

It turns out\cite{GrossEtAlCMP2021} that, for an {\it arbitrary} (meaning: not a `stabilizer state') state 
$|\psi\rangle \in {\cal H}_n$
the  average
\begin{eqnarray}
\label{LabelEqCliffordAverageNonStabilizerStates}
&&\sum_{{\hat V} \in {\rm Cliff}(n,p)}
\left [{\hat V} |\psi\rangle \ \langle \psi| {\hat V}^\dagger\right ]^{\otimes Q},
\\ \nonumber
&&
\end{eqnarray}
 will, while still (obviously) expressible as a linear combination of the elements
${\hat R}(T)$ of the
commutant, in general not simply be an equal-weight superposition of the latter, but may 
depend\footnote{Certain linear combinations of expressions (\ref{LabelEqCliffordAverageNonStabilizerStates})
with different non-stablilizer states $|\psi\rangle$ can even be shown to represent Haar averages~\cite{GrossEtAlCMP2021}.}
on details of the
state $|\psi\rangle$ itself.
However, if we take $|\psi\rangle$ to be any {\it stabilizer state} 
$|{\cal S}_0\rangle\in {\rm Stab}(n,p) \in {\cal H}_n$, the sum 
can be expressed as an equal-weight sum over the commutant, namely
\begin{eqnarray}
\label{LabelEqSumOverClifford}
&&
{1\over
{\cal N}_{\rm norm}
}
\sum_{{\hat V} \in {\rm Cliff}(n,p)}
\left [{\hat V} |{\cal S}_0\rangle \ \langle {\cal S}_0| {\hat V}^\dagger\right ]^{\otimes Q}
=
\\ \nonumber
&&
=
\sum_{|{\cal S}\rangle \in {\rm Stab}(n,p)}
\Big [|{\cal S}\rangle \ \langle {\cal S}|\Big]^{\otimes Q}
=
{1\over Z_{n,p,Q} }
\sum_{T \in \Sigma_Q(p)} {\hat R}(T),
\qquad
\end{eqnarray}
where ${\cal N}_{\rm norm}$ and 
\begin{eqnarray}
\nonumber
Z_{n,p,Q}
=p^n\prod_{k=1}^{(Q-2)} (p^k+p^n),
\end{eqnarray}
are normalization
factors. The first  equality follows because the set ${\rm Stab}(n,p)$
of stabilizer states is (see item (6) above)  a single orbit under the action of the Clifford 
group on stabilizer states.\footnote{If the first sum is over the projective Clifford group ${\overline{\rm Cliff}}(n,p)$, which gives the same result up to possibly an overall multiplicative constant,
it follows from
(\ref{LabelEqOrderProjectiveCliffordGroup})
that the normalization factor would in this case read
${\overline{\cal N}}_{\rm norm}=$
$p^n p^{n^2}\prod_{i=1}^n(p^i-1)$, because~\cite{Gro06}
every element of ${\overline{\rm Cliff}}(n,p)$ is a product of an element of the Pauli group (modulo phases)
and a (projective) representation of the symplectic group $Sp_{2n}(p)$. The result
follows because  a subgroup of order $p^n$
of the Pauli group
is the stabilizer group of the stabilizer state, and the length of the orbit is $|{\rm Stab}(n,p)|=$
$p^n \prod_{i=1}^n (p^i+1)$ (as mentioned above).
}

\section{Weingarten 
 functions for Haar} 
 \label{AppendixWHaar}
Let us consider the 
operator\footnote{The last two lines are obtained from the 2nd line by inserting a complete set of
$D$ states in each of the $Q$ tensor factors, e.g., ${\hat U}=$
$|\bar{i}_1\rangle \ U_{ \bar{i}_1, j_1} \ \langle j_1|$, etc. -- we use the notation
$|\bar{i}_1, ..., \bar{i}_Q\rangle=$
$|\bar{i}_1\rangle \otimes  ... \otimes |\bar{i}_Q\rangle$, etc..}
(repeated indices summed)
\begin{eqnarray}
\nonumber
&&{\hat X} :=
\mathbb{E}_{{\hat U}\in U(D)} \ 
\big [{\hat U} ( ...) {\hat U}^\dagger\big ]^{\otimes Q}=
\\ \nonumber
&&
=\int_{\rm Haar} d \mu ({\hat U})
\ 
\big [{\hat U} ( ...) {\hat U}^\dagger\big ]^{\otimes Q}
=
\\ \nonumber
&&
=\int_{\rm Haar} d\mu(U) \ 
U_{\bar{i}_1, j_1}
...
U_{\bar{i}_Q, j_Q}
U^\dagger_{\bar{j}_1, i_1}
...
U^\dagger_{\bar{j}_Q, i_Q} \times
\qquad 
\\
\nonumber
&&
\times
\big (
|\bar{i}_1, ..., \bar{i}_Q\rangle \langle i_1, ..., i_Q|
\big )
\otimes
\big (
|\bar{j}_1, ..., \bar{j}_Q\rangle \langle j_1, ..., j_Q|
\big )
\\ 
\label{LabelEqDEF-X-hat}
\end{eqnarray}
where 
$i_1, ...,\bar{i}_Q, j_1, ... \bar{j}_Q \in \{1, ..., D\}$.
We can view the second tensor  factor in the 4th line, $\big (
|\bar{j}_1, ..., \bar{j}_Q\rangle \langle j_1, ..., j_Q|
\big )$, as defining an orthonormal basis of the $Q$-fold tensor product
of an ``in'' vector space of operators
${\cal H}^{\rm op}_{\rm in}$, i.e. of $\Big ({\cal H}^{\rm op}_{\rm in}\Big )^{\otimes Q}$.
In the same way,  we can view the first tensor factor in the 4th line,
$\big (
|\bar{i}_1, ..., \bar{i}_Q\rangle \langle i_1, ..., i_Q|
\big )$,  as defining an orthonormal basis
of the $Q$-fold tensor product an ``out'' vector space of operators
${\cal H}^{\rm op}_{\rm out}$, i.e. of $\Big ({\cal H}^{\rm op}_{\rm out}\Big )^{\otimes Q}$.
In other words, we can view ${\hat X}$ as an element of the tensor product of these two vector spaces
of operators:
\begin{eqnarray}
\label{LabelEqXhatinTensorProductOfOperatorVectorspaces}
&&
{\hat X}
\in
\Big ({\cal H}^{\rm op}_{\rm out}\Big )^{\otimes Q} \otimes \Big ({\cal H}^{\rm op}_{\rm in}\Big )^{\otimes Q}.
\end{eqnarray}

Let us now consider an unitary operator
${\hat W}_1$
acting
{(by conjugation)}
solely on ${\cal H}_{\rm out}$, while we act with the identity operator
${\bf I}$ on
${\cal H}_{\rm in}$. For such an operator we 
obtain\footnote{More explicity, the left hand side is the Haar average of
$U_{\bar{i}_1, j_1}
...
U_{\bar{i}_Q, j_Q}
U^\dagger_{\bar{j}_1, i_1}
...
U^\dagger_{\bar{j}_Q, i_Q}$~
$\big (
{\hat W}_1^{\otimes Q}|\bar{i}_1, ..., \bar{i}_Q\rangle \langle i_1, ..., i_Q|
{{\hat W}_1^\dagger} {}^{\otimes Q}
\big )$
$\otimes
\big (
|\bar{j}_1, ..., \bar{j}_Q\rangle \langle j_1, ..., j_Q|\big )$, 
equal to ${\hat X}$ upon letting $ U \to W_1 U $.}
\begin{eqnarray}
\label{LabelEqInvarianceIn}
&&
\Big ({{\hat W}_1}^{\otimes Q} \otimes {\bf I}^{\otimes Q}\Big )
{\hat X}
\Big (
{{{\hat W}_1}^{\dagger \otimes Q}} \otimes {\bf I}^{\otimes Q}
\Big )
= {\hat X}
\end{eqnarray}
which follows from  the invariance of the Haar measure under left-multiplication
of ${\hat U}$ 
with  ${\hat W}_1$.
Similarly, for a unitary operator
${\hat W}_2$ acting
{(by conjugation)}
solely on ${\cal H}_{\rm in}$,  while we act with the the identity operator on ${\cal H}_{\rm out}$,
we obtain
\begin{eqnarray}
\label{LabelEqInvarianceOut}
&&\Big ({\bf I}^{\otimes Q} \otimes {{{\hat W}_2}^{\dagger \otimes Q}}\Big )
{\hat X}
\Big (
{ {\bf I}^{\otimes Q} \otimes {{\hat W}_2}}^{\otimes Q}
\Big )
= {\hat X}
\end{eqnarray}
due to invariance of the Haar measure under right-multiplication
of ${\hat U}$ by ${\hat W}_2$.
Owing to (\ref{LabelEqInvarianceIn}) and (\ref{LabelEqInvarianceOut}),
the operator ${\hat X}$ can be expanded as a tensor product of operators 
${\hat R}(\sigma)$
which form a basis of 
the
commutant acting on $\Big ({\cal H}^{\rm op}_{\rm in}\Big )^{\otimes Q}$,
and 
operators
${\hat R}(\tau)$
which form a basis of the commutant acting on $\Big ({\cal H}^{\rm op}_{\rm out}\Big )^{\otimes Q}$, i.e.
\begin{eqnarray}
\label{LabelEqExpansionTensorProductCommutants}
{\hat X}
=
\sum_{\sigma, \tau \in S_Q}
\ \ 
\text{Wg}_{D}(\sigma, \tau) \ {\hat R}(\sigma) \otimes {\hat R}(\tau),
\end{eqnarray}
where $\text{Wg}_{D}(\sigma, \tau)$ are 
expansion
coefficients\footnote{{Which are real because ${\hat X}$ is Hermitian, and $\text{Wg}_{D}(\sigma, \tau) = \text{Wg}_{D}(\sigma^{-1}, \tau^{-1})$ .}}
called Weingarten functions.
-- Clearly, reading
(\ref{LabelEqDEF-X-hat})
in conjunction with (\ref{LabelEqExpansionTensorProductCommutants}),
(\ref{eqRmatrixHaarvertex-vertex-edge}), 
and (\ref{LabelEqDefActionOfPermutationsEdge}), equation (\ref{LabelEqDEF-X-hat})
is equivalent to the statement
(\ref{LabelEqWeingartenFunctionInComponents}), and constitutes a proof of the latter.

We close this discussion with a slightly  more compact, basis-independent formulation of the same facts.
In particular, if we apply ${\hat X}$ to an operator in
$\Big ({\cal H}^{\rm op}_{\rm in}\Big )^{\otimes Q}$ that is a $Q$-fold
tensor product of an operator ${\hat O}_{\rm in}$ which is an element in the Hilbert space
${\cal H}^{\rm op}_{\rm in}$ of ``in'' operators, we obtain the following result which
is an element of the $Q$-fold tensor product in the Hilbert space
${\cal H}^{\rm op}_{\rm out}$ of ``out'' operators:
\begin{eqnarray}
\label{LabelEq-X-hat-Mapping-SuperOperator}
&
{\hat X}: \ \Big ({\cal H}^{\rm op}_{\rm in}\Big )^{\otimes Q} &\to \Big ({\cal H}^{\rm op}_{\rm out}\Big )^{\otimes Q}
\\ 
\nonumber
& \left ({\hat O}_{\rm in}\right)^{\otimes Q}
&\to
{\hat X} \left[\left ({\hat O}_{\rm in}\right)^{\otimes Q} \right]
\end{eqnarray}
where
\begin{eqnarray}
\nonumber
{\hat X}\left[\left ({\hat O}_{\rm in}\right)^{\otimes Q} \right]
=\int_{\rm Haar} d \mu ({\hat U})
\ 
\big [{\hat U} {\hat O}_{\rm in} {\hat U}^\dagger\big ]^{\otimes Q}.
\end{eqnarray}
The operator ${\hat X}$ is (evidently) invariant under the action of independent unitary operators ${\hat W}_1$
and ${\hat W}_2$, implemented by right and left multiplication of  the (integrated) 
Haar unitary~${\hat U}$
\begin{eqnarray}
\nonumber
&& 
{\hat X} \left[\left ({\hat O}_{\rm in}\right)^{\otimes Q} \right]
=
{\hat X} \left[
{\hat W}_2^{\otimes Q} {\hat O}_{\rm in}^{\otimes Q} {{\hat W}_2^{\dagger \otimes Q}} \right]=
\\
\nonumber
&&
={\hat W}_1^{\otimes Q}
\left (
{\hat X} \left[\left ({\hat O}_{\rm in}\right)^{\otimes Q} \right]
\right )
{{\hat W}_1^{\dagger \otimes Q}}.
\end{eqnarray}
This is the symmetry giving rise to the expansion
(\ref{LabelEqExpansionTensorProductCommutants})
of ${\hat X}$
in terms of a tensor product of elements of the commutant.
The same logic will apply in the Clifford case, the only difference being that the
commutant will then be  different.

We note for future use the fact that we can apply the operator ${\hat X}$ to any element ${\hat R}(\tau)$
of the commutant, viewed as an element of the vector space of operators
$\Big ({\cal H}^{\rm op}_{\rm in}\Big )^{\otimes Q}$. Using the fact that ${\hat R}(\tau)$
commutes by definition with ${\hat U}^{\otimes Q}$, we see that
${\hat X}\left [{\hat R}(\tau)
\right] = {\hat R}(\tau)$ where,
by
(\ref{LabelEq-X-hat-Mapping-SuperOperator}),
the right hand side is an element of the vector space 
$\Big ({\cal H}^{\rm op}_{\rm out}\Big )^{\otimes Q}$.

\subsection{Key properties of  the Weingarten functions
$\text{Wg}_{D}(\sigma, \tau)$ for the Haar case}
\label{LabelEqKeyPropertiesWeingartenHaar}

Let us perform the trace over the second tensor factor
$\Big ({\cal H}^{\rm op}_{\rm in}\Big )^{\otimes Q}$
of ${\hat X}$
defined in
(\ref{LabelEqDEF-X-hat}) and (\ref{LabelEqXhatinTensorProductOfOperatorVectorspaces}); in the following we denote the trace over the 
1st\footnote{i.e. over $\Big ({\cal H}^{\rm op}_{\rm out}\Big )^{\otimes Q}$}
and 2nd\footnote{i.e. over $\Big ({\cal H}^{\rm op}_{\rm in}\Big )^{\otimes Q}$} tensor factor
by ${\rm tr}^{(1)}$
and
${\rm tr}^{(2)}$.
When denoting the identity operator in the Hilbert space underlying the first tensor factor,
 i.e. in ${\cal H}_{\rm out}$,
by ${\bf I}$ we obtain
for any permutation $\tau \in S_Q$ 
\begin{eqnarray}
\nonumber
&&
{\rm tr}^{(2)} {\hat X} \left (
{\bf I}^{\otimes Q} \otimes {\hat R(\tau) } 
\right ) =
{\hat X} \left [ {\hat R}(\tau)
\right]
={\hat R}(\tau)=
\\ \nonumber 
&&
=\sum_{\sigma_o, \sigma_i \in S_Q}
\ \ 
\text{Wg}_{D} (\sigma_o, \sigma_i) \ {\hat R}(\sigma_o) {\rm tr} \left[ {\hat R}^\dagger (\sigma_i) {\hat R}(\tau)  \right]=
\\ 
\label{LabelEqInverseOfWeingarten-Haar}
&&
=\sum_{\sigma_o, \sigma_i \in S_Q}
\ \ 
\text{Wg}_{D} (\sigma_o, \sigma_i) \ {\hat R}(\sigma_o) \  D^{Q - |\sigma_i,\tau |} \in
\\ \nonumber
&&
\in 
\Big ({\cal H}^{\rm op}_{\rm out}\Big )^{\otimes Q},
\end{eqnarray}
where in the first line we used the fact that ${\hat R}(\tau)$ is an element of the commutant
[and the definition 
(\ref{LabelEqDEF-X-hat}) of
${\hat X}$, as well as of
the metric,
(\ref{LabelEq-Haar-RTN-PartitionFunction}) and
(\ref{LabelEqHaarEdgeBoltzmannWeight})],
and in the second line we used (\ref{LabelEqExpansionTensorProductCommutants}); here $D=d^2$ is the dimension
of the Hilbert space
on which the 2-site unitary gate acts. Since ${\hat R}(\sigma_o)$ form a basis of the commutant, (\ref{LabelEqInverseOfWeingarten-Haar})
implies that
\begin{eqnarray}
\label{LabelEqLeftInverseHaar}
\sum_{\sigma_i}
\text{Wg}_{D} (\sigma_o, \sigma_i) \   D^{Q - |\sigma_i,\tau |} = \delta_{\sigma_o, \tau}.
\end{eqnarray}
This equation says that the Weingarten function is the left-inverse of $D^{Q - |\sigma_i,\tau |}$
under matrix multiplication when both are viewed as matrices.

The analogous argument
applied to the equation ${\rm tr}^{(1)} {\hat X} \left (
 {\hat R(\tau) \otimes{\bf I}^{\otimes Q}} 
\right ) 
=$ ${\hat R}(\tau)$ implies a similar equation for the right inverse,
\begin{eqnarray}
\label{LabelEqRightInverseHaar}
\sum_{\sigma_o}
D^{Q - |\tau,\sigma_o |} \ 
\text{Wg}_{D} (\sigma_o, \sigma_i) 
= \delta_{ \tau, \sigma_i}.
\end{eqnarray}

A simplification occurs in the Haar case because the metric satisfies
$|\sigma, \tau|=$ $|\sigma^{-1}\tau|$, i.e. it doesn't depend separately on
$\sigma$ and $\tau$.
This has two implications: (i) First, and most importantly, the Weingarten function, being the inverse of
$D^{Q-|\sigma,\tau|}=$ $D^{Q-|\sigma^{-1}\tau|}$ is also a function of the combined variable,
namely 
$\text{Wg}_{D} (\sigma_o, \sigma_i)=$ $ \text{Wg}_{D} (\sigma^{-1}_o \sigma_i)$.
(ii) Second, without loss of generality, we may replace $\tau \to 1$ by the identity permutation $1$ in
(\ref{LabelEqLeftInverseHaar}) and (\ref{LabelEqRightInverseHaar}).

\subsection{Consequences of (\ref{LabelEqHaarAverageOneVertex}) and
(\ref{LabelEqHaarSumRtau})}

Set 
$j_1=...=j_Q=\bar{j}_1= ... = \bar{j}_Q=1$ in those equations.  
If we define the projection operator onto this state in $({\cal H}_{\rm in})^{\otimes Q}$,
\begin{eqnarray}
\nonumber
 {\hat {\bf Q}^0} := 
 {|1\rangle}^{\otimes Q}
\ 
{\langle 1|}^{\otimes Q},
\end{eqnarray}
we conclude that
\begin{eqnarray}
\nonumber
&&
{\rm tr}^{(2)} 
\big [
{\hat X} 
\  
\left (
{\bf I}^{\otimes Q}
\otimes
{\hat {\bf Q}^0}
\right )
\big ]
=
\\ \nonumber
&&
=\int_{\rm Haar} d\mu({\hat U})
\ 
\big [{\hat U}|1\rangle
\ 
{\langle 1|}
{\hat U}^\dagger\big ]^{\otimes Q}=
\\ \nonumber
&&
= {\rm const.} \sum_{\tau \in S_Q} {\hat R}(\tau)
\end{eqnarray}
where we used (\ref{LabelEqHaarAverageOneVertex}), (\ref{LabelEqHaarSumRtau}).

On the other hand, using the form (\ref{LabelEqExpansionTensorProductCommutants})
for
${\hat X}$, the same expression is equal to
\begin{eqnarray}
\nonumber
=
\sum_{\sigma, \tau \in S_Q}
\ \ 
\text{Wg}_{D}(\sigma, \tau) \ {\hat R}(\sigma) 
\ \ 
\left (
{\rm tr} \big [
{\hat R}(\tau) \ {\hat {\bf Q}^0}
\big ]
\right )
\end{eqnarray}
Owing to
(\ref{eqRmatrixHaar})
the trace in the above equation is unity
independent\footnote{
Because the state $|1\rangle^{\otimes Q}$ is invariant under permutation of the $Q$ tensor factors.} 
of the element $\tau$ of the commutant $S_Q$,
\begin{eqnarray}
\nonumber
\left (
{\rm tr} \big [
{\hat R}(\tau) \ {\hat {\bf Q}^0}
\big ]
\right ) =1,
\quad {\rm for \ all} \ \tau\in S_Q.
\end{eqnarray}
We therefore conclude that
\begin{eqnarray}
\nonumber
&&
{\rm const.} \sum_{\sigma \in S_Q} {\hat R}(\sigma)
=
\\ \nonumber
&&
=\sum_{\sigma \in S_Q}
\ \ 
\left (
\sum_{\tau\in S_Q}
 \text{Wg}_{D} (\sigma, \tau)
\right )
\ {\hat R}(\sigma) 
\end{eqnarray}
which means that the expression in parenthesis is a constant,
{i.e.} independent of $\sigma$,
\begin{eqnarray}
\label{LabelEqConstraintWeingartenCoeffClifford-from-StabilizeHaar}
\sum_{\tau \in S_Q}
\text{Wg}_{D}(\sigma, \tau) 
= {\rm const.}, \ \ {\rm for \ all} \ \sigma\in S_Q.
\qquad \ \ \ 
\end{eqnarray}
Of course, here in the Weingarten case for Haar under consideration here, $\text{Wg}_{D}(\sigma, \tau)$
is a function of only $\sigma\tau^{-1}$ and so this condition is obviously satisfied.

\section{Weingarten 
functions for Clifford} \label{AppendixWClifford}

Let us start with the analog of (\ref{LabelEqDEF-X-hat}) for Clifford:
\begin{eqnarray}
\label{LabelEqDEF-Y-hat}
&&{\hat Y} :=
\mathbb{E}_{{\hat V}\in {\rm Cliff}(n,p)}
\big [{\hat V} (...) {\hat V}^\dagger\big ]^{\otimes Q}=
\\ \nonumber
&&
={1\over {\cal N}_{\rm Cliff}}
\sum_{{\hat V} \in {\rm Cliff}(n,p)}
\ 
\big [{\hat V} (...) {\hat V}^\dagger\big ]^{\otimes Q}=
\\ \nonumber
&&
={1\over {\cal N}_{\rm Cliff}}
\sum_{{\hat V} \in {\rm Cliff}(n,p)} 
V_{\bar{i}_1, j_1}
...
V_{\bar{i}_Q, j_Q}
V^\dagger_{\bar{j}_1, i_1}
...
V^\dagger_{\bar{j}_Q, i_Q} \times
\qquad 
\\ \nonumber
&&
\times
\big (
|\bar{i}_1, ..., \bar{i}_Q\rangle \langle i_1, ..., i_Q|
\big )
\otimes
\big (
|\bar{j}_1, ..., \bar{j}_Q\rangle \langle j_1, ..., j_Q|
\big ),
\\ \nonumber
&&
{\rm where}
\\ \nonumber
&&
{\hat Y}
\in
\Big ({\cal H}^{\rm op}_{\rm out}\Big )^{\otimes Q} \otimes \Big ({\cal H}^{\rm op}_{\rm in}\Big )^{\otimes Q}.
\end{eqnarray}
Note that here $|i\rangle \in {\cal H}_n = ({\bf C}_p)^n$ 
{denotes elements of}
an {\it orthonormal basis}
of ${\cal H}_n$, having  no  relationship with `stabilizer states' in ${\cal H}_n$
(but see
Sect.
\ref{LabelSubSectionConnectionWithStabilizerStates} below).
Now, the analogues of 
(\ref{LabelEqInvarianceIn})
and (\ref{LabelEqInvarianceOut}) follow upon the replacement ${\hat X}\to {\hat Y}$,
for 
any {\it Clifford}
unitary operators ${\hat W}_1$ and ${\hat W}_2$. Furthermore, the analog of
(\ref{LabelEqExpansionTensorProductCommutants}) for the expansion of ${\hat Y}$
as a tensor product of operators in the commutant goes 
through in the same way as for Haar, leading to
\begin{eqnarray}
\label{LabelEqExpansionTensorProductCommutantsClifford}
{\hat Y}
=
\sum_{T_o, T_i \in \Sigma_Q(p)}
\ \ 
\text{Wg}_{D} (T_o, T_i) \ {\hat R}(T_o) \otimes {\hat R}(T_i),
\qquad
\end{eqnarray}
where $\text{Wg}_{D}(T_o, T_i)$ are generalized Weingarten functions (see also Ref.~\cite{PhysRevLett.121.170502} for explicit expressions in the case of $Q=4$ replicas).


\subsection{Key properties of the Weingarten functions
$\text{Wg}_{D}(T_o, T_i)$ for the Clifford case}

We proceed in complete analogy with the Haar case, discussed in Sect.
\ref{LabelEqKeyPropertiesWeingartenHaar}.
For any element of the commutant $T \in \Sigma_Q(p)$, we have
\begin{eqnarray}
\nonumber
&&
{\rm tr}^{(2)} {\hat Y} \left (
{\bf I}^{\otimes Q} \otimes {\hat R(T) } 
\right ) 
={\hat R}(T)=
\\ \nonumber 
&&
=\sum_{T_o, T_i \in \Sigma_Q(p)}
\ \ 
\text{Wg}_{D} (T_o, T_i) \ {\hat R}(T_o) {\rm tr} \left[ {\hat R}^\dagger (T_i) {\hat R}(T)  \right]=
\\ 
\label{LabelEqInverseOfWeingarten-Clifford}
&&
=\sum_{T_o, T_i \in \Sigma_Q(p)}
\ \ 
\text{Wg}_{D} (T_o, T_i) \ {\hat R}(T_o)  D^{Q - |T_i, T |},
\end{eqnarray}
where in the first line we used the fact that $T$ belongs to the commutant. We thus have
\begin{equation}
\label{LabelEqCliffWeinLeftInverse}
\sum_{ T_i \in \Sigma_Q(p)}
\ \ 
\text{Wg}_{D} (T_o, T_i) \  D^{Q - |T_i, T |} = \delta_{T_o, T}.
\end{equation}
The last equation says that the Weingarten function
is the left-inverse of $D^{Q-|T_i, T|}$ (both viewed as a matrix).
Similarly, by acting in the first line of
(\ref{LabelEqInverseOfWeingarten-Clifford})
with ${\hat R}(T)$ on the first tensor factor and performing the trace over the first,
we get a similar equation for the right-inverse.

\subsection{Connection with Stabilizer States - consequences of
(\ref{LabelEqSumOverClifford})}
\label{LabelSubSectionConnectionWithStabilizerStates}
To this end, we 
pick an orthonormal basis in $({\bf C}^p)^n$
so that one element of this basis is the state $|0\rangle^{\otimes n} \in {\cal H}_n = ({\bf C}^p)^n$
where $0$ denotes the number zero in the finite number field ${\bf F}_p$; this is
the state that is described in items (6) and (7) of
Appendix \ref{LabelAppendixPropertiesCliffordGroupStabilizerStates}. Having done so,
we use this orthonormal basis in
(\ref{LabelEqDEF-Y-hat})
and set 
$j_k=j^0_k$, $\bar{j}_k = \bar{j^0}_k$ for $k=1, ..., Q$, where
$j^0_1=...=j^0_Q=\bar{j^0}_1= ... = \bar{j^0}_Q=0^{\otimes n}$.  (In better notation,
$|j^0_1\rangle= ... = |j^0_Q\rangle =$ $|0\rangle^{\otimes n}$, etc..)

If we define the projection operator onto this state in $({\cal H}_{\rm in})^{\otimes Q}$
\begin{eqnarray}
\nonumber
 {\hat {\bf P}^0} := 
 \Big [{|0\rangle}^{\otimes n}\Big]^{\otimes Q}
\ \ \ 
\Big [{\langle 0|}^{\otimes n}\Big]^{\otimes Q},
\end{eqnarray}
we conclude that
\begin{eqnarray}
\nonumber
&&
{\rm tr}^{(2)} 
\big [
{\hat Y} 
\  
\left (
{\bf I}^{\otimes Q}
\otimes
{\hat {\bf P}^0}
\right )
\big ]
=
\\ \nonumber
&&
={1\over {\cal N}_{\rm Cliff}}
\sum_{{\hat V} \in {\rm Cliff}(n,p)}
\ 
\big [{\hat V}(|0\rangle^{\otimes n})
\ \ 
({\langle 0|}^{\otimes n})
{\hat V}^\dagger\big ]^{\otimes Q}=
\\ \nonumber
&&
= {\rm const.} \sum_{T \in \Sigma_Q(p)} {\hat R}(T)
\end{eqnarray}
where we used (\ref{LabelEqSumOverClifford}).

On the other hand, using the form (\ref{LabelEqExpansionTensorProductCommutantsClifford}) for
${\hat Y}$, the same expression is equal to
\begin{eqnarray}
\nonumber
=
\sum_{T_o, T_i \in \Sigma_Q(p)}
\ \ 
\text{Wg}_{D}(T_o, T_i) \ {\hat R}(T_o) 
\ \ 
\left (
{\rm tr} \big [
{\hat R}(T_i) \ {\hat {\bf P}^0}
\big ]
\right ).
\end{eqnarray}
Owing to Eq. (4.10) of \cite{GrossEtAlCMP2021}, the trace in the above equation is unity
independent of the element $T_i$ of the commutant $\Sigma_Q(p)$,
\begin{eqnarray}
\nonumber
\left (
{\rm tr} \big [
{\hat R}(T_i) \ {\hat {\bf P}^0}
\big ]
\right ) =1,
\quad {\rm for \ all} \ T_i\in\Sigma_Q(p).
\end{eqnarray}
We therefore conclude that
\begin{eqnarray}
\nonumber
&&
{\rm const.} \sum_{T \in \Sigma_Q(p)} {\hat R}(T)
=
\\ \nonumber
&&
=\sum_{T_o\in \Sigma_Q(p)}
\ \ 
\left (
\sum_{T_i \in \Sigma_Q(p)}
\text{Wg}_{D} (T_o, T_i)
\right )
\ {\hat R}(T_o) 
\end{eqnarray}
which means that the expresssion in parenthesis is a constant, independent of $T_o$,
\begin{eqnarray}
\label{LabelEqConstraintWeingartenCoeffClifford-from-Stabilizer}
\sum_{T_i \in \Sigma_Q(p)}
\text{Wg}_{D} (T_o, T_i) 
= {\rm const.}, \ \ {\rm for \ all} \ T_o\in \Sigma_Q(p).
\qquad \ \ \ 
\end{eqnarray}
This is the analog of (\ref{LabelEqConstraintWeingartenCoeffClifford-from-StabilizeHaar})
of the Haar case, where it is tied to the group structure of the commutant.

\subsection{Generalization to other groups}

Note that the generalized Weingarten formulas of the previous section do not rely on any specific property of the commutant of the Clifford group. As such, most of our results can be generalized to derive statistical mechanics models for RTNs and monitored quantum circuits involving averages over other subgroups of the unitary group, expressed 
solely in terms of the commutant of such subgroups. Once this algebraic object, the commutant, is known, the link and Weingarten weights derived for the Clifford group carry over without any change, with the ``spins'' $T_i$ now belonging to this new commutant. Previously, Weingarten functions had been derived for unitary~\cite{Collins1}, orthogonal and symplectic groups~\cite{Collins2,Collins3} --- See Ref.~\cite{2021arXiv210914890C} for a recent review. 
The formulas in 
the present appendix are completely general, and extend the notion of the Weingarten function to all situations for which the commutant is known.   

\section{Metric on the commutant of the Clifford group }
\label{appendixMetric}

The Boltzmann weights of the Clifford Stat Mech model rely on the existence of a 
{\it metric}
on the commutant: for any two elements $T_a$ and $T_b$ of the commutant $\Sigma_Q(p)$,
there exists a {\it metric} 
which arises from the trace
\begin{eqnarray}
\label{LabelEq-MetricRTC-Clifford2}
W(T_a, T_b) := {\rm tr} \left [
{{\hat R}^\dagger}(T_a)
{\hat R}(T_b) 
\right ]= (p^{N})^{Q- |T_a, T_b|}, \ \ \ \  
\end{eqnarray}
for any $T_a, T_b \in \Sigma_Q(p)$, 
where ${\hat R}(T)$ is defined in (\ref{LabelEq-DEF-R-T}).
Our main results rely only on the standard properties satisfied by this metric: $|T_a, T_b| \geq 0,$
$|T_a, T_b|= |T_b, T_a|$, and $|T_a, T_b|=0$ if and only if $T_a=T_b$. In this appendix, we provide additional details on the definition of this metric following closely Ref.~\onlinecite{GrossEtAlCMP2021}. 

Consider the double coset representation of the commutant defined in eq.~\eqref{LabelEq-DoubleCosetDecomposition}. The simplest elements are the orthogonal group elements themselves, $T_1=O_1, T_2=O_2 \in {\cal O}_Q(p)$, 
which, as noted in the line below\eqref{LabelEq-DEF-Delta}, correspond to the first element of the disjoint union of double cosets in~\eqref{LabelEq-DoubleCosetDecomposition}: When applied to those, ${\hat r}$ is  a representation of the group ${\cal O}_Q(p)$ [see our eq. (A1)], and thus we have eq.~\eqref{LabelEq-Clifford-edge-Metric}. 
When $O_1=O_2$, we have 
$ \langle {\vec \mu} | (O_1^t O_2) |{\vec \mu}\rangle =1$ for all ${\vec \mu}$, thus
$e^{- \ln (p) \ |O_1, O_2|}=1$ and $|O_1, O_2|=0$. When $O_1 \not = O_2$, we have 
$ \langle {\vec \mu} | (O_1^t O_2) |{\vec \mu}\rangle <1$ for at least some ${\vec \mu}$, thus
$e^{- \ln (p) \ |O_1, O_2|} < 1$ and $|O_1, O_2|> 0$.

Now consider two general elements $T_1, T_2$ of the commutant  $\Sigma_Q(p)$ (which may or may not be elements of the same double coset). According to 
Proposition 4.17, as well as eq.~(4.18) and Lemma 4.18 of Ref.~\onlinecite{GrossEtAlCMP2021}, we can transform those by right and left multiplication with group elements in ${\cal O}_Q(p)$ to elements ${T'}_1$ and ${T'}_1$,
$$
T_j \to {T'}_j \equiv O_j T_j {O'}_j,  \ \ {\rm with} \ 
O_j, {O'}_j \in {\cal O}_Q(p),
\ \  (j=1,2)
$$
so that  the right and left defect subspaces
{${T'}_{j,LD}$ and ${T'}_{j, RD}$}
of each ${T'}_j$ 
become equal, ${N'}_j \equiv
 {T'}_{j, LD} ={T'}_{j, RD}$, {\it i.e.}
\begin{eqnarray}
\label{LabelEqLeftRightDefectEqual}
O_j \  T_{j,LD} =  {O'}^t_j \ T_{j, RD} \ 
 \qquad (j=1,2).
\end{eqnarray}
\vskip .2cm
Since the last equality in eq. (\ref{LabelEqLeftRightDefectEqual}) above can be achieved  by choosing\footnote{
letting $O_j \to {\mathbf 1}$, and ${O'}_j^t \to {O_j}^t \ {O'}_j^t= ({O'}_j \ {O_j})^t$}
for example ${O}_1 = {O}_2 = {\mathbf 1}$, let us make that choice so that
\begin{eqnarray}
\label{LabelEqLeftRightDefectEqualSimplified}
{T'}_1 =  T_1 \  {\cal O'}_1, \quad {T'}_2= T_2 \ {O'}_2. 
\end{eqnarray}
Then, using  eq.~\eqref{LabelEq-Representation-Property} we get for the trace in~\eqref{LabelEq-Clifford-edge-Metric}
\begin{eqnarray}
\label{LabelEqTraceForMetric}
{\rm tr} \left (
[{\hat r}(T_1)]^\dagger \ {\hat r}(T_2)
\right )
&=
{\rm tr} \left (
{\hat r}({O'}_1)\ 
[{\hat r}({T'}_1)]^\dagger \  \ {\hat r}({T'}_2) \ {\hat r}({O_2'}^t)
\right ) \notag \\
& \qquad = 
tr \left (
{\hat r}({O_2'}^t {O'}_1)\ 
[{\hat r}({T'}_1)]^\dagger \  \ {\hat r}({T'}_2)
\right ). \qquad \qquad
\end{eqnarray}

\vskip .1cm
Now, owing to (\ref{LabelEqLeftRightDefectEqual})
above and Theorem 4.24 of Ref.~\onlinecite{GrossEtAlCMP2021}, the operators ${\hat r}({T'}_j)$ are both proportional to (CSS)  projectors,
\begin{eqnarray}
\nonumber
{\hat r}({T'}_j)
= p^{{\rm dim}  {N'}_j} \ P_{{\rm CSS}({N'}_j)}.
\end{eqnarray}
Note in particular their Hermiticity, especially $[{\hat r}({T'}_1)]^\dagger=$ ${\hat r}({T'}_1)$. Next, eq. (4.24) of Ref.~\onlinecite{GrossEtAlCMP2021} proves the result
\begin{equation}
\label{LabelEqSemigroup}
[{\hat r}({T'}_1)]^\dagger  {\hat r}({T'}_2)
=
{\hat r}({T'}_1)  {\hat r}({T'}_2)
=p^{{\rm dim}  ({N'}_1\cap {N'}_2)} {\hat r} ({T'}_1  {T'}_2)
\end{equation}
which (at the same time)  defines the meaning of the product of ${T'}_1$ and ${T'}_2$ on the right hand side (the ``semigroup property'').

Using these  results, the trace in eq.~(\ref{LabelEqTraceForMetric}) above
then leads to the following expression for the metric $e^{-\ln(p) \ |T_1, T_2|}$
\begin{eqnarray}
\label{LabelEqMetricGeneralCase}
e^{-\ln(p)  [Q-{\rm dim}  ({N'}_1\cap {N'}_2)]}
{\rm tr}
\left (
{\hat r}({O_2'}^t {O'}_1) \
{\hat r} ({T'}_1  {T'}_2)
\right ). \ \  \ 
\end{eqnarray}
Note that in the special case  where $T_1={T'}_1$ and $T_2={T'}_2$, i.e. when both are already  proportional to (CSS) projectors, thus when
both already have equal right and left defect 
subspaces, we also have
${O'}_j={\mathbf 1}$ for $j=1,2$ (see, e.g.,
eq.(\ref{LabelEqLeftRightDefectEqualSimplified})).
In this case, using
Lemma 4.25 of Ref.~\onlinecite{GrossEtAlCMP2021}, we 
find\footnote{in particular, using the unnumbered equation below eq. (4.26) of Ref.~\onlinecite{GrossEtAlCMP2021}}
\begin{eqnarray}
\nonumber
|T_1, T_2|=
[Q - {\rm dim} (N_1 \cap N_2) - {\rm dim} (N_1^\perp \cap N_2^\perp)]
\end{eqnarray}
and we have
$|T_1, T_2| \geq 0$ since $(N_1 \cap N_2)^\perp \supseteq N_1^\perp + N_2^\perp \supseteq N_1^\perp \cap N_2^\perp$.

\vskip .3cm

In the general case, eq.(\ref{LabelEqMetricGeneralCase}) above, where ${\hat r}({O'}_2^t {O'}_1)$ is not the identity, 
a closed form expression is presently
{not known, even though an explicit  elementary expression for the metric is 
given by (\ref{LabelEqElementaryDEFMetric})
and (\ref{LabelEq-Clifford-edge-Metric}),
as already mentioned below (\ref{LabelEqElementaryDEFMetric}).}
Finally, we note that since the permutation group $S_Q$ is a 
subgroup\footnote{recall, 
e.g., \eqref{LabelEqSubSpace-T-for-Permutation}}  
of ${\cal O}_Q(p)$, it follows from its definition that the metric between two permutations is 
the same as that familiar from the Haar circuits and expressable in terms of the cycle-counting function, a property used in the main text.

\section{Some details on multifractal scaling}
\label{LabelAppendixSectionDetailsMultifractal}



In this Appendix we present more details on the multifractal scaling of the purity, discussed in Sect.~\ref{LabelSubSectionMultifractalScaling}. 
This is a special case of the general discussion of the scaling of a correlation function (at disorder dominated critical points) in disordered systems, as discussed in Refs.~\cite{LudwigHierarchies1990,DuplantierLudwigPRL1991,Pixley2022Multifractality}.

We start from the scaling of all moments of the purity at the transition, Eq.~(\ref{LabelEqScalingOfMoments}),
 \begin{eqnarray}
 \label{LabelEqScalingOfMomentsCOPY}
\overline{
[G_(x_1, x_2) ]^k
}
\sim
{B_k\over
R_{12}^{2 X_k}},
\end{eqnarray}
where the purity in a fixed realization of circuit disorder is a random variable, viewed as a 2-point correlation function
$G(x_1, x_2)=$~${\rm tr} \left ( {\hat \rho}_{A,\mathbf{m}} \right )^2$ associated with the boundary condition changing (bcc) operator determining entanglement properties~\cite{VasseurPotterRTN2018,JianVasseurMeasurement2019}.
Here $R_{12}$ is the chord distance
\begin{eqnarray}
\nonumber
R_{12} := {L\over \pi} \sin\left (
{\pi\over L} |x_1-x_2|
\right )
=
{L\over \pi} \sin\left (
{\pi\over L} |A|
\right ).
 \end{eqnarray}
Since upon recalling  (\ref{LabelEqGsecondRenyiEntropy}), 
$\ln G(x_1, x_2)=-
S_A^{(2)}$, the cumulant expansion of the left hand side of 
$(\ref{LabelEqScalingOfMomentsCOPY})$
reads
\begin{eqnarray}
\label{LabelEqAppCumulantExpansion}
&&
\overline{
[G_(x_1, x_2) ]^k
}=
\\ \nonumber
&&
=
\exp
\bigl \{
-k \  \overline{S_A^{(2)}}
+
{k^2\over 2!}
\
\overline{
\left (
S_A^{(2)} - \overline{S_A^{(2)}}
\right )^2
}
-
{k^3 \over 3!} \ 
\kappa_3[S_A^{(2)}]
+\dots
\bigr \}
\end{eqnarray}
where
$\kappa_3[S_A^{{(2)}}]$ denotes the 3rd cumulant of the random variable $S_A^{(2)}$.
We compare this with the right hand side
of
$(\ref{LabelEqScalingOfMomentsCOPY})$, using the Taylor expansion
(\ref{LabelTaylorExpansionXN})
of the exponents $X_k$ in powers of $k$,
\begin{eqnarray}
\label{LabelTaylorExpansionXNCOPYApp}
X_k = k  \ x^{(1)} +{k^2 \over 2!} \  x^{(2)}
+ {k^3\over 3!} \ x^{(3)} + \dots,
\end{eqnarray}
yielding
\begin{eqnarray}
\label{LabelEqExpansionrhsCOPYApp}
&&
B_k 
\exp \{
- 2 X_k \ln R_{12}
\}
=
\\ \nonumber
&&
=
B_k \ 
\exp \{
- \bigl [ k \  2 x^{(1)}
+
{k^2\over 2!} \ 2 x^{(2)}
+
{k^3\over 3!} \ 2 x^{(3)}
+ ...
\bigr ]\ln R_{12}
\}
\end{eqnarray}
Comparison of
(\ref{LabelEqAppCumulantExpansion}) with (\ref{LabelEqExpansionrhsCOPYApp}) shows that all cumulants grow\footnote{\andreas{The analogous expansion of $B_k=$~$\exp\{ k\ b_1 +(k^2/2!) \ b_2 + ...\}$ shows that
a non-universal constant offset $(-1)^k \ b_k$ appears in the $k$-th cumulant $\kappa_k[S_A^{(2)}]$ of the entanglement entropy $S_A^{(2)}$, but these offsets will of course just represent subleading contributions in the regime of interest of large values of $R_{12}$.}} proportional to $\ln R_{12}$, the constants of proportionality being universal and equal to $(-1)^{k-1}$  times twice the expansion coefficients $x^{(k)}$ appearing the Taylor expansion
(\ref{LabelTaylorExpansionXNCOPYApp}),
\begin{eqnarray}
\label{LabelFirstCumulantCOPYApp}
\overline{
S_A^{(2)}}
\ && \ 
\sim  \ 2 \ x^{(1)} \ln R_{12}
\\ 
\label{LabelSecondCumulantCOPYApp}
\overline{
\left (
S_A^{(2)} - \overline{S_A^{(2)}}
\right )^2
}
\ && \ 
\sim \ -  2  \ x^{(2)} \ln R_{12}
\\ 
\label{LabelThirdCumulantCOPYApp}
\kappa_3[S_A^{(2)}]
\ && \ 
\sim \ 
2 \ x^{(3)} \ln R_{12}
\\ \nonumber
&& \dots,
\end{eqnarray}
These are Eq.s~(\ref{LabelFirstCumulant})-(\ref{LabelThirdCumulant}) of the main text.

\newpage
\bibliography{references}

\end{document}